\PassOptionsToPackage{unicode}{hyperref}
\PassOptionsToPackage{hyphens}{url}
\PassOptionsToPackage{dvipsnames,svgnames,x11names}{xcolor}
\documentclass[
  12pt]{article}

\usepackage{amsmath,amssymb}
\usepackage{iftex}
\ifPDFTeX
  \usepackage[T1]{fontenc}
  \usepackage[utf8]{inputenc}
  \usepackage{textcomp} 
\else 
  \usepackage{unicode-math}
  \defaultfontfeatures{Scale=MatchLowercase}
  \defaultfontfeatures[\rmfamily]{Ligatures=TeX,Scale=1}
\fi
\usepackage{lmodern}
\ifPDFTeX\else  
\fi
\IfFileExists{upquote.sty}{\usepackage{upquote}}{}
\IfFileExists{microtype.sty}{
  \usepackage[]{microtype}
  \UseMicrotypeSet[protrusion]{basicmath} 
}{}

\usepackage{xcolor}
\makeatletter
\ifx\paragraph\undefined\else
  \let\oldparagraph\paragraph
  \renewcommand{\paragraph}{
    \@ifstar
      \xxxParagraphStar
      \xxxParagraphNoStar
  }
  \newcommand{\xxxParagraphStar}[1]{\oldparagraph*{#1}\mbox{}}
  \newcommand{\xxxParagraphNoStar}[1]{\oldparagraph{#1}\mbox{}}
\fi
\ifx\subparagraph\undefined\else
  \let\oldsubparagraph\subparagraph
  \renewcommand{\subparagraph}{
    \@ifstar
      \xxxSubParagraphStar
      \xxxSubParagraphNoStar
  }
  \newcommand{\xxxSubParagraphStar}[1]{\oldsubparagraph*{#1}\mbox{}}
  \newcommand{\xxxSubParagraphNoStar}[1]{\oldsubparagraph{#1}\mbox{}}
\fi
\makeatother

\usepackage{longtable,booktabs,array}
\usepackage{calc} 
\usepackage{etoolbox}
\makeatletter
\patchcmd\longtable{\par}{\if@noskipsec\mbox{}\fi\par}{}{}
\makeatother
\usepackage{graphicx}
\makeatletter
\def\maxwidth{\ifdim\Gin@nat@width>\linewidth\linewidth\else\Gin@nat@width\fi}
\def\maxheight{\ifdim\Gin@nat@height>\textheight\textheight\else\Gin@nat@height\fi}
\makeatother
\makeatletter
\def\fps@figure{htbp}
\makeatother

\usepackage[margin=1in]{geometry}

\makeatletter
\@ifpackageloaded{caption}{}{\usepackage{caption}}
\AtBeginDocument{%
\ifdefined\contentsname
  \renewcommand*\contentsname{Table of contents}
\else
  \newcommand\contentsname{Table of contents}
\fi
\ifdefined\listfigurename
  \renewcommand*\listfigurename{List of Figures}
\else
  \newcommand\listfigurename{List of Figures}
\fi
\ifdefined\listtablename
  \renewcommand*\listtablename{List of Tables}
\else
  \newcommand\listtablename{List of Tables}
\fi
\ifdefined\figurename
  \renewcommand*\figurename{Figure}
\else
  \newcommand\figurename{Figure}
\fi
\ifdefined\tablename
  \renewcommand*\tablename{Table}
\else
  \newcommand\tablename{Table}
\fi
}
\@ifpackageloaded{float}{}{\usepackage{float}}
\floatstyle{ruled}
\@ifundefined{c@chapter}{\newfloat{codelisting}{h}{lop}}{\newfloat{codelisting}{h}{lop}[chapter]}
\floatname{codelisting}{Listing}

\makeatother
\makeatletter
\@ifpackageloaded{caption}{}{\usepackage{caption}}
\@ifpackageloaded{subcaption}{}{\usepackage{subcaption}}
\makeatother

\ifLuaTeX
  \usepackage{selnolig}  
\fi
\usepackage[]{natbib}
\usepackage{bookmark}

\IfFileExists{xurl.sty}{\usepackage{xurl}}{} 
\usepackage{xcolor}
\definecolor{mutedblue}{RGB}{30, 100, 200} 

\hypersetup{
  pdftitle={Title},
  pdfauthor={Author 1; Author 2},
  pdfkeywords={3 to 6 keywords, that do not appear in the title},
  colorlinks=true,
  linkcolor={mutedblue},
  filecolor={mutedblue},
  citecolor={mutedblue},
  urlcolor={mutedblue},
  pdfcreator={LaTeX via pandoc}}

\newcommand{\ms}{\mathsmaller}
\newcommand{\squish}[1]{\!#1\!}
\newcommand{\tm}{\squish{-}}
\newcommand{\txm}{\textrm{-}}
\newcommand{\txp}{\textrm{+}}
\newcommand{\tp}{\squish{+}}
\newcommand{\teq}{\squish{=}}
\newcommand{\Zrd}{Z^{\ms{(r,D)}}}
\newcommand{\frd}{f^{\ms{(r,D)}}}
\newcommand{\Frd}{F^{\ms{(r,D)}}}

\newcommand{\Poisrd}{\textrm{Pois}_\mu^{\ms{(r,D)}}}
\newcommand{\Pois}{\textrm{Pois}}
\newcommand{\defeq}{\overset{\textrm{def}}{=}}

\newcommand{\iidsim}{\overset{\textrm{iid}}{\sim}}
\newcommand{\indsim}{\overset{\textrm{ind.}}{\sim}}

\newcommand{\pdgey}{p_d^{\ms{(>Y)}}}
\newcommand{\pdley}{p_d^{\ms{(<Y)}}}
\newcommand{\pdeqy}{p_d^{\ms{(=Y)}}}
\newcommand{\ngey}{N^{\ms{(>Y)}}}
\newcommand{\nley}{N^{\ms{(<Y)}}}

\newcommand{\neqy}{N^{\ms{(=Y)}}}
\newcommand{\ndgey}{N_d^{\ms{(>Y)}}}
\newcommand{\ndley}{N_d^{\ms{(<Y)}}}
\newcommand{\ndeqy}{N_d^{\ms{(=Y)}}}
\newcommand{\ndpgey}{N_{d+1}^{\ms{(>Y)}}}
\newcommand{\ndpley}{N_{d+1}^{\ms{(<Y)}}}
\newcommand{\ndpeqy}{N_{d+1}^{\ms{(=Y)}}}

\newcommand{\Zrmdm}{Z^{\ms{(r\txm 1,\, D \txm1)}}}
\newcommand{\Zrdm}{Z^{\ms{(r,\, D \txm 1)}}}

\newcommand{\Zdd}{Z^{\ms{(D,\, D)}}}
\newcommand{\Zdmdm}{Z^{\ms{(D \txm 1,\, D \txm1)}}}

\newcommand{\Zoned}{Z^{\ms{(1,\,D)}}}
\newcommand{\Zonedm}{Z^{\ms{(1,\,D \txm 1)}}}
\newcommand{\Zoneone}{Z^{\ms{(1,\,1)}}}

\newcommand{\Frmdm}{F_\theta^{\ms{(r\txm 1,\, D\txm1)}}}
\newcommand{\Frdm}{F_\theta^{\ms{(r,\, D\txm1)}}}
\newcommand{\Frmd}{F_\theta^{\ms{(r\txm1,\, D)}}}

\newcommand{\Fdmdm}{F_\theta^{\ms{(D\txm 1,\,D\txm 1)}}}
\newcommand{\Frmdkm}{F_\theta^{\ms{(r\txm1,\, D\txm k \txm 1)}}}
\newcommand{\Frdk}{F_\theta^{\ms{(r,\, D\txm k)}}}
\newcommand{\Frdkm}{F_\theta^{\ms{(r,\, D\txm k\txm 1)}}}

\newcommand{\ex}{\operatorname{\mathbb{E}}}

\def\ind{\mathbb{1}}

\usepackage{algorithm}
\usepackage{algorithmic}

\counterwithin{algorithm}{section}
\usepackage{setspace} 

\usepackage[capitalize,nameinlink]{cleveref}
\crefrangelabelformat{equation}{(#3#1#4--#5#2#6)}
\usepackage{nicefrac}
\usepackage{relsize}

\newtheorem{theorem}{Theorem}[section]         
\newtheorem{lemma}{Lemma}[section]             
\newtheorem{proposition}{Proposition}[section] 
\newtheorem{corollary}{Corollary}[section]
\newtheorem{fact}{Fact}[section]

\newtheorem{remark}{Remark}[section]

\usepackage{bbold}

\usepackage{hyperref}
\usepackage{lipsum}

\crefname{section}{section}{sections} 
\crefname{appendix}{appendix}{appendices}
\crefalias{appsec}{section}

\usepackage{chngcntr}
\usepackage{authblk} 

\title{\bf Modeling Latent Underdispersion\\with Discrete Order Statistics}
\author[1]{Jimmy Lederman}
\author[1]{Aaron Schein}
\affil[1]{Department of Statistics, University of Chicago}

\begin{document}

\def\spacingset#1{\renewcommand{\baselinestretch}%
{#1}\small\normalsize} \spacingset{1}


\maketitle

\bigskip
\begin{abstract}
The Poisson distribution is the default choice of likelihood for probabilistic models of count data. However, due to the equidispersion contraint of the Poisson, such models may have predictive uncertainty that is artificially inflated. While overdispersion has been extensively studied, conditional underdispersion---where latent structure renders data more regular than Poisson---remains underexplored, in part due to the lack of tractable modeling tools. We introduce a new class of models based on discrete order statistics, where observed counts are assumed to be an order statistic (e.g., minimum, median, maximum) of i.i.d.~draws from some discrete parent, such as the Poisson or negative binomial. We develop a general data augmentation scheme that is modular with existing tools tailored to the parent distribution, enabling parameter estimation or posterior inference in a wide range of such models. We characterize properties of Poisson and negative binomial order statistics, exposing interpretable knobs on their dispersion. We apply our framework to four case studies---i.e., to commercial flight times, COVID-19 case counts, Finnish bird abundance, and RNA sequencing data---and illustrate the flexibility and generality of the proposed framework. Our results suggest that order statistic models can be built, used, and interpreted in much the same way as commonly-used alternatives, while often obtaining better fit, and offer promise in the wide range of applications in which count data arise.
\end{abstract}

\noindent%
\vfill

\newpage
\spacingset{1.6} 

\section{Introduction}\label{sec-intro}

Since the early work of~\citet{von1898gesetz} on deaths by horse-kick in the Prussian cavalry, the Poisson distribution has served as the canonical model for frequencies of “pure chance” events. Such frequencies are characterized by \textit{equidispersion}, where their mean and variance are equal. Departures from Poisson behavior---most commonly in the form of \textit{overdispersion}---are then often taken as evidence of latent structure or unobserved heterogeneity. This view has become standard in applied statistics, motivating the widespread use of alternatives like the negative binomial~\citep{greenwood1920inquiry}, which accommodate variability beyond the Poisson, and the frequent textbook advice to check for overdispersion before analyzing count data (e.g.,~\citet{mccullagh1989generalized}).

Much less attention has been paid to the phenomenon of \textit{underdispersion}. Yet just as overdispersion might arise from clustering or rate heterogeneity, among other aspects of events that do not occur by ``pure chance'', underdispersion can arise from regularity or repulsion, as with the periodic timing of subscription purchases~\citep{platzer2016ticking}, or the ``inhibitory dependence''~\citep{cox1966statistical} commonly observed in the spatial distribution of competing species~\citep{clark1954distance,hosie1969native}. While documented in several domains, understanding the types of data-generating mechanisms that give rise to underdispersion is still an active area of research~\citep{puig_mechanisms_2024}.

\begin{figure}
\begin{center}
\includegraphics[width=\linewidth]{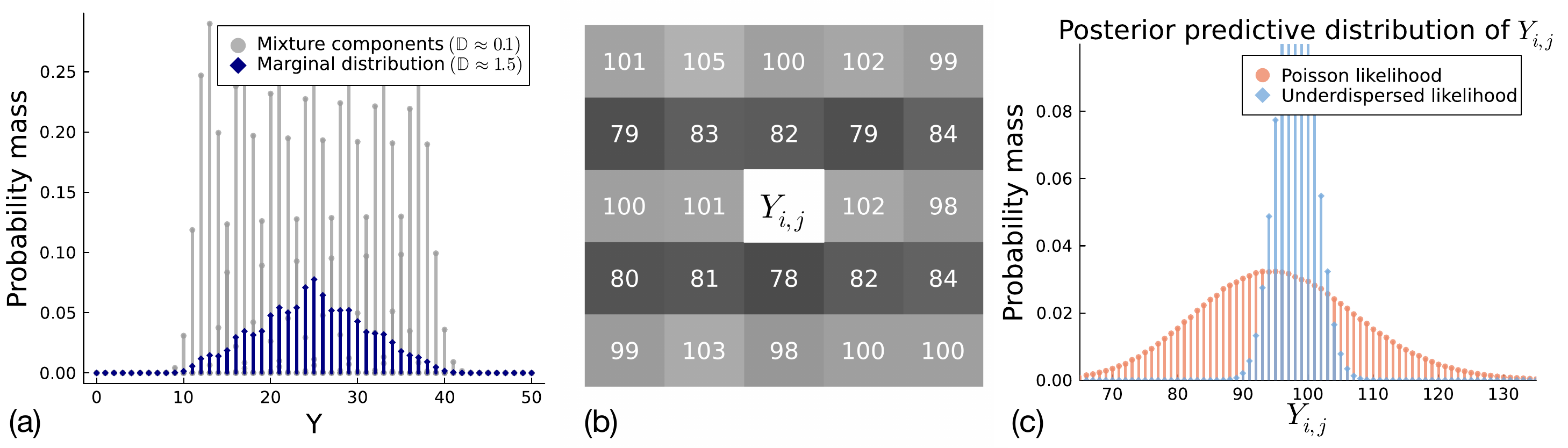}
\end{center}
\caption{(a) Underdispersed conditional distributions can yield an overdispersed marginal. (b) A count matrix where each row is underdispersed. (c) Posterior predictive distribution over the missing data point for a Poisson model and an underdispersed model; the Poisson model's predictive distribution is artificially unable to contract around its mode.
\label{fig:intro}}
\end{figure}

Underdispersion may be so overlooked because it is fundamentally difficult to diagnose. An example of this is given in~\Cref{fig:intro}a, where a mixture of underdispersed distributions gives rise to an overdispersed marginal. Each count-valued data point $Y_i \in \mathbb{N}_0$ is associated with a mixture component $C_i \in [K]$, and sampled as $Y_i \mid C_i=k \indsim f_k$ where each component distribution $f_k$ is underdispersed with \textit{index of dispersion} $\mathbb{D}[Y \mid C = k] \approx 0.1$, defined as $\mathbb{D}[Y \mid C = k] \defeq \nicefrac{\textrm{Var}(Y \mid C=k)}{\mathbb{E}[Y\mid C=k]}$. However, an analyst with only access to the observed counts $Y_i$, and not the latent labels $C_i$, might compute the variance-to-mean ratio across all data points, which in this case would yield evidence of \textit{marginal overdispersion}
$\mathbb{D}[Y] \approx 1.5$, where $\mathbb{D}[Y] \defeq \nicefrac{\textrm{Var}(Y)}{\mathbb{E}[Y]}$ is the dispersion index of the marginal mixture distribution $Y_i \indsim \sum_{k=1}^K P(C=k) \, f_{k}$. Marginal overdispersion $\mathbb{D}[Y] > 1$ arises in this case despite \textit{conditional underdispersion} $\mathbb{D}[Y \mid C] < 1$, and without building and fitting an appropriate latent variable model, the analyst may remain unaware of the regularity in the data that is revealed only after conditioning on latent structure.

The benefit of modeling conditional underdispersion is illustrated in~\Cref{fig:intro}b-c. We observe a count matrix with one missing entry $Y_{i,j}$. The observed counts are marginally overdispersed but conditionally underdispersed within rows. We posit two models, both of which assume a likelihood $Y_{i,j} \indsim f_{\theta_{i,j}}$ where $f_{\theta_{i,j}}$ is a discrete distribution with parameter $\theta_{i,j}$. The parameters factorize as $\theta_{i,j} \defeq \theta^{\ms{(1)}}_i  \theta^{\ms{(2)}}_j$ into two sets of row- and column-specific scalar parameters, which can be estimated from the observed data. One model is conditionally equidispersed, with likelihood $f_{\theta_{i,j}} \equiv \Pois(\theta_{i,j})$, while the second is conditionally underdispersed, with likelihood defined as an underdispersed count distribution that this paper will introduce. We fit both models to the observed counts and then compute the \textit{posterior predictive distribution} over the missing data point $P(Y_{i,j} \mid \mathbf{Y}_{\backslash ij}) = \int f_{\theta_{i,j}}(Y_{i,j}) \, \mathbf{d}P(\theta_{i,j} \mid \mathbf{Y}_{\backslash ij})$, where $P(\theta_{i,j} \mid \mathbf{Y}_{\backslash ij})$ is the posterior distribution over $\theta_{i,j}$ given observed data $\mathbf{Y}_{\backslash ij}$. We see in~\Cref{fig:intro}c that the posterior predictive of the underdispersed model is highly concentrated around the values of $Y_{i,j}$ which seem likely given data, while the conditionally Poisson model cannot concentrate, due simply to its equidispersion constraint, which inflates the uncertainty of its predictions beyond a level reasonably licensed by the data. An even more detailed example of this simple phenomenon is presented in~\Cref{appsec:toy}.

There may be no straightforward way to diagnose conditional underdispersion beyond ``building, fitting, and critiquing''~\citep{box1976,blei2014} perhaps many probabilistic models until arriving at one that fits. The main premise of this paper is that such a process is hindered by the lack of tools for models of conditionally underdispersed counts. Underdispersed count distributions have been introduced in the literature as alternatives to the Poisson, including the Conway-Maxwell-Poisson (CMP)~\citep{conway1961queueing,shmueli_useful_2005}, double Poisson~\citep{efron_double_1986}, gamma count distribution~\citep{winklemann_duration_1995}, and generalized Poisson~\citep{consul2006generalized}, among others~\citep{sellers_underdispersion_2017}. However, while Bayesian and frequentist approaches have been developed for parameter estimation in simple i.i.d.~settings, the development of more complex models based on these distributions is hampered by analytic intractability, such as the lack of any closed-form conjugate priors, and by a general lack of mature inference techniques tailored to them.

This paper introduces a new approach for modeling conditional underdispersion by way of \textit{discrete order statistics}. An order statistic $\Zrd$ is the $r^{\textrm{th}}$ smallest of $D$ random variables that are drawn $Z_1,\dots, Z_D \iidsim f_\theta$ from \textit{parent distribution} $f_\theta$ with (possibly multivariate) parameter $\theta$. Recent work shows that order statistics of the Poisson distribution are  underdispersed~\citep{badiella_ultra_2023}---i.e., if $Y = \Zrd$ for $Z_1,\dots,Z_D \iidsim \textrm{Pois}(\mu)$, then $\mathbb{D}[Y] \leq 1$. Motivated by this fact, this paper develops the theory and practice of building and fitting probabilistic models that assume the data are order statistics of discrete distributions.

\begin{figure}
\begin{center}
\includegraphics[width=\linewidth]{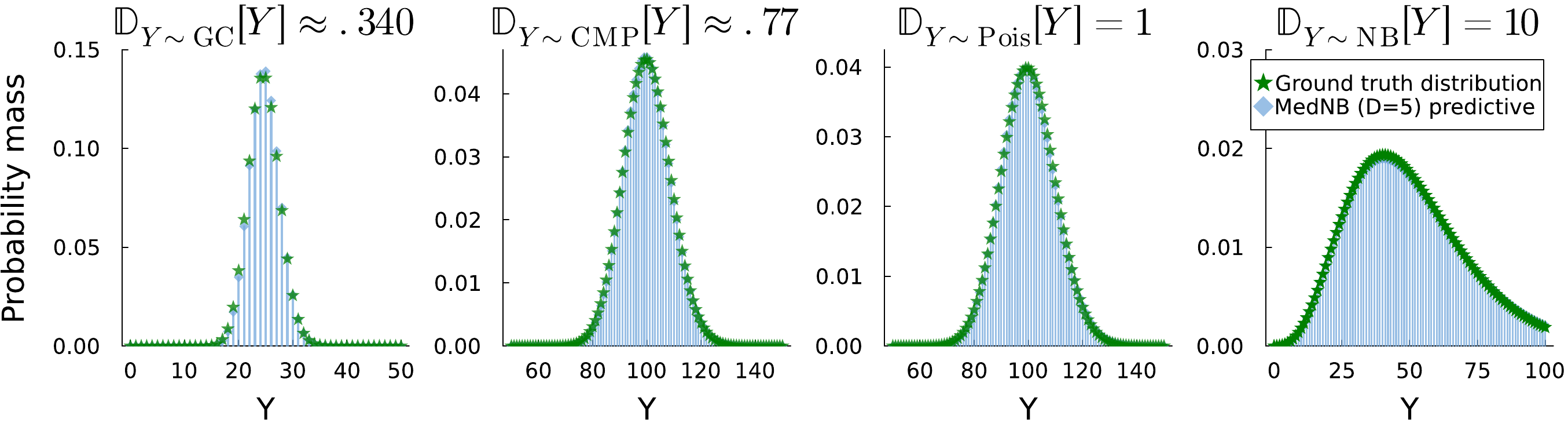}
\end{center}
\caption{Order statistic distributions can fit a variety of existing under- and ovedispersed distributions. Visualized is the posterior predictive distributions of a median negative binomial model, fit to samples from a gamma count, Conway-Maxwell-Poisson, Poisson, and negative binomial distribution. See further details in~\Cref{sec:MCMC}. \label{fig:postpredNB_choice}}
\end{figure}

A key strength of this approach is its modularity: models based on order statistics inherit the full suite of tools already developed for the parent distribution. This is particularly useful for Poisson order statistic models, as there exists a large literature developing complex latent variable models based on a Poisson likelihood, such as factor models~\citep{canny2004gap,dunson2005bayesian,cemgil2009bayesian,zhou2011beta,gopalan2014content}, tensor decompositions~\citep{kolda2009tensor,schein2015bayesian}, state-space models~\citep{charlin2015dynamic,schein2016pgds}, spatial models~\citep{kaiser1997modeling,zhai2025modeling}, or deep belief networks~\citep{ranganath2015deep,zhou2015poisson}, among many others. As we show, the tools and techniques these papers contribute for scalable parameter estimation and posterior inference in Poisson-based models can be repurposed for fitting models based on Poisson order statistics.~\looseness=-1

Our main technical contribution enabling this approach is a general scheme for ``data augmentation''~\citep{tanner1987calculation} which applies broadly to any setting where an integer-valued data point $Y \in \mathbb{Z}$ is assumed to be an order statistic $Y\teq\Zrd$ of a discrete parent distribution $Z_1,\dots, Z_D \iidsim f_\theta$. Our scheme treats the variables $\mathbf{Z}_{1:D}$ as \textit{missing data}.  Estimation of $\theta$ can then proceed by alternating between a data augmentation step, which updates $\mathbf{Z}_{1:D}$ from their \textit{complete conditional distribution} $P(\mathbf{Z}_{1:D} \mid \theta, \Zrd\teq Y)$, and a parameter update step, which updates the parameters from $P(\theta \mid \mathbf{Z}_{1:D})$. Since $\mathbf{Z}_{1:D}$ are assumed i.i.d.~from the parent $f_\theta$, the latter step follows directly from parameter estimation in models that simply assume $Y \sim f_\theta$. For parent distributions like the Poisson, for which parameter estimation techniques in a wide range of models are already well-developed, the only technical barrier to fitting an analogously wide range of order statistic models is in updating $\mathbf{Z}_{1:D}$ from its complete conditional. This distribution exhibits complex dependence due to the positive probability of ties $P(Z_1 = \dots = Z_D \mid \theta, \Zrd \teq Y) > 0$, a property of discrete order statistics that is notoriously difficult and, despite an extensive literature on them (see e.g.,~\cite{nagaraja1992order}), generally hampers their application~\citep[p.g.~116]{johnson2005univariate}. Nevertheless, despite the complete conditional's analytic intractability, we develop an algorithm that can efficiently generate exact samples, which is general to any \textit{order} $D$, \textit{rank} $r$ and discrete parent $f_\theta$. This algorithm can be modularly incorporated into an EM-like algorithm~\citep{dempster1977maximum} for MLE/MAP estimation, or into Markov chain Monte Carlo (MCMC) for full posterior inference, among other schemes.

Equipped with a general strategy for inference in any discrete order statistic model, this paper then develops the theory for two families of order statistics, based on the Poisson and negative binomial, which we show are particularly well-suited to modeling conditionally underdispersed count data. Building on the work of~\citet{badiella_ultra_2023}, we characterize the dispersion properties of a Poisson order statistic, $Y\teq\Zrd$ for $Z_d \iidsim \textrm{Pois}_\mu$, as a function of the order $D$, rank $r$, and parameter $\mu$. We show that $D$ acts as a \textit{``pseudo-index'' of dispersion}, monotonically determining the dispersion $\mathbb{D}[Y]$ for most settings of $\mu$ and $r$. We focus particular attention to minimum, median, and maximum statistics, corresponding to ranks of $r=1$, $r=\lceil \nicefrac{D}{2}\rceil$ for odd $D$, and $r=D$, respectively. We show particularly appealing properties of Poisson medians, whose probability mass concentrates around $\mu$ as $D$ increases, and thus can be roughly used as a discrete analogue to location-scale families. We then develop analogous results for the properties of negative binomial order statistics---i.e., $Y\teq\Zrd$ where $Z_d \iidsim \textrm{NB}_{\alpha, p}$. In particular, we show that negative binomial order statistics can be either over- or underdispersed, and can express finer-grained levels of underdispersion. For both models, we show how $r$ and $D$ play a central role in determining the shape of the distribution and develop a data-adaptive approach for model selection based on placing a prior and approximating the posterior $P(r,D \mid \theta, \Zrd\teq Y)$.

In four case studies---on commercial flight times, COVID-19 case counts, Finnish bird abundance, and RNA sequencing---we illustrate the framework's flexibility by building discrete order statistic models tailored to various aspects of the data at hand, such as relational structure, time series, and covariates. We show how the proposed data augmentation strategy combines easily with existing model-fitting techniques and other augmentation strategies to obtain elegant closed-form inference algorithms. The order statistic models we build often obtain substantially better fit than the corresponding parent baseline. We also show that such models maintain the highly interpretable ``parts-based'' representation~\citep{lee1999learning,gillis2020nonnegative} that Poisson models often enjoy. Overall, we find that order statistic models can be built, used, and interpreted in much the same way as commonly-used alternatives, while often obtaining better fit.~\looseness=-1

\section{Preliminaries}
\label{sec:prelim}

An \textit{order statistic} $\Zrd$ with \textit{rank} $r$ and \textit{order} $D$ is defined as the $r^{\textrm{th}}$ smallest of $D$ i.i.d.~random variables $Z_1,\dots,Z_D \overset{\textrm{iid}}{\sim} f_\theta$ that are drawn from \textit{parent distribution} $f_\theta$ with (possibly multivariate) \textit{parameter} $\theta$. We denote the probability mass function (PMF) and cumulative distribution function (CDF) of $\Zrd$ as $\frd_\theta$ and $\Frd_\theta$ respectively. 

For a discrete parent distribution---i.e., $\Zrd \in \mathbb{Z}$---the CDF of the order statistic can be expressed as $\Frd_\theta(z) = \sum_{t=r}^D \binom{D}{t} \, F_\theta(z)^t \, \big(1-F_\theta(z)\big)^{D-t}$ where $F_\theta$ is the CDF of the parent distribution. The PMF $\frd_\theta$ can thus be expressed as $\frd_\theta(z) = \Frd_\theta(z)- \Frd_\theta(z-1)$.

\section{Poisson Order Statistics for Underdispersed Counts}
\label{sec:pois}
Consider a count-valued datapoint $Y \in \mathbb{N}_0$ assumed to be a Poisson order statistic:
\begin{equation}
\label{eq:poisorderstat}
Y \sim \Poisrd \,\,\Leftrightarrow\,\, Y\teq\Zrd \textrm{ where } Z_1,\dots,Z_D \iidsim \textrm{Pois}_\mu
\end{equation}
where $\mu > 0 $ is the parameter of the Poisson parent---i.e., $\mathbb{E}[Z_d] = \mu$. The following lemma then provides the starting motivation for building models around such an assumption.
\begin{lemma}
\label[lemma]{lemma:badiella}
    (\textbf{Poisson order statistics are underdispersed~\citep{badiella_ultra_2023}})
 For any $\mu$, $r$, and $D 
 \geq 2$, $Y \sim \textrm{Pois}_\mu^{\ms{(r,D)}}$ is underdispersed with index of dispersion $\mathbb{D}[Y] < 1$.
\end{lemma}
\begin{figure}[t!]
\begin{center}
\includegraphics[width=\linewidth]{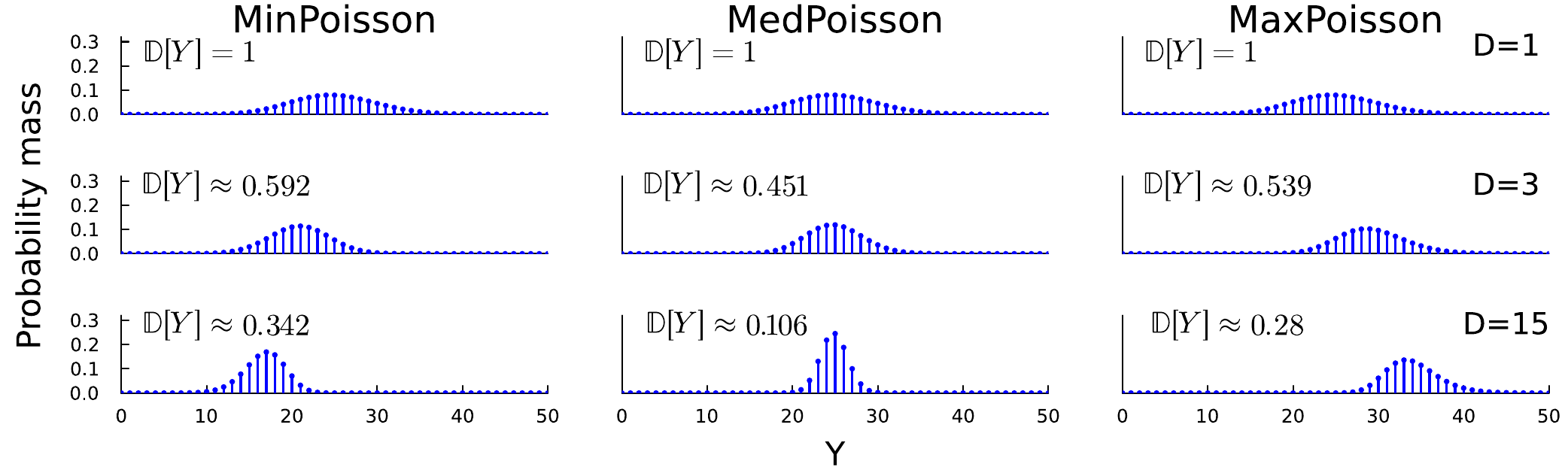}
\end{center}
\caption{PMFs for the Min, Med, and MaxPoisson distributions for $\mu \teq 25$ and $D \in \{1,3,15\}$. \label{fig:AllPoisHist}}
\end{figure}

Following the general formula given in~\Cref{sec:prelim}, the PMF for $Y \sim \textrm{Pois}_\mu^{\ms{(r,D)}}$ is simple to compute and can be written in terms of the CDF of the Poisson---i.e., $F_{\textrm{Pois}_\mu}(y) = \nicefrac{\Gamma(\lfloor y\tp1\rfloor,\,\mu)}{\lfloor y\rfloor !}$---where $\Gamma(\cdot, \cdot)$ is the upper incomplete gamma function. \Cref{fig:AllPoisHist} visualizes the PMF for several choices of $r$ and $D$. We focus on three choices of $r$ corresponding to the minimum ($r\teq1$), median ($r\teq \lceil \nicefrac{D}{2}\rceil$ for $D$ odd), and maximum ($r\teq D$), which for Poisson order statistics we refer to as MinPoisson, MedPoisson, and MaxPoisson, respectively. 

All choices of $r$ and $D \geq 2$ yield underdispersion, as given by~\Cref{lemma:badiella}, and seen in~\Cref{fig:AllPoisHist}. It is evident that the choice of $r$ impacts the shape of the PMF: while the mode of the Min and Max shifts away from $\mu$ as $D$ grows, the mode of the MedPoisson remains near $\mu$. The MedPoisson is also more symmetric than the Min and Max, which both become increasingly skewed as $D$ grows. \Cref{fig:AllPoisHist} also suggests that dispersion decreases in $D$.  Overall, the rank $r$ and order $D$ play an important role in shaping the distribution. In what follows, we formalize that role to provide guidance in selecting $D$ and $r$ when building models. Most importantly, this section develops the notion of $D$ as a \textit{``pseudo-index'' of dispersion}.

\begin{figure}[t!]
\begin{center}
\includegraphics[width=\linewidth]{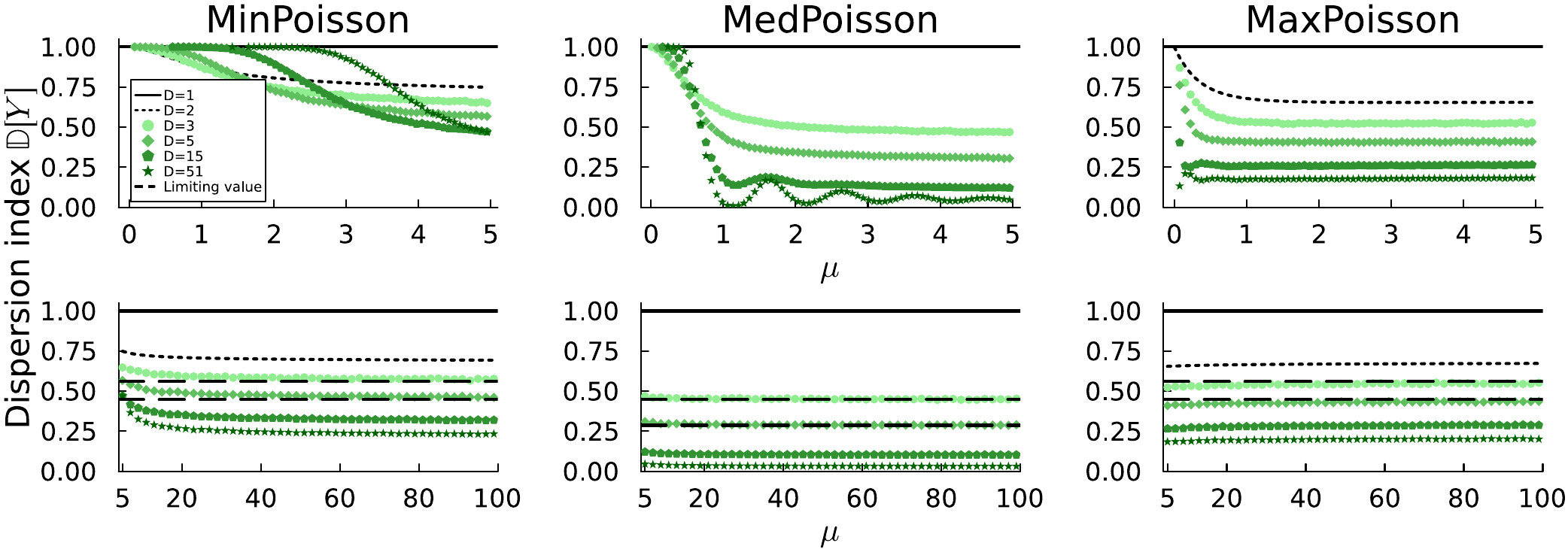}
\end{center}
\caption{Estimated dispersion for the Min, Max, and MedPoisson for $D \!\in\! \{3,5,15,51\}$ and $\mu \!\in\! (0,5]$ (top row) or $\mu \!\in\! [5,100]$ (bottom). Dotted lines indicate exact values of dispersion, availble only for $D=2$. Dashed lines indicate exact values for $\mu \rightarrow \infty$.
\label{fig:poissonfano}}
\end{figure}

The functional relationship between $\mu$, $D$, $r$ and the dispersion level $\mathbb{D}_{Y\sim \textrm{Pois}_\mu^{\ms{(r,D)}}}[Y]$ is complex and analytically intractable in general. For the simple case of $D=2$, \citet{steutel1989gamma} give the first two moments, whose ratio then defines the dispersion index:
$$\mathbb{D}_{Y \sim \textrm{Pois}_\mu^{\ms{(r,2)}}}[Y] = \frac{\exp(2\mu) - \mu \exp(-2\mu)\big(I_0(2\mu) + I_1(2\mu)\big)^2 + I_0(2\mu)(-1)^{r}}{\exp(2\mu) + \big(I_0(2\mu) + I_1(2\mu)\big)(-1)^{r}}$$ where $I_v(x)$ is the modified Bessel function of the first kind. Though analytic expressions are not available for $D>2$, we show we can still characterize the interplay between $\mu$, $D$, $r$ and the dispersion level. We visualize this interplay in~\Cref{fig:poissonfano}, where Monte Carlo estimates of the dispersion are shown for a range of $\mu$ and $D$. In the top versus bottom panels we range over small versus large values of $\mu \in (0, 5]$ and $\mu \in [5, 100]$, respectively. For all cases of $r$ and $D$ the dispersion level seems to converge to a constant as $\mu$ grows large; we formalize this observation in the proposition below.

\begin{proposition}
\label[proposition]{prop:constantdisp}
    (\textbf{Constant dispersion as $\mu \rightarrow \infty$}) Consider $Y \sim \textrm{Pois}_\mu^{\ms{(r,D)}}$ and a standard normal order statistic $U \sim \mathcal{N}_{0,1}^{\ms{(r,D)}}$. Then for any $D \in \mathbb{N}$ and $r \in [D]$ the dispersion level of $Y$ converges in $\mu$ to the variance of $U$: $\lim_{\mu \rightarrow \infty} \mathbb{D}_{Y \sim \textrm{Pois}_\mu^{\ms{(r,D)}}}[Y] = \textrm{Var}_{U \sim \mathcal{N}_{0,1}^{\ms{(r,D)}}}[U]$.
\end{proposition}

Since analytic expressions for the variance of standard Gaussian order statistics are known for $D \leq 5$~\citep{jones1948exact,godwin1949some}, a consequence of~\Cref{prop:constantdisp} is that we obtain analytic expressions for the limiting dispersion of Poisson order statistics for $D \leq 5$. For example,  $\lim_{\mu \rightarrow \infty} \mathbb{D}_{Y \sim \textrm{Pois}_\mu^{\ms{(2,2)}}}[Y] \teq \nicefrac{(\pi-1)}{\pi}$ and $\lim_{\mu \rightarrow \infty} \mathbb{D}_{Y \sim \textrm{Pois}_\mu^{\ms{(2,3)}}}[Y] \teq \nicefrac{(\pi-\sqrt{3})}{\pi}$. We give these as dashed lines in~\Cref{fig:poissonfano}. Another consequence of~\Cref{prop:constantdisp} is given below.
\begin{corollary}
    \textbf{(Symmetry of limiting dispersion)} Consider two standard normal order statistics $U \sim \mathcal{N}_{0,1}^{\ms{(r,D)}}$ and $V \sim \mathcal{N}_{0,1}^{\ms{(D-r+1,D)}}$ with symmetrically opposing ranks (e.g., min and max, $2^{\textrm{nd}}$ smallest and $2^{\textrm{nd}}$ largest, etc.). Since the normal is symmetric, we have that $V \overset{d}{=} -U$, and hence: $\lim_{\mu \rightarrow \infty} \mathbb{D}_{Y \sim \textrm{Pois}_\mu^{\ms{(r,D)}}}[Y] = \lim_{\mu \rightarrow \infty} \mathbb{D}_{Y \sim \textrm{Pois}_\mu^{\ms{(D-r+1,D)}}}[Y]$.   
\end{corollary}
While there are other considerations in selecting between Min and Max models, this corollary suggests their dispersion properties are similar when $\mu$ is large (e.g., $\mu > 20$; see~\Cref{fig:poissonfano}).

Another pattern evident in~\Cref{fig:poissonfano} when $\mu$ is large (bottom row) is that the dispersion appears non-increasing in $D$, as darker lines corresponding to larger $D$ are always upper-bounded by lighter lines. While this pattern is difficult to prove for the Min and Max models, we show it for the median model in~\Cref{cor:nonincpois} below, building again on~\Cref{prop:constantdisp}. These facts imbue $D$ with a particular interpretation, which we give afterward.

\begin{corollary} 
\label[corollary]{cor:nonincpois}
(\textbf{Non-increasing limiting dispersion for MedPoisson}) Let $Y \sim \textrm{MedPois}^{\ms{(D)}}_\mu$ for odd $D$. As the variance of Gaussian sample medians is non-increasing for odd $D$~\citep{lin1989variances}, it holds that: $
\lim_{\mu \rightarrow \infty} \mathbb{D}_{Y \sim \textrm{MedPois}_\mu^{\ms{(D)}}}[Y] \geq \lim_{\mu \rightarrow \infty} \mathbb{D}_{Y \sim \textrm{MedPois}_\mu^{\ms{(D+1)}}}[Y]
$.
\end{corollary}

\begin{remark}
\label[remark]{remark:pseudoindex}
\textbf{($D$ as a pseudo-index of dispersion)} \Cref{prop:constantdisp,cor:nonincpois} cast $D$ as a \textit{``pseudo-index'' of dispersion} for Poisson order statistic models, as the underdispersion is 1) roughly constant for a given $D$ across large values of $\mu > 5$, and 2) monotonic in $D$, provably for the median model, and empirically for the min and max models.
\end{remark}

The role of $D$ is less interpretable in the case of sufficiently small $\mu < 5$, where the functional relationship between $D$, $r$, $\mu$, and the degree of underdispersion is complex, as seen in~\Cref{fig:poissonfano}. For instance, the level of underdispersion oscillates in $\mu$ for the MedianPoisson with large $D$ (top middle) and is non-monotonic in $D$ for the Min (top left). While unappealing, we argue that the small-$\mu$ regime is less important to characterize, since underdispersion in small count data typically corresponds to zero inflation, for which many approaches already exist and could be combined with the model class we propose here.

One other limitation of Poisson order statistic models is that the range of underdispersion levels they can express is limited and contains large gaps. We see in~\Cref{fig:poissonfano} that the maximum index of dispersion the Median model can achieve in the large-$\mu$ regime is $\lim_{\mu \rightarrow \infty} \mathbb{D}_{Y \sim \textrm{MedPois}^{\ms{(3)}}_{\mu}}[Y]=\nicefrac{(\pi - \sqrt{3})}{\pi} \approx 0.44$. We will show in~\Cref{sec:nb} that replacing the Poisson with the negative binomial as the parent distribution generates a model class which covers a broader range of underdispersion levels. However, we note that it is for very low levels of dispersion (e.g., $\mathbb{D}[Y] < 0.5$) that new models are particularly motivated and it is in exactly this range that Poisson order statistics happen to cover many levels. In particular, the Median model smoothly covers a range of dispersion below 0.25, achieving the  minimum level of dispersion possible as $D$ grows large, a fact which we state below.
\begin{proposition}\label{prop:largDmedpois} (\textbf{Large $D$ limit of the MedPoisson}) 
The MedPoisson achieves minimal dispersion in the limit, where it either converges to a point mass around $\mu$---i.e., $\lim_{D \rightarrow \infty}\textrm{MedPois}^{\ms{(D)}}_{\mu}(y) = \delta_{\lfloor \mu \rceil}(y)$ where $\lfloor \mu \rceil$ denotes ``either $\lfloor \mu \rfloor$ or $\lceil \mu \rceil$''---or to a discrete uniform distribution on two consecutive integers, at least one of which is $\lfloor \mu \rfloor$ or $\lceil \mu \rceil$.

\end{proposition}

To summarize the results of this section, Poisson order statistic models are well-tailored to highly underdispersed count data (e.g., $\mathbb{D}[Y] < 0.5$) of at least moderate magnitude (e.g., $\mu >5$). In this regime, $D$ acts as a pseudo-index of dispersion, which mononotically controls the degree of underdispersion and does so independently of $\mu$. It is thus important to select $D$ when building models, which one could do via cross validation, or data-adaptively by placing a prior over $D$ and inferring it as a latent variable. We embrace the latter  and describe an approach that is general for any discrete order statistic model in~\Cref{sec:inference}.

Selecting the rank $r$ of the model is also important. Having explored the properties of the min, median, and max models, we suggest the median model as a default choice, as it has the appealing property of achieving maximum underdispersion and concentrating around $\mu$ in the limit of the pseudo-index $D \rightarrow \infty$. Thus, the MedianPoisson can be viewed as a discrete analogue to location-scale families whose scale controls how concentrated the distribution is around a fixed location. While the MinPoisson model seems to have few if any advantages over the other two, the MaxPoisson model does admit a computational advantage when the data are very sparse, which we highlight in~\Cref{sec:inference} where we develop the methodology for parameter estimation and posterior inference in these models.

\section{Negative Binomial Order Statistics for Over and Underdispersed Counts}\label{sec:nb}

Poisson order statistic models have two limitations, first being their inability to express a fine array of underdispersion levels, particularly for the range $\mathbb{D}[Y] \in (0.5, 1)$, and the second being their inability to express any overdispersion $\mathbb{D}[Y] > 1$. This section replaces the Poisson with the negative binomial as the parent distribution, to obtain a richer family of order statistic models which can express both a finer range of underdispersion and an arbitrary range of overdispersion. The cost to this greater flexibility is greater complexity, both in terms of parameters and in terms of analytic and computational tractability.
\begin{figure}
\begin{center}
\includegraphics[width=\linewidth]{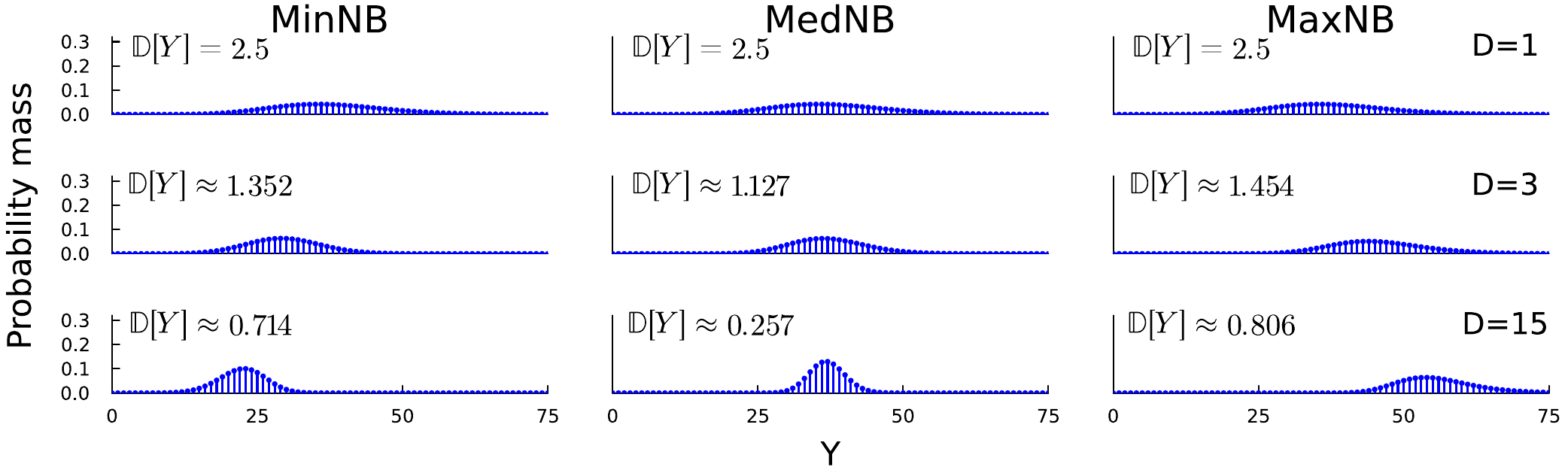}
\end{center}
\caption{PMFs for the M, Med, and MaxNB distributions for $\alpha = 25$, $p=.4$, and $D \in \{1,3,15\}$. \label{fig:NBhist}}
\end{figure}

The models we consider in this section assume the data $Y$ are drawn as
\begin{equation}
\label{eq:nborderstat}
Y \sim \textrm{NB}_{\alpha,p}^{\ms{(r,D)}}\,\,\Leftrightarrow Y\teq\Zrd \textrm{ where } Z_1,\dots,Z_D \iidsim \textrm{NB}_{\alpha,p}
\end{equation}
where $\alpha>0$ and $p \in (0,1)$ are the \textit{stopping} and \textit{failure probability} parameters of the negative binomial parent distribution. Following the general formula given in~\Cref{sec:prelim}, the PMF can be written in terms of the CDF of the negative binomial---i.e., $F_{\textrm{NB}_{\alpha,p}}(y) = \nicefrac{\textrm{B}(p, y+1,\alpha)}{\textrm{B}(y+1,\alpha)}$---where $\textrm{B}(\cdot, \cdot)$ is the upper incomplete beta function. \Cref{fig:NBhist} visualizes the PMF for several $D$ and the three values of $r$ corresponding to the Min, Med, and MaxNB. 

Negative binomial order statistics can attain both over and underdispersion depending on the choices of $D$, $r$, $\alpha$ and $p$, as shown in~\Cref{fig:NBhist}. Below we formalize the role of each of these parameters in shaping the distribution of $Y \sim \textrm{NB}_{\alpha,p}^{\ms{(r,D)}}$. We find that $r$ and $D$ affect dispersion in a similar way as they did for the Poisson, while $\alpha$ plays a similar role to $\mu$, which we formalize in~\Cref{prop:constantdispnb,cor:nonincnb}. The extra parameter $p$ then adds greater flexibility in attaining a wider range of dispersion levels, thus acting as another knob on dispersion and qualifying the role of $D$ to be a``pseudo-index'' of \textit{minimal} dispersion. ~\looseness=-1

\begin{figure}
\begin{center}
\includegraphics[width=\linewidth]{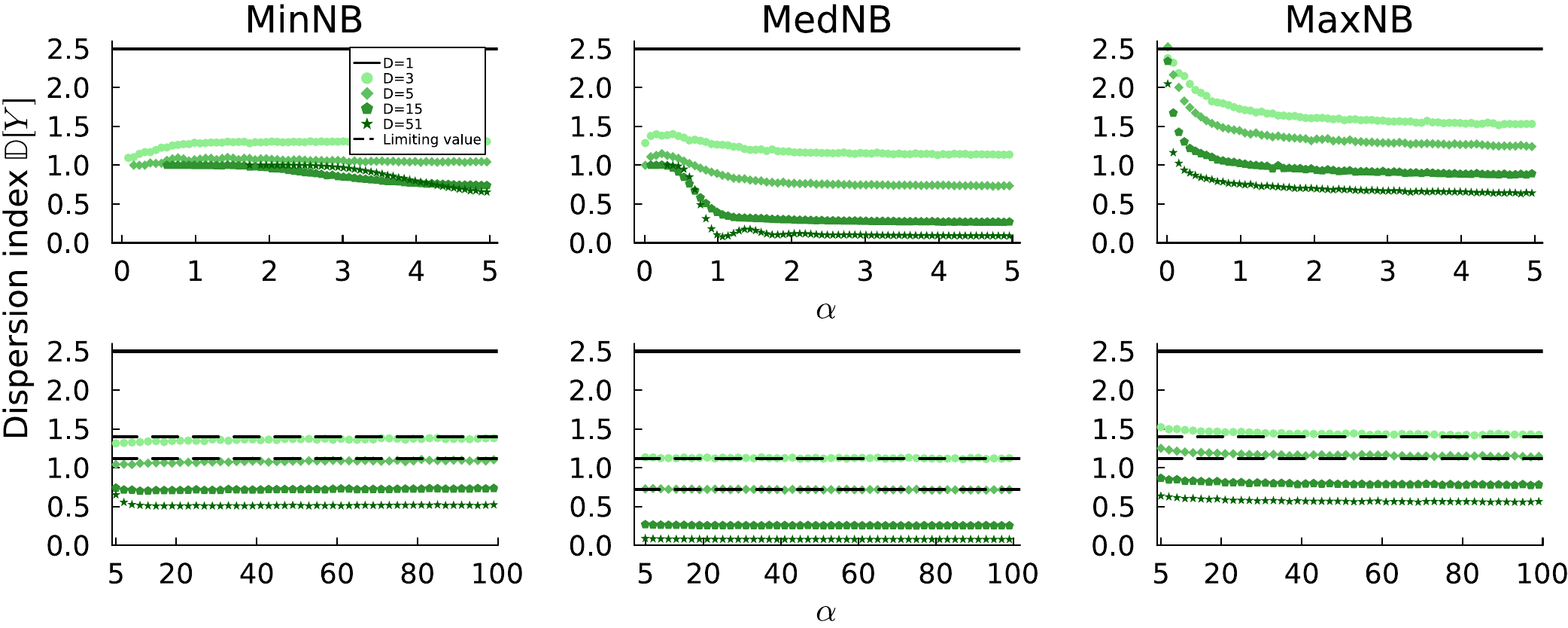}
\end{center}
\caption{Estimated dispersion for Min, Med, and MaxNB distributions for $D \in \{3,5,15,51\}$ for $p=.6$ and $\alpha \in (0,5]$ (top row) or  $\alpha \in [5,100]$ (bottom). Dashed lines indicate exact values for $\alpha \to \infty$.\label{fig:fanoNB}}
\end{figure}
In~\Cref{fig:fanoNB} we show Monte Carlo estimates of the dispersion index for each of the MinNB, MedNB, and MaxNB across choices of $D$ and $\alpha$. Like in the small-$\mu$ regime for Poisson order statistics, we see that the relation between small values of $\alpha \in (0,5]$ (top panel) and dispersion is irregular, whereas for large $\alpha \in (5,100]$ (bottom panel) the dispersion seems to converge to a constant in $\alpha$, which we confirm below.
\begin{proposition}
\label[proposition]{prop:constantdispnb}
    (\textbf{Constant dispersion as $\alpha \rightarrow \infty$}) Consider $Y \sim \textrm{NB}_{\alpha,p}^{\ms{(r,D)}}$ and a normal order statistic $U \sim \mathcal{N}_{0,1}^{\ms{(r,D)}}$. Then for any $D \in \mathbb{N}$, $r \in [D]$, and $p\in (0,1)$ the dispersion level of $Y$ converges in $\alpha$: $\lim_{\alpha \rightarrow \infty} \mathbb{D}_{Y \sim \textrm{NB}_{\alpha,p}^{\ms{(r,D)}}}[Y] = \tfrac{1}{p}\,\textrm{Var}_{U \sim \mathcal{N}_{0,1}^{\ms{(r,D)}}}[U]$.
\end{proposition}
The dispersion level for fixed $p$ and large $\alpha$ is also non-increasing in $D$, which can be seen empirically in the bottom panel of~\Cref{fig:fanoNB} and formally shown for the median model below.
\begin{corollary} 
\label[corollary]{cor:nonincnb}
(\textbf{Non-increasing limiting dispersion for MedNB}) Let $Y \!\sim\! \textrm{MedNB}^{\ms{(D)}}_{\alpha,p}$ for odd $D$ and $p \!\in\! (0,1)$. As the variance of Gaussian sample medians is non-increasing for odd $D$~\citep{lin1989variances}: $
\lim_{\alpha \rightarrow \infty} \mathbb{D}_{Y \sim \textrm{MedNB}_{\alpha,p}^{\ms{(D)}}}[Y] \geq \lim_{\alpha \rightarrow \infty} \mathbb{D}_{Y \sim \textrm{MedNB}_{\alpha,p}^{\ms{(D+1)}}}[Y].
$
\end{corollary}
The limiting dispersion for negative binomial and Poisson order statistics differ by only a factor of $\nicefrac{1}{p}$, which is the index of dispersion for the negative binomial itself. Since $p$ is continuous, unlike $D$, this extra dispersion parameter grants negative binomial order statistics a greater range of dispersion. For example, $\lim_{\alpha \to \infty} \mathbb{D}_{Y \sim \textrm{NB}_{\alpha,p}^{\ms{(2,3)}}}[Y] = \nicefrac{(\pi - 1)}{p\pi}$ is determined both by $p$ and the choice $D=3$. 
Overall, larger values of $p$ decrease dispersion for large $\alpha$. Moreover, in the large-$\alpha$ regime, a negative binomial order statistic with fixed $D$ is able to express an arbitrary level of overdispersion by decreasing $p$ toward 0. However, the minimal index of dispersion encoded for large $\alpha$ is $\textrm{Var}_{U \sim \mathcal{N}_{0,1}^{\ms{(r,D)}}}[U]$. From this perspective, $D$ can be seen as controlling the \textit{minimum} index of dispersion that a negative binomial order statistic can obtain for a given $p$.

\section{Data Augmentation for Discrete Order Statistics}
\label{sec:inference}
This section develops a flexible framework for parameter estimation and posterior inference in discrete order statistic models. Such models assume that data points are drawn $Y \sim \frd_\theta$ for some parent distribution $f_\theta$ whose support is discrete $\textrm{supp}(f_\theta) \subseteq \mathbb{Z}$. Our particular focus has been on Poisson and negative binomial order statistic models where $f_\theta \equiv \Pois_{\mu}$ and $f_\theta \equiv \textrm{NB}_{\alpha,p}$, respectively; however, the inference framework presented in this section is general for any discrete parent distribution. 

Given a discrete dataset $\mathbf{Y}_{1:n}$, the models we consider assume $Y_i \indsim f^{\ms{(r_i,D_i)}}_{\theta_i}$ where data points have potentially different orders $D_i$, ranks $r_i$, and parent parameters $\theta_i$, which may be functions of a shared set of global parameters---i.e., $\theta_i \equiv \theta_i(\Theta)$. Our main goal is then to estimate the parameters $\Theta$ given the data $\mathbf{Y}_{1:n}$. This may be done via maximum likelihood, or in the case where priors are imposed, via approximation to the posterior distribution $P(\Theta \mid \mathbf{Y}_{1:n})$. We focus on the latter, but the framework we develop is modular and enables approaches for MLE, MAP, and full posterior inference.

To derive a general framework for inference, we adopt the auxiliary-variable representation of the order statistic likelihood, where the draw $Y \sim \frd_\theta$ is equivalent to $Y = \Zrd$ with $Z_1,\dots,Z_D \iidsim f_\theta$. Our inference approach centers around a ``data augmentation''~\citep{tanner1987calculation} scheme that explicitly instantiates $\mathbf{Z}_{1:D}$ as latent variables. The focus of this framework is then around the auxiliary variables' complete conditional distribution: 
\begin{align}
\label{eq:compcond}
    P_\theta(\mathbf{Z}_{1:D} \mid \Zrd\teq Y) \,\propto\, \mathbb{1}(Y\teq\Zrd) \,\prod_{d=1}^D f_\theta(Z_d)
\end{align}
which conditions on the model parameters $\theta$ and the observed event $\Zrd\teq Y$. If this complete conditional distribution is tractable---e.g., to sample from or to form expectations under---then we can modularly build algorithms for approximate inference by building on existing work for parameter estimation under the original parent distribution. For example, for a Poisson likelihood $Z_d \iidsim \Pois_\mu$ there are a number of priors $\mu \sim g$ for which the posterior $P(\mu \mid \mathbf{Z}_{1:D}) \propto g(\mu) \prod_{d=1}^D \Pois_\mu(Z_d)$ is tractable, or for which approximate inference schemes already exist. To estimate $\mu$ under the Poisson order statistic model $Y \sim \Poisrd$, the idea is to then construct an algorithm that alternates between a data augmentation step, which updates the auxiliary variables under $P_\mu(\mathbf{Z}_{1:D} \mid \Zrd\teq Y)$, and the existing step which updates $\mu$ under $P(\mu \mid \mathbf{Z}_{1:D})$. Many algorithms can be characterized by this alternation, including forms of Markov chain Monte Carlo (MCMC) for posterior inference or expectation-maximization (EM) for MLE and MAP.

The complete conditional in~\Cref{eq:compcond} is complex due to the dependence among $\mathbf{Z}_{1:D}$ induced by conditioning on the event $Y\teq\Zrd$ and the positive probability of ties---i.e., $P(Z_1= \cdots = Z_D \mid \Zrd\teq Y)>0$. Except in special cases, this distribution is intractable analytically. However, we show that exact samples $\mathbf{Z}_{1:D} \sim P_\theta(\mathbf{Z}_{1:D} \mid \Zrd\teq Y)$ can be efficiently simulated, and  we provide sampling algorithms that are general for any order $D \in \mathbb{N}$, rank $r \in [D]$ and discrete parent distribution $f_\theta$.

By Bayes rule, and by noting that $P_\theta(Z_d \mid \mathbf{Z}_{1:d-1}) = P_\theta(Z_d) \equiv f_\theta(Z_d)$ we have that
\begin{align}
\label{eq:condzd}
P_\theta(Z_d \mid \mathbf{Z}_{1:d-1},\, \Zrd =Y) &= \frac{P_\theta(\Zrd =Y \mid \mathbf{Z}_{1:d})}{P_\theta(\Zrd =Y \mid \mathbf{Z}_{1:d-1})} \, f_\theta(Z_d)
\end{align}
Forming~\Cref{eq:condzd} thus reduces to forming $P_\theta(\Zrd =Y \mid \mathbf{Z}_{1:d})$ which is the probability that $\Zrd$---i.e., the $r^{\textrm{th}}$ smallest of $\mathbf{Z}_{1:D}$---takes the value $Y$ given observed values of $\mathbf{Z}_{1:d}$. Intuition might suggest that this will closely relate to how many of $\mathbf{Z}_{1:d}$ are greater, less than, or equal to $Y$. For example, if $\Zrd$ is the minimum ($r=1$) and we observe that the first variable is $Z_1 < Y$, then $\Zrd$ cannot equal $Y$---i.e., $P_\theta(Z^{\ms{(1,D)}} =Y \mid Z_1 < Y) =0$. As it happens, this is the \textit{only} relevant information in $\mathbf{Z}_{1:d}$, as we formalize below.~\looseness=-1

\begin{lemma} 
\label[lemma]{lemma:suffstat}(\textbf{Sufficient statistics and analytic form of $P_\theta(\Zrd =Y \mid \mathbf{Z}_{1:d})$}) 
\noindent Define $\mathbf{N}_{d} \equiv [\ndley,\,\ndeqy,\,\ndgey]$ to be the following three summary statistics of $\mathbf{Z}_{1:d}$:\vspace{0.2em} 

$\ndley \defeq \sum_{t=1}^d \ind(Z_t < Y),\,\,\, \ndeqy \defeq \sum_{t=1}^d \ind(Z_t = Y),\,\,\,\textrm{and } \ndgey \defeq \sum_{t=1}^d \ind(Z_t > Y)$.\vspace{0.2em} 

\noindent These statistics are sufficient for $\Zrd$---i.e., $P_\theta(\Zrd\teq Y \mid \mathbf{Z}_{1:d}) = P_\theta(\Zrd\teq Y \mid \mathbf{N}_{d})$. \,\, For given values $[n_1,n_2,n_3]$, this probability $P_\theta(\Zrd\teq Y \mid \mathbf{N}_{d} \teq [n_1,n_2,n_3])$ then equals: 
\begin{align*}
    &= \begin{cases}
     0 &\textrm{if }n_1 \geq r \textrm{ or }n_3 \geq D - r + 1\\
     F_\theta^{\ms{(r \txm n_1 \txm n_2,\, D \txm d + 1)}}(Y) - F_\theta^{\ms{(r \txm n_1,\, D \txm d + 1)}}(Y\tm 1) 
     &\textrm{else if } n_2  < \textrm{min}(r \tm n_1,\, D \tm n_3 \tm r \tp 1)\\
        1- F_\theta^{\ms{(r \txm n_1,\, D \txm d + 1)}}(Y \tm 1) &\textrm{else if }r - n_1 \leq n_2 < D- n_3 -r+1\\
        F_\theta^{\ms{(r\txm n_1 \txm n_2,\, D \txm d + 1)}}(Y) & \textrm{else if } D \tm n_3 \tm r \tp 1\leq n_2 < r \tm n_1\\
        1 & \textrm{otherwise}
\end{cases}
\end{align*}

\end{lemma}
The key idea behind the form of the probability in~\Cref{lemma:suffstat} is that conditioning the event $\Zrd = Y$ on $Z_1 < Y$ or $Z_1 > Y$ effectively gives rise to a new order statistic---i.e., $P_\theta(\Zrd \teq Y \mid Z_{1} <Y) \teq P_\theta(\Zrmdm \teq Y)$ and  $P_\theta(\Zrd \teq Y \mid Z_{1} > Y) \teq P_\theta(\Zrdm \teq Y)$. This observation can be successively applied for the number of auxiliary variables observed to be less than or greater than $Y$. Note that with $\ndeqy = 0$, the probability reduces to the PMF of the order statistic $Z^{\ms{(r-\ndley,D-\ndley-\ndgey)}}$. Due to the dependency induced by ties, however, if $\ndeqy > 0$, the probability does not reduce to the PMF of a new order statistic but rather a function of order statistic CDFs, depending on the value of $\ndeqy$ in relation to $r$, $D$, $\ndley$, and $\ndgey$. Carefully accounting for the possibility of $\ndeqy$ exact ties with $\Zrd = Y$ leads to the complex form of this probability. 

The sufficiency property in~\Cref{lemma:suffstat} leads to simple and efficient schemes for sampling from $P_\theta(\mathbf{Z}_{1:D} \mid \Zrd\teq Y)$. The idea is that the complex dependencies among $\mathbf{Z}_{1:D} \mid \Zrd$ are broken when further conditioning on a set of variables $\mathbf{C}_{1:D}$, each of which is a three-level categorical $C_d$ that denotes whether $Z_{d}$ is greater, less than, or equal to $Y$:
\begin{equation}
\label{eq:cdef}
C_d \in \{\ms{\ms{<Y}},\, \ms{\ms{=Y}},\, \ms{\ms{>Y}}\} \,\,\textrm{ where }\,\, C_d=c \,\Leftrightarrow\, Z_d \in \mathbb{Z}_{c} \,\cap\, \textrm{supp}(Y)
\end{equation}
For example, $C_d = \textrm{``}\ms{> Y}\textrm{''}$ denotes that $Z_d \in \mathbb{Z}_{\ms{> Y}} \equiv \{Y\tp 1,\,Y\tp 2, \dots\}$ while $C_d=\textrm{``}\ms{=Y}\textrm{''}$ corresponds to the singleton set $\mathbb{Z}_{\ms{=Y}} \equiv \{Y\}$ and thus denotes the event that $Z_d=Y$. Note that a weaker form of~\Cref{lemma:suffstat} is that $P_\theta(\Zrd\teq Y \mid \mathbf{Z}_{1:D})=P_\theta(\Zrd\teq Y \mid \mathbf{C}_{1:D})$, and the sufficient statistics can be defined as $N^{\ms{(c)}}_d = \sum_{t=1}^d \ind(C_t = c)$ for $c \in \{\ms{\ms{<Y}},\, \ms{\ms{=Y}},\, \ms{\ms{>Y}}\}$. Schemes for sampling $\mathbf{Z}_{1:D} \mid \Zrd$ then follow from the following theorem.

\begin{theorem} 
\label{theorem:sampling}(\textbf{Exact simulation of $\mathbf{Z}_{1:D} \sim P_\theta(\mathbf{Z}_{1:D} \mid \Zrd\teq Y)$}) Conditional on $\mathbf{C}_{1:D}$, $\mathbf{Z}_{1:D}$ are independent of $\Zrd$. Moreover, as a consequence, they are mutually conditionally independent, each distributed according to a truncation of the parent distribution:
\begin{align*}
P_\theta(\mathbf{Z}_{1:D} \mid \mathbf{C}_{1:D}, \Zrd =Y) &= P_\theta(\mathbf{Z}_{1:D} \mid \mathbf{C}_{1:D}) = \prod_{d=1}^D \underset{\mathbb{Z}_{\ms{C_d}}}{\textrm{trunc}} \,f_\theta(Z_d)
\end{align*}
where $\underset{\mathcal{S}}{\textrm{trunc}} \,f$ denotes the distribution $f$ renormalized to have support over $\mathcal{S} \cap \textrm{supp}(f)$. The following procedure thus generates a joint sample of $\mathbf{Z}_{1:D} \mid \Zrd\teq Y$:
\begin{align*}
\mathbf{C}_{1:D} &\sim P_\theta(\mathbf{C}_{1:D} \mid \Zrd\teq Y) \,\,\textrm{ and }\,\,
Z_d \mid C_d \indsim \underset{\mathbb{Z}_{\ms{C_d}}}{\textrm{trunc}} \,f_\theta
\end{align*}
\end{theorem}

Since routines for efficiently simulating from truncations of common distributions (e.g., Poisson) are readily available, the above algorithm is easily implementable provided a strategy for sampling the categorical variables $\mathbf{C}_{1:D} \mid \Zrd$. A natural strategy is to sample each $C_d$ sequentially from $P_\theta(C_d \mid \mathbf{C}_{1:d-1}, \Zrd\teq Y)$, which ultimately has convenient analytic structure in terms of the sufficient statistics $\mathbf{N}_{d-1}$. By Bayes rule, for $c \in \{\ms{\ms{<Y}},\, \ms{\ms{=Y}},\, \ms{\ms{>Y}}\}$
\begin{equation}
\label{eq:catprob}
p_d^{\ms{(c)}} \defeq P_\theta(C_d =c \mid \mathbf{C}_{1:d-1}, \Zrd\teq Y) = \frac{P_\theta(\Zrd = Y \,\mid\, \mathbf{N}_{d-1}, C_d = c)}{P_\theta(\Zrd = Y \,\mid\, \mathbf{N}_{d-1})} P_\theta(C_d = c)
\end{equation}
where $P_\theta(C_d = c)$ corresponds to a simple evaluation of the parent distribution's CDF and the remaining two terms can both be computed via~\Cref{lemma:suffstat}.

The algorithm in~\Cref{theorem:sampling}, which samples $C_d \mid \mathbf{C}_{1:d-1}, \Zrd$ sequentially for $d=1,\dots,D$, can be further improved by recognizing ``break conditions'' under which observing only a subset $\mathbf{C}_{1:d}$ for $d <D$ is sufficient to render all variables $\mathbf{Z}_{1:D}$ conditionally independent of $\Zrd$ and each other. The break conditions, which are defined in terms of the sufficient statistics $\mathbf{N}_d$, are described below.

\begin{theorem} \label{theorem:break}(\textbf{Break conditions of $C_d \sim P_\theta(C_d \mid \mathbf{C}_{1:d-1},\, \Zrd \teq Y)$}) After sampling $\mathbf{C}_{1:d}$ for $d <D$ according to~\Cref{theorem:sampling}, three conditions on $\mathbf{N}_d$ may occur, after which $\mathbf{Z}_{1:D} \mid \mathbf{C}_{1:d}$ are independent of $\Zrd$.  By~\Cref{theorem:sampling}, $\mathbf{Z}_{1:d} \mid \mathbf{C}_{1:d}$ are mutually conditionally independent, distributed according to truncations of the parent distribution. Moreover, $\mathbf{Z}_{d+1:D} \mid \mathbf{C}_{1:d}$ are mutually conditionally independent and identically distributed according to
\begin{align*}
\mathbf{Z}_{d+1:D} \mid \mathbf{C}_{1:d},\Zrd \teq Y\,\, &\iidsim 
\begin{cases}
\underset{\ms{[Y,\,\infty)} \hspace{1em}}{\textrm{trunc}\,f_\theta}  &\textrm{if } \ndeqy \geq 1 \textrm{ and } \ndley = r -1\\
\underset{\ms{(-\infty,\, Y]} \hspace{1em}}{\textrm{trunc}\,f_\theta}  &\textrm{if } \ndeqy \geq 1 \textrm{ and } \ndgey = D -r\\
f_\theta &\textrm{if } \ndeqy = \max\left(r- \ndley,\,\, D-r-\ndgey +1 \right)
\end{cases}
\end{align*}
\end{theorem}

\Cref{theorem:sampling,theorem:break} both describe algorithms for sampling $\mathbf{Z}_{1:D} \mid \Zrd\teq Y$. We provide pseudocode in~\Cref{alg:generalsample} of the appendix, where lines 9--17 optionally exploit the break conditions. We empirically validate the algorithms using Geweke tests~\citep{geweke2004getting} and provide proofs of their correctness in~\Cref{supp:proofsection}.

For $I$ data points, \Cref{alg:generalsample} has a worst-case computational complexity of $\mathcal{O}(ID)$ as it involves a $D$-length for-loop over $\mathcal{O}(1)$ operations for each data point. For sparse count data, the max model $r=D$ has the particular benefit that data augmentation need only be run for the non-zero data points $Y>0$, since $P(Z_1=\dots=Z_D =0\mid Z^{\ms{(D,D)}}=Y=0)=1$. Thus, for the max model with count data, the worst-case complexity is $\mathcal{O}(I_{\textrm{nz}D})$ where $I_{\textrm{nz}} \ll I$ for sparse data. More interesting is the practical or average-case complexity, which can be improved by taking several things into account. The algorithm requires $D$ serial calls to a procedure for computing the probabilities and sampling from a three-level categorical distribution, and then $D$ calls to a procedure for sampling from the truncated parent, the latter of which can be parallelized. Exploiting the break conditions then reduces the average-case number of serial categorical steps from $D$, sometimes substantially.

\section{Case Studies}
\label{sec:casestudies}
Here we provide several case studies where we build probabilistic models based on discrete order statistics for real count datasets in a range of different applications. We tailor the structure the models to the particular application and illustrate how to select the family of order statistic (e.g., MedPois versus MaxNB) based on relevant aspects of the data.~\looseness=-1

\subsection{Study Design \& Evaluation}
Each model will define a likelihood for observation $i$ that can be written $Y_i  \sim f^{\ms{(r_i,D_i)}}_{\theta_i}$ where the order $D_i$, rank $r_i$, parameter $\theta_i$ may differ across observations. The parameter may be a function of global parameters $\theta_i \equiv \theta_i(\Theta)$ and the rank $r_i \in \{1, \lceil \nicefrac{D_i}{2}\rceil, D_i\}$ will always be a function of $D_i$ corresponding to the min, median, or max, respectively. 

Each model will specify a prior over the global parameters $\Theta$ which may be hierarchical. Some models will further place a prior over $D_i$ and infer it as a latent variable. In such cases, for the minimum and maximum models where $D_i \geq 1$, we will set some maximum $D^{\textrm{max}}$ as a hyperparameter and then place a shifted binomial prior, which we can write as $(D_i-1) \sim \textrm{Binomial}(D^{\textrm{max}}-1,\, q)$ for some probability parameter that we can further model as, for example, $q \sim \textrm{Beta}(a, b)$. For the median models, where $D_i \in \{1, 3, 5, \dots, \}$ must be odd, we then define $D_i \sim \textrm{OddBinomial}(D_{\textrm{max}},\, q)$ for odd $D_{\textrm{max}}$ to be equal in distribution to $D_i \overset{d}{=} 2X_i + 1$ where $X_i$ is a standard binomial random variable $X_i \sim \textrm{Binomial}(\nicefrac{(D_{\textrm{max}}-1)}{2},\, q)$. 

For all models, we derive a Gibbs sampler for MCMC-based posterior inference which returns a set of $S$ posterior samples of the parameters and latent variables. For instance, a portion of the output would be $S$ posterior samples of the global parameters $\Theta$---i.e., $\{\Theta^{\ms{(s)}}\}_{s=1}^S$ such that $\Theta^{\ms{(s)}} \iidsim P(\Theta \mid \, \mathbf{Y}_\textrm{obs})$ where $\mathbf{Y}_\textrm{obs} \equiv \{Y_i\}_{i \in \mathcal{I}_\textrm{obs}}$ is the set of observed data. Each Gibbs sampler alternates between the data augmentation step described in~\Cref{sec:inference}---i.e., $\mathbf{Z}_{i,1:D_i}^{(s)} \sim P_{\theta^{\ms{(s\textrm{-}1)}}_i}(\mathbf{Z}_{i,\,1:D_i}  \mid Z_i^{\ms{(r_i,D_i)}} = Y_i)$ for each observed datapoint $i$---and updates to the model parameters which then condition on the auxiliary variables---i.e., $\Theta^{(s)} \sim P(\Theta \mid \mathbf{Z}^{(s)})$.

We emphasize that by conditioning on the auxiliary variables, the updates to the parameters will take the same form as those for models whose likelihood is the parent distribution $f_\theta$.

Unless stated otherwise, for each experiment, we fit models using four MCMC chains, collecting $500$ samples from each ($S \teq 2000$), thinned by 20, after 4000 warmup iterations.~\looseness=-1 

To assess model fit, we will holdout a portion of the data $\mathbf{Y}_\textrm{held} \equiv \{Y_i\}_{i \in \mathcal{I}_\textrm{held}}$ such that the full data set is $\mathbf{Y}_\textrm{obs} \cup \mathbf{Y}_\textrm{held}$ and the sets $\mathcal{I}_\textrm{obs}$ and $\mathcal{I}_\textrm{held}$ define a partition of data indices. Following~\citet{gelman2014understanding}, we will use \textit{log pointwise predictive density (LPPD)} of the heldout data as a measure of model fit. We use a function of LPPD called the \textit{information rate (IR)} which measures the average ``surprise'' (i.e., Shannon information) of new data:
\begin{align}
\textrm{IR} = -\tfrac{1}{|\mathcal{I}_\textrm{held}|
}\underbrace{\sum_{i \in \mathcal{I}_\textrm{held}}\log P(Y_i \mid \mathbf{Y}_\textrm{obs})}_{=\textrm{LPPD}} &= -\tfrac{1}{|\mathcal{I}_\textrm{held}|
}\sum_{i \in \mathcal{I}_\textrm{held}} \log \int \frd_{\theta_i}(Y_i)\, P(\theta_i \mid \, \mathbf{Y}_\textrm{obs}) \mathbf{d}\theta_i
\end{align}
This defines a posterior expectation which we approximate, in practice, with Monte Carlo as $\textrm{IR} \approx -\tfrac{1}{|\mathcal{I}_\textrm{held}|
}\sum_{i \in \mathcal{I}_\textrm{held}} \log \left[\tfrac{1}{S} \sum_{s=1}^S \frd_{\theta_i^{\ms{(s)}}}(Y_i)\right]$ using the samples $\theta_i^{\ms{(s)}} \iidsim P(\theta_i \mid \, \mathbf{Y}_\textrm{obs})$ returned by MCMC. This will involve posterior samples of $(r_i,D_i)$ in cases where they are inferred. 

Since we will be interested in comparing model fit to the $D=1$ baseline, which corresponds to the parent distribution $f_\theta$, we also report \textit{information gain (IG)} defined as $\text{IG} = \textrm{IR}_{\textrm{baseline}} - \textrm{IR}$ where $\textrm{IR}$ and $\textrm{IR}_{\textrm{baseline}}$ are the information rate of a given model and the baseline, respectively. This measures the reduction in average surprise from the baseline, where higher is better.

\subsection{Case Study I: Frontier Flight Times}\label{sec:flights}

We first analyze a dataset of $42,773$ flight times from the airline Frontier which covers $583$ routes between $99$ US airports. Each observation $Y_i \in \mathbb{N}_0$ is the number of minutes of flight $i$. The empirical index of dispersion across all flights is $\widehat{\mathbb{D}}[Y_i] \approx 4.56$, indicating marginal overdispersion, but the average within-route index of dispersion is $\tfrac{1}{583}\sum_{k \in \textrm{routes}} \widehat{\mathbb{D}}[Y_i \mid \textrm{route}[i]=k] \approx 0.64$, indicating conditional underdispersion. We should thus expect a model with an underdispersed likelihood, conditional on route, to improve in fit over a Poisson model. We posit the following generative model, which assumes $Y_i$ is a median Poisson, conditional on certain covariates, latent variables, and parameters:
\begin{align}
   Y_i &\indsim \textrm{MedPois}^{\ms{(D_{\textrm{route}[i]})}}_{\mu_i} \textrm{  where  }\mu_i \defeq a_{\textrm{orig}[i]} + b_{\textrm{dest}[i]} + c_{\textrm{route}[i]}\, \textrm{dist}_{\textrm{route}[i]}
\end{align}
where $\textrm{orig}[i]$ and $\textrm{dest}[i]$ denote the origin and destination airport for flight $i$, respectively, $\textrm{route}[i] \equiv (\textrm{orig}[i],\textrm{dest}[i])$ denotes the route, and $\textrm{dist}_{\textrm{route}[i]}$ is its distance in miles. Each unique airport $j$ is associated with the positive intercepts $a_j$ and $b_j$, for origin and destination respectively, which we assume are gamma-distributed:  $a_j, b_j \iidsim \Gamma(1,1)$. Each unique route $k$ is associated with a coefficient on distance which we also assume is drawn $c_k \iidsim \Gamma(1,1)$. Finally, we model the dispersion heterogeneously across routes as
$D_k \iidsim \textrm{OddBinomial}(D_{\textrm{max}},\, \rho)$ such that $D_k \in \{1,3,5,\dots,D_{\textrm{max}}\}$ and further assume a uniform prior on $\rho \sim \textrm{Beta}(1,1)$. 

After data augmentation, all priors in the model are conditionally conjugate and yield closed-form complete conditionals, which we give, along with those of all models in this section, in~\Cref{sec:MCMC}. In this case, we require one additional form of data augmentation (to the main one given in~\Cref{sec:inference}) which, in general terms, represents $Z \sim \Pois(\sum_{k=1}^K \mu_k)$ as $Z=\sum_{k=1}^K Z_k$ where $Z_k \indsim \Pois(\mu_k)$. This augmentation is standard in the literature on Poisson latent variables models (e.g., see~\citet{schein2019allocative} for a survey).

We set $D_{\textrm{max}} = 9$ and also consider models with fixed values of $D_k\in \{1,3,5,7,9\}$. To evaluate predictive performance, we hold out a random 20\% of all flights and compute information gain (IG), as described above, over the corresponding Poisson baseline ($D=1$).

\begin{table}
\footnotesize
\begin{center}
\begin{tabular}{lccccc}
\toprule
model &$D_k$ inferred & $D_k=3$ & $D_k=5$ & $D_k=7$ & $D_k=9$ \\
\midrule
info gain (IG) $(\uparrow)$ & $\mathbf{0.070} \pm .013$ &
$0.028 \pm .006$ &
$-0.120 \pm .015$ &
$-0.335 \pm 0.024$ &
$-0.585 \pm 0.034$ \\
\bottomrule
\end{tabular}
\end{center}
\caption{Information gain of underdispersed MedPoisson models over the Poisson baseline.\label{info_flight}}
\end{table}

\begin{figure}
\begin{center}
\includegraphics[width=\linewidth]{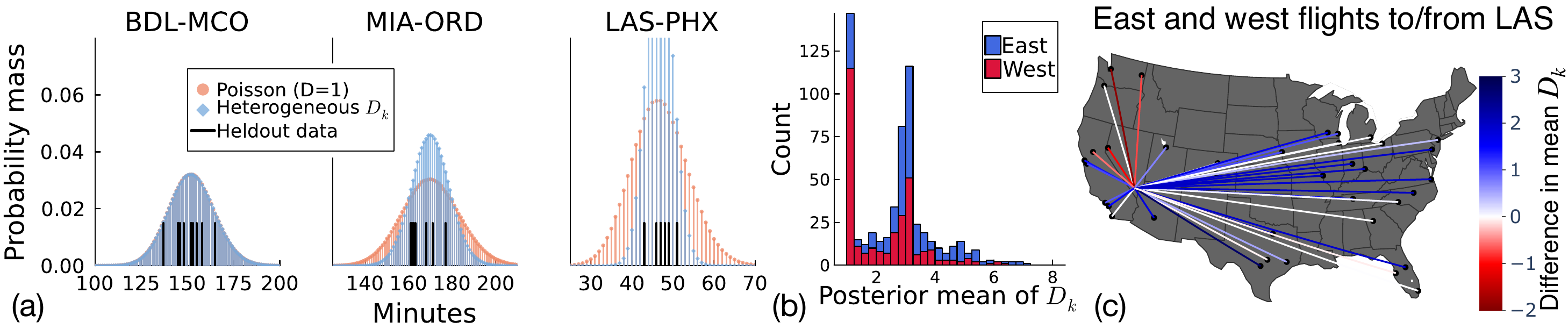}
\end{center}
\caption{(a) Posterior predictive distributions for heldout flights from three example routes. (b) Westbound routes are more likely to have small mean $D_k$, indicating more uncertainty. (c) The difference in mean $D_K$ for routes departing Las Vegas (LAS) and the same routes arriving in LAS. Many transcontinental flights have larger mean $D_k$ when eastbound.}
\label{fig:flight_postpreds}
\end{figure}

\Cref{info_flight} reports the results. The model with heterogeneous $D_k$ across routes shows the greatest gain in predictive performance over the Poisson baseline, while the model with a fixed level of modest underdispersion at $D_k=3$ also exhibits improvement. However, the models with fixed $D_k \geq 5$ exhibit a decrease in performance since forcing an excess level underdispersion not licensed by the data results in overconfident predictive distributions.

The heterogeneity of $D_k$ inferred across routes is visualized in~\Cref{fig:flight_postpreds}. For routes with more variability in flight times, like between Hartford (BDL) and Orlando (MCO), the model places large posterior probability on $D=1$, and the posterior predictive distribution mirrors that of the Poisson model. In contrast, for routes with less variability, such as Las Vegas (LAS) to/from Phoenix (PHX), the model places posterior probability on $D > 1$, thus inducing a predictive interval more concentrated than the equidispersed baseline. We note that these predictive distributions are not artificially concentrated: the $95$\% posterior predictive credible intervals across all routes for the model with heterogeneous $D_k$ have coverage of $95.7$\% on average across all train-test splits, compared to $98.1$\% for the Poisson model.~\looseness=-1

We also find interpretable structure in the pattern of $D_k$ across routes. \Cref{fig:flight_postpreds} shows that the posterior over $D_k$ among westbound routes is substantially shifted toward lower values than that of eastbound routes. This finding accords with recent research by~\citet{kim2020impact} showing that increased climate variability interacts with the jet stream to asymmetrically impact flight times for westbound versus eastbound US domestic flights.

\subsection{Case Study II: Cumulative COVID-19 Case Counts}\label{sec:covid}

Next we present an approach for forecasting reports of cumulative COVID-19 case counts with a conditionally underdispersed model. Our dataset of COVID-19 case counts is derived from that of~\citet{garcia2022public} and covers $C=$3,004 US counties (or county equivalents) for $T=295$ days from March 11, 2020 to December 31, 2020. We consider the following autoregressive model which conditions on the reported case counts at $t=0$ (in this case corresponding to March 10, 2020) and then assumes the counts in county $i$ are drawn sequentially as median Poissons for $t=1 \dots T$ as:
\begin{equation}
\label{eq:covidlikelihood}
Y_{i,t} \sim \textrm{MedPois}^{\ms{(D_{i,t})}}_{\mu_{i,t}} \hspace{0.25em}\textrm{ where }\hspace{0.25em}\mu_{i,t} \,\defeq\, Y_{i,t-1} \,\,+ \,\,  \log (\textrm{population}_i)\Big(\varepsilon + \alpha\sum_{k=1}^K \theta_{i,k}\phi_{k,t}\Big)
\end{equation}
where the rate $\mu_{i,t}$ depends on the previous case count $Y_{i,t-1}$ and the log-population of county $i$, which is multiplied by a function of model parameters intended to represent the latent ``growth rate'' at that time. Although, in principle, cumulative counts should only ever increase, we find many instances in the data of $Y_{i,t} < Y_{i,t-1}$ corresponding to corrections and other reporting irregularities---thus a likelihood tightly concentrated around $Y_{i,t-1}$ plus some \textrm{``growth''} is appropriate. The degree to which the likelihood concentrates is then given by the order $D_{i,t}$ which we allow to vary across all $i$ and $t$. We then assume the following prior which allows the model to share information across units when inferring $D_{i,t}$:
\begin{equation}
\label{eq:covidbinomial}
D_{i,t} \sim \textrm{OddBinomial}\left(D_{\textrm{max}},\, \rho_{i,t}\right) \textrm{ where } \rho_{i,t} \defeq \textrm{logit}^{-1}\big(\sum_{q=1}^Q \beta_{i,q}\tau_{q,t}\big)
\end{equation}
As given, the model has two sets of parameters indexed by time. To allow the model to be used for forecasting, we assume the following prior which samples them for $t=2 \dots T$ as
\begin{align}
\label{eq:phikt}
\phi_{k,t} \sim \Gamma(a^{\ms{(\phi)}} +b^{\ms{(\phi)}}\phi_{k,t-1},\,\,b^{\ms{(\phi)}}) \,\,\,\textrm{ and }\,\,\,
\tau_{q,t} \sim \mathcal{N}(a^{\ms{(\tau)}}+b^{\ms{(\tau)}}\tau_{q,t-1},\,\nicefrac{1}{\lambda_t^{\ms{(\tau)}}})
\end{align}
where $a^{\ms{(\phi)}} \geq 0$ and $b^{\ms{(\phi)}} >0$ are hyperparameters that control the drift and autocorrelation of the evolution of $\phi_{k,t}$ over time and $a^{\ms{(\tau)}} \in \mathbb{R}$ and $b^{\ms{(\tau)}} \in \mathbb{R}$ control the drift and autocorrelation of $\tau_{k,t}$. We further treat the special case of $t=1$ as $\phi_{k,1} \sim \Gamma(1,1)$ and $\tau_{q,1} \sim \mathcal{N}(0,\nicefrac{1}{\lambda_1^{\ms{(\tau)}}})$ .  Finally we place the following simple priors over the remaining parameters: 
$\theta_{i,k} \iidsim \Gamma(1,1)$, $\beta_{i,q} \iidsim \mathcal{N}(0,\nicefrac{1}{\lambda^{\ms{(\beta)}}})$, $\lambda^{\ms{(\beta)}} \sim \Gamma(1,1)$, $\lambda^{\ms{(\tau)}}_t \sim \Gamma(1,1)$, $\varepsilon \sim \Gamma(\nicefrac{1}{2},\,1)$, and $\alpha \sim \Gamma(2,\,1)$.

All priors are conditionally conjugate and yield closed-form conditional updates, but only after three additional forms of data augmentation. The first is the same as in the previous study and is necessary for obtaining closed-form updates to the parameters in~\Cref{eq:covidlikelihood}. The second is the scheme of~\citet{acharya2015nonparametric} involving Chinese Restaurant Table random variables, which is necessitated by the non-conjugate Markov chain over $\phi_{k,t}$ in~\Cref{eq:phikt}. The third is the P\'olya-gamma augmentation of~\citet{polson2013bayesian}, which allows for closed-form updates to Gaussian-distributed parameters in a binomial probability via the sigmoid link, as in~\Cref{eq:covidbinomial}. Derivations are provided in~\Cref{sec:MCMC}.~\looseness=-1

We consider $K\in \{1,5,10\}$ and set $D_{\textrm{max}} \teq 9$, $Q=25$, $a^{\ms{(\phi)}} \teq 0.01$, $b^{\ms{(\phi)}} \teq 100$, $a^{\ms{(\tau)}} \teq 0$, and $b^{\ms{(\tau)}} \teq 1$. We also fit models with a fixed $D_{i,t} \in \{1,3,5,7,9\}$. To evaluate the predictive performance of each model, we hold out the last ten days of cases for a randomly selected $20$\% of counties and calculate information gain over the $D_{i,t}=1$ model.

\begin{figure}
\begin{center}
\includegraphics[width=\linewidth]{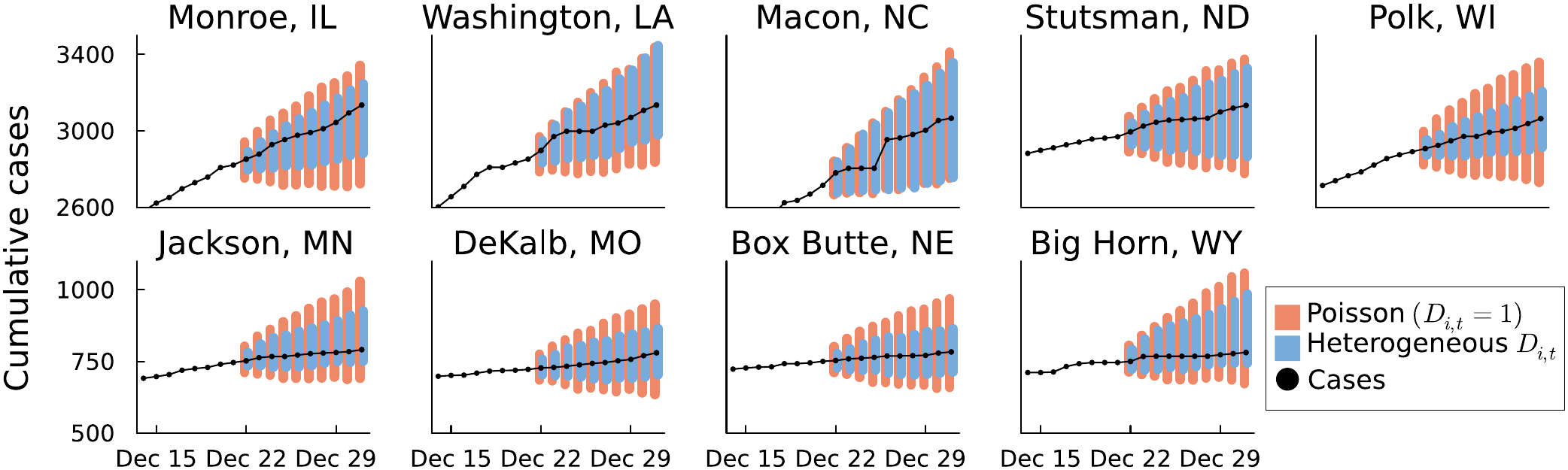}
\end{center}
\caption{$95$\% posterior predictive credible intervals for the 10-step forecast of COVID-19 case counts in 9 US counties. Forecasts from an underdispersed model (blue) result in tighter credible intervals around the true counts than those of the Poisson model (red). \label{COVID_forecast}}
\end{figure}

An underdispersed likelihood vastly improves probabilistic forecasts, as illustrated in~\Cref{COVID_forecast}. This improvement is substantiated fully in~\Cref{table:COVIDinfo} of~\Cref{sec:addexpcovid}, where the model with inferred order $D_{i,t}$ achieves the best performance.

\begin{figure}
\begin{center}
\includegraphics[width=\linewidth]{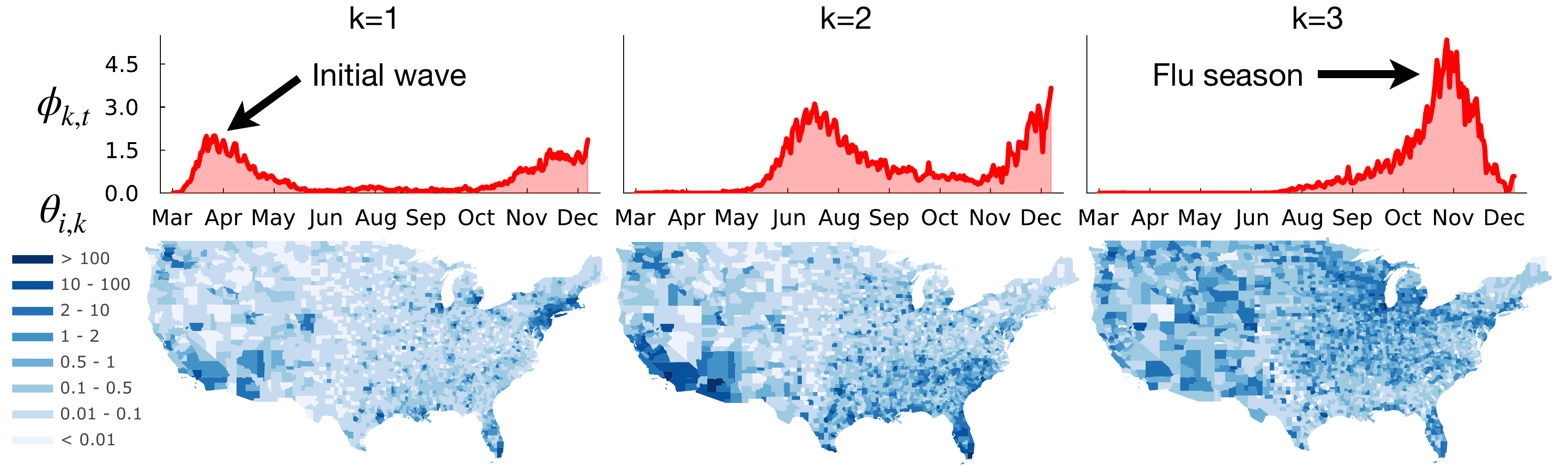}
\end{center}
\caption{Inferred factors correspond to waves of the COVID-19 pandemic, with $k=1$ being the ``initial wave'' that was reported heavily in New York City, Seattle, and Los Angeles, $k=2$ being the ``second wave'' corresponding to the Pandemic's introduction to the Southeast, and $k=3$ being the resurgence that occurred during the typical flu season. \label{COVID_structure}}
\end{figure}

We further explore the inferred parameters in the ``growth rate'' in~\Cref{eq:covidlikelihood} which are 1) $\theta_{i,k}$ the amount that county $i$ loads onto \textit{factor} $k$ and 2) $\phi_{k,t}$ the rate of factor $k$ at time $t$. Most non-negative factorizations like this one admit a ``parts-based representation'' that is highly interpretable~\citep{lee1999learning,gillis2020nonnegative}. Indeed, we find that the inferred factors correspond to different waves of the pandemic, as illustrated in~\Cref{COVID_structure}, which occurred at different times in 2020 and affected different geographic regions of the US. 

\subsection{Case Study III: Finnish Bird Abundance}\label{sec:birds}
We next consider the dataset of bird abundance in Finland analyzed by~\citet{scherting2024} and \citet{piirainen2023species}. As with most species abundance data, this dataset is very sparse. We thus opt for a maximum order statistic model, which benefits computationally from sparsity, as described in~\Cref{sec:inference}. Specifically, we assume the number of bird species $j$ observed at sampling site $i$ is drawn
\begin{align}
     Y_{i,j} &\sim \textrm{MaxPois}_{\mu_{i,j}}^{\ms{(D_{i,j})}} \textrm{ where } \mu_{i,j} \defeq \sum_{k=1}^K \theta_{i,k}\phi_{kj}
\end{align}
We assume gamma priors over the non-negative factors: $\theta_{i,k} \iidsim \Gamma(1,1)$ and $\phi_{k,j} \iidsim \Gamma(1,1)$. To model $D_{i,j}$, we first assumed a prior similar to the one in the previous case study. As fully described in~\Cref{sec:addexpbirds}, this model exhibited substantial heterogeneity in predictive performance across test points, while showing no improvement over the Poisson baseline in aggregate. To explore that heterogeneity, we built a model that incorporates $Q=21$ covariates $X_{i,1},\dots,X_{i,Q}$ (including a constant) for each sampling site $i$, modeling $D_{i,j}$ as
\begin{align}
    \label{eq:birdbinomial2}
     (D_{i,j}-1) &\indsim \textrm{Binomial}\left(D_{\textrm{max}}-1,\,\, \sigma\Big(\sum_{q=1}^Q X_{i,q} \beta_{q,j}\Big)\right)
\end{align}
where the $\beta_{q,j}$ terms act as regression coefficients; we further modify the model to place a shrinkage prior over them---i.e., $\beta_{q,j} \indsim \mathcal{N}(m_q,\,\nicefrac{1}{\lambda_q^{\ms{(\beta)}}}),\,\,m_q \iidsim \mathcal{N}(0,1) $, and $\lambda_q^{\ms{(\beta)}} \iidsim \Gamma(1,1)$---where each species $j$ has its own set of coefficients $\beta_{q,j}$ that are shrunk toward $m_q$.

\begin{figure}
\begin{center}
\includegraphics[width=\linewidth]{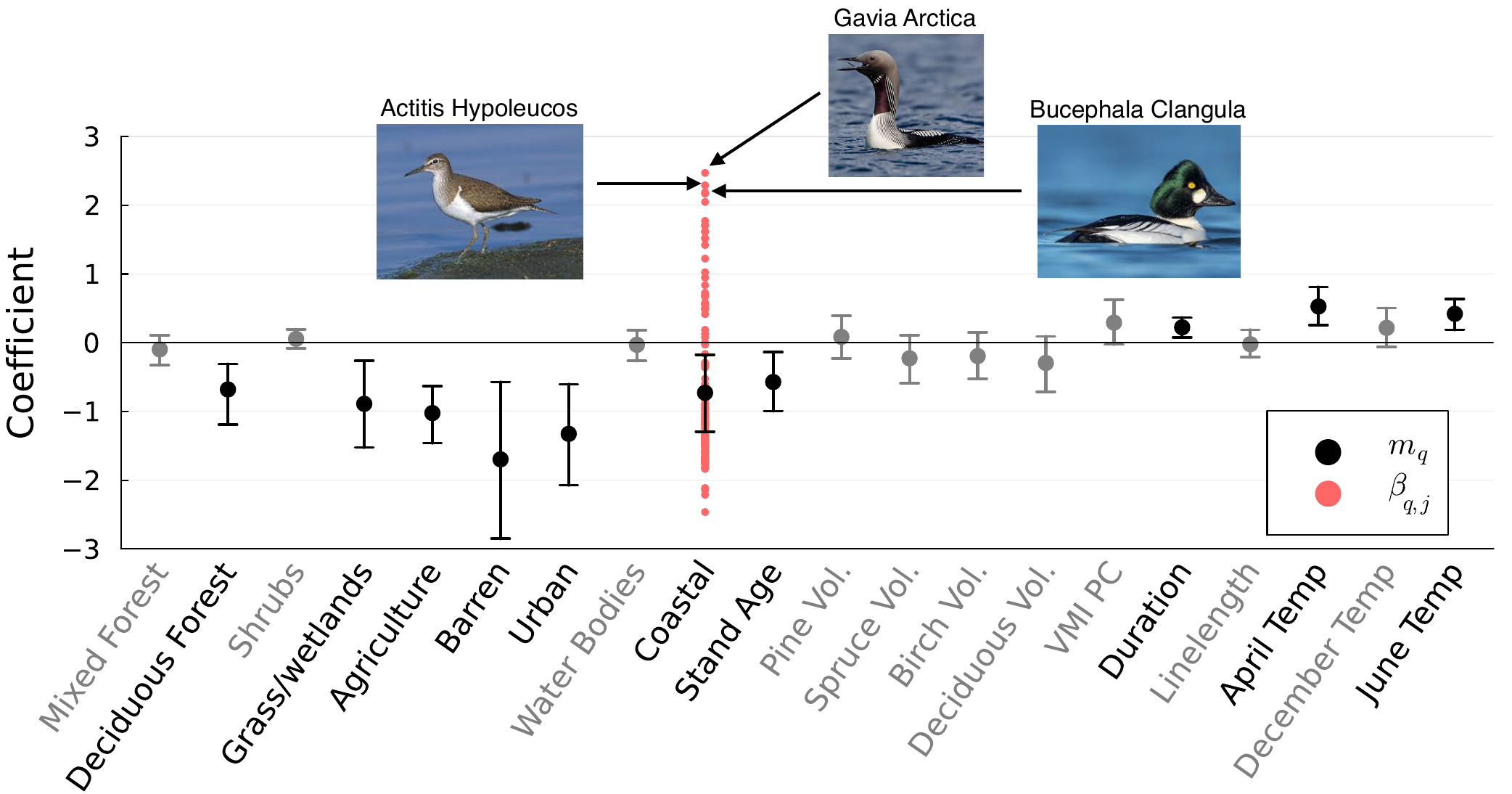}
\end{center}
\caption{Posterior mean values of $m_q$ with 95\% credible intervals. $\beta_{q,j}$ is also shown for each species $j$ and for the "Coastal" covariate.\label{birds_coeffs}}
\end{figure}

\Cref{birds_coeffs} shows 95\% posterior credible intervals for the mean effect $m_q$ of each sampling-site covariate $q$ across species $j$, where ten covariates have significantly non-null effect sizes. We find substantial heterogeneity across species, where the posterior mean of $\beta_{q,j}$ for some species $j$ has the opposite sign of the corresponding mean $m_q$. We highlight one such example for the ``Coastal'' covariate---i.e., $X_{i,q}=\ind(\textrm{site $i$ is coastal})$---where the global mean $m_q$ is significantly negative, while some species-specific effects are large and positive, specifically for water-dwelling bird species. The takeaway is highly interpretable: in coastal areas, the presence of water-dwelling birds is more predictable and less dispersed than the presence of other birds.

\subsection{Case Study IV: RNA-Sequencing}\label{sec:genes}

Negative binomial models are frequently used to analyze read-count data of RNA sequencing (RNASeq), which is widely known to be marginally overdispersed~\citep{anders2010differential}.  Recent work questions whether such data is \textit{conditionally} dispersed enough to license the use of such models~\citep{sarkar2021separating}, highlighting that there may be significant heterogeneity across genes, some of which may even be underdispersed~\citep{gao2018dreamseq}. In this study, we build a negative binomial order statistic model that can be conditionally over- or underdispersed, and allow the parameters controlling dispersion to vary across subjects and genes. Fitting this model then allows us to explore questions about the patterns of conditional dispersion hidden beneath the marginal overdispersion of RNASeq data.~\looseness=-1

We created a subject-by-gene matrix of read-counts using data from the Prostate Adenocarcinoma (PRAD) cohort of The Cancer Genome Atlas (TCGA)~\citep{zuley2016cancer}. We selected $309$ subjects, grouped into six sub-types of cancer. We then selected the 1,000 genes exhibiting the largest variance across sub-types while also exhibiting a sparsity level of at least $25$\%. The resultant $309 \times 1000$ count matrix is $40$\% sparse and has a marginal index of dispersion of $\widehat{\mathbb{D}}[Y_{i,j}]\approx$ 130,000. However, the per-gene marginal dispersion varies widely across columns, ranging from around 50 to 600,000, suggesting that some entries could conceivably be underdispersed when further conditioned on latent structure.

We assume the read-count $Y_{i,j}$ of gene $j$ for subject $i$ is a median negative binomial:
\begin{align}
    Y_{i,j} &\sim \textrm{MedNB}_{\alpha_{i,j},\, p_j}^{\ms{(D_{i,j})}} \hspace{0.25em} \textrm{ where } \alpha_{i,j} \defeq \sum_{k=1}^K \theta_{i,k}\phi_{k,j}\\
    D_{i,j} &\sim \textrm{OddBinomial}\left(D_{\textrm{max}},\,\rho_{i,j}\right)  \hspace{0.25em}\textrm{ where } \rho_{i,j} \defeq \textrm{logit}^{-1}\big(c_j + \sum_{q=1}^Q \tau_{i,q}\beta_{q,j}\big)
\end{align}
We set $D_{\textrm{max}}=9$, $K=10$, and $Q=25$ and then adopt simple priors for the remaining parameters---$p_j \iidsim \textrm{Beta}(1,1)$, $\theta_{i,k},\phi_{k,j} \iidsim \Gamma(1,1)$, $\beta_{q,j} \iidsim \mathcal{N}(0,\nicefrac{1}{\lambda^{\ms{(\beta)}}_j})$, $\tau_{i,q} \iidsim \mathcal{N}(0,\nicefrac{1}{\lambda^{\ms{(\tau)}}_i})$, $\lambda^{\ms{(\beta)}}_j, \lambda^{\ms{(\tau)}}_i \iidsim \Gamma(1,1)$, and  $c_j \iidsim \mathcal{N}(c,\sigma^2)$. The negative binomial can be linked to a latent Poisson random variable via an augmentation scheme described by~\citet{zhou2018nonparametric} which represents $Z \sim \textrm{NB}(\alpha, p)$ as a compound Poisson random variable with $Z = \sum_{\ell=1}^L Z_\ell$ where $Z_\ell \iidsim \textrm{LogSeries}(p)$ and $L \sim \textrm{Pois}(\alpha\ln(\nicefrac{1}{1-p}))$. Using this additional augmentation, along with Poisson thinning, P\'olya-gamma, and the scheme described in~\Cref{sec:inference}, we obtain closed-form complete conditionals for all parameters; see~\Cref{sec:MCMC}.

\begin{figure}[h!]
\begin{center}
\includegraphics[width=6in]{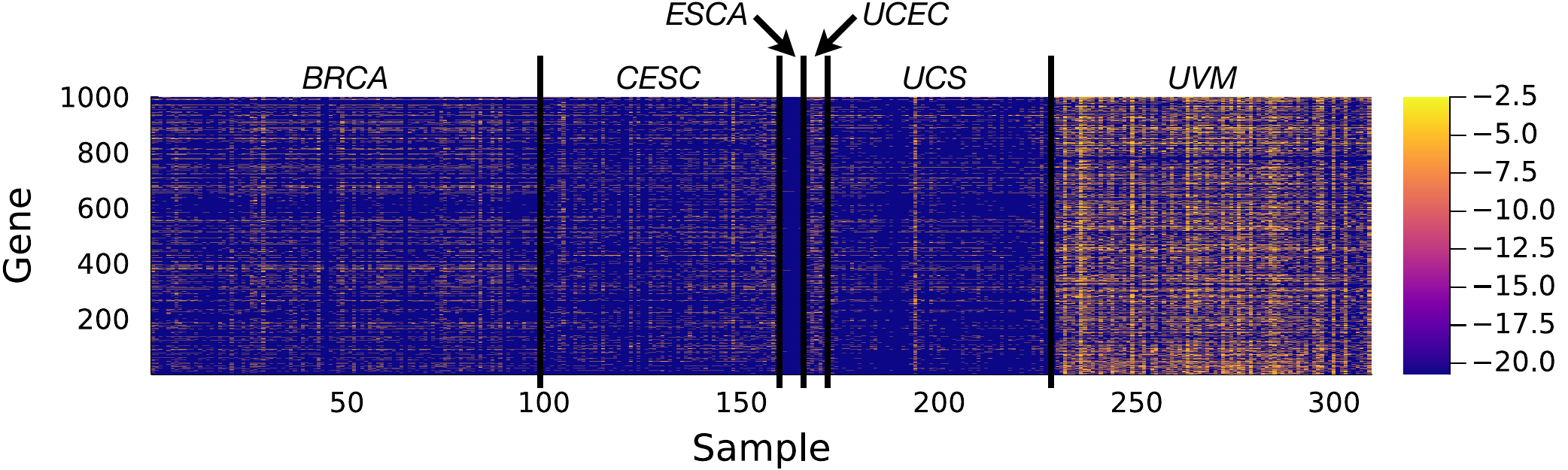}
\end{center}
\caption{Posterior (log-)probabilities of underdispersion for all entries in the gene-by-sample matrix of RNA-sequencing read-count data. While there is little evidence of conditional underdispersion overall, with a maximum posterior probability of around 0.1, the conditional dispersion shows interesting heterogeneity across cancer type.
\label{fig:genes_dispersion}}
\end{figure}

For each posterior sample, we Monte Carlo estimate the dispersion index, and use those to estimate the posterior probability of underdispersion---i.e., $P\Big(\mathbb{D}[Y_{i,j} \mid D_{i,j},\alpha_{i,j}, p_j] < 1 \mid \mathbf{Y}\Big)$. The largest posterior probability of underdispersion is only about 0.1, providing little evidence of conditional underdispersion, and suggesting that negative binomial models for read-counts may indeed be appropriate. Interestingly, these patterns are highly heterogeneous across samples and predictive of cancer sub-type as seen in~\Cref{fig:genes_dispersion}, where there is substantially more evidence of underdispersion for \textit{Uveal Melanoma (UVM)} samples compared to the other cancers, such as \textit{Esophageal Carcinoma (ESCA)}, for which there is very little evidence.

\section{Discussion}
Underdispersion ``flies under the radar''~\citep{sellers_underdispersion_2017} and is fundamentally more difficult to detect than overdispersion, as marginal dispersion, which can be estimated directly, is always greater than conditional dispersion, which requires a model to detect. This paper introduces a fundamentally new way to model underdispersion, via discrete order statistics. This framework is highly modular, allowing any existing techniques tailored to the parent distribution to be combined with a simple data augumentation strategy, thus enabling the iterative cycle of model-building and -fitting. 

Most similar is the work of \citet{canale2011bayesian}, who introduce a flexible framework based on rounded continuous distributions---e.g., $Y_i = \lfloor Z_i\rfloor$ where $Z_i \indsim \mathcal{N}(\mu_i, \sigma_i^2)$---which allow for models with arbitrarily dispersed count likelihoods. Inference proceeds similarly as in our case. For a given continuous parent $f_\theta$, a simple data augmentation step first samples $Z_i \mid Y_i \sim \underset{\ms{(Y_i,\, Y_i+1]} \hspace{1em}}{\textrm{trunc}\,f_\theta}$. The parameters can then be updated as if the likelihood were the parent, exploiting conjugacy, and any other convenient forms. The main benefit of such an approach is to employ the toolbox Gaussian latent variable modeling toward count data. However, latent variable models based on non-negative parameter representations---which often emit a highly interpretable ``parts-based'' interpretation---largely rely on Poisson or multinomial likelihoods for closed-form inference schemes, which cannot be easily adopted into this framework.~\looseness=-1

The alternative is to build models directly around one of the handful of count distributions that have been recognized as underdispersed. While there is a large literature on such distributions, few are accompanied by analytic convenience. The Conway-Maxwell-Poisson (CMP), for example, is an exponential family, but its conjugate prior has no closed-form, and inference methods in CMP models must contend with this~\citep{chanialidis2018efficient,huang2021bayesian,benson2021bayesian}. In general, building complex latent variable models on top of the CMP, gamma count~\citep{winklemann_duration_1995}, generalized Poisson~\citep{consul2006generalized} or other such distributions, is hampered by intractability.

As we have illustrated in four case studies, the proposed framework offers great flexibility in tailoring underdispersed models to a variety of data involving covariates, relational stucture, and time series, among other complex structures. In some of these cases, the ``right'' model reveals conditional underdispersion hidden by empirical overdispersion. Such results suggest the application of the proposed framework holds promise in the wide range of scientific and industrial applications in which discrete data arise.

\setlength{\bibsep}{8.5pt}
\renewcommand{\bibpreamble}{\linespread{0.9}\selectfont}
\bibliographystyle{plainnat}
\bibliography{biblio_no_url.bib}

\newpage

\appendix

\renewcommand{\thefigure}{A.\arabic{figure}}
\setcounter{figure}{0}

\counterwithin{figure}{section}
\renewcommand{\thefigure}{\thesection.\arabic{figure}}

\counterwithin{table}{section}
\renewcommand{\thetable}{\thesection.\arabic{table}}

\section{Proofs for Sections 3 and 4}
\label{sec:poisproof}

\textit{Proof of~\Cref{prop:constantdisp}}. Denote $Y \sim \textrm{Pois}_\mu^{\ms{(r,D)}}$ and $U \sim \mathcal{N}_{0,1}^{\ms{(r,D)}}$. We apply limiting forms of the expectation and variance for Poisson order statistics from~\citet{steutel1989gamma}:
\begin{align*}
    \lim_{\mu\to\infty} \frac{\textrm{Var}[Y]}{\ex[Y]} 
        &= \lim_{\mu\to\infty}\frac{\mu \textrm{Var}[U] + \nicefrac{\sqrt{\mu}}{3}
        \textrm{Cov}(U,U)}{\mu + \sqrt{\mu}\ex[U] + \nicefrac{1}{6}(\ex[U^2]-1)}\\
            &= \lim_{\mu\to\infty}\frac{\textrm{Var}[U] + \nicefrac{1}{3\sqrt{\mu}}\textrm{Cov}(U,U^2)}{1 + \nicefrac{1}{\sqrt{\mu}}\ex[U] + \nicefrac{1}{6\mu}(\ex[U^2]-1)}\\
    &= \textrm{Var}[U]
\end{align*}

\textit{Proof of~\Cref{prop:largDmedpois}}. Denote $Y \sim \textrm{MedPois}^{\ms{(D)}}_{\mu}$. Recall the Poisson CDF $F_{\textrm{Pois}_\mu}(y) = \nicefrac{\Gamma(\lfloor y\tp1\rfloor,\,\mu)}{\lfloor y\rfloor !}$. Note that if for some $y \in \mathbb{N}_0$, $F_{\textrm{Pois}_\mu}(y) = .5$, then there are an infinite number of solutions to the equation $F_{\textrm{Pois}_\mu}(y) = .5$ in the interval $[y,y+1)$. Otherwise, there are no solutions. When there are no solutions, the limiting distribution of $Y$ as $D\to\infty$ is degenerate, and when there are infinite solutions, then the limiting distribution is a discrete uniform distribution with two-point support on two consecutive integers~\citep{nagaraja1992order}.

Define the population median of the Poisson as $\textrm{Med}(\mu) = \inf \{y \in \mathbb{R}: F_{\textrm{Pois}_\mu}(y) \geq \nicefrac{1}{2}\}$. For any $\mu$, $\textrm{Med}(\mu) \in [\mu - \log 2, \mu + \nicefrac{1}{3})$~\citep{adell2005median}. The empirical distribution of the Poisson converges almost surely to the population distribution as $D\to\infty$. Therefore, as $D\to\infty$, the sample median must also be in the range $[\mu - \log 2, \mu + \nicefrac{1}{3})$, and therefore either of the integers $\lceil \mu \rceil$ and $\lfloor \mu \rfloor$. When the distribution of the median is degenerate in the limit of $D$, it then must be degenerate at either $\lceil \mu \rceil$ or $\lfloor \mu \rfloor$. When the distribution is discrete uniform on a two point support, one of the points must be either $\lceil \mu \rceil$ or $\lfloor \mu \rfloor$.

\vspace{15pt}

\textit{Proof of~\Cref{prop:constantdispnb}}. Denote $Y \sim \textrm{NB}_{r,p}^{\ms{(r,D)}}$ and $U \sim \mathcal{N}_{0,1}^{\ms{(r,D)}}$.~\citet{young1970order} gives the following limit in distribution:
$$\lim_{r\to\infty}\frac{pY - (1-p)r}{\sqrt{rp}} \overset{d}{=} U$$
Therefore, limiting forms for the expectation and variance can be written as
$$\lim_{r\to\infty}\ex[Y] = \lim_{r\to\infty} \frac{\sqrt{r}(1-p)}{p}\ex[U] + \frac{r(1-p)}{p}\hspace{1cm}\lim_{r\to\infty}\textrm{Var}[Y] = \lim_{r\to\infty} \frac{r(1-p)}{p^2}\textrm{Var}[U]$$
and the limit of the ratio yield the desired result
\begin{align*}
    \lim_{r\to\infty} \frac{\textrm{Var}[Y]}{\ex[Y]} 
    &= \lim_{r\to\infty}\frac{\frac{r(1-p)}{p^2}\textrm{Var}[U]}{\frac{\sqrt{r}(1-p)}{p}\ex[U] + \frac{r(1-p)}{p}}\\
    &= \lim_{r\to\infty}\frac{\frac{(1-p)}{p^2}\textrm{Var}[U]}{\frac{(1-p)}{\sqrt{r}p}\ex[U] + \frac{(1-p)}{p}}\\
    &= \frac{\frac{(1-p)}{p^2}\textrm{Var}[U]}{\frac{(1-p)}{p}}\\
    &= \frac{1}{p}\textrm{Var}[U]
\end{align*}

\newpage

\phantomsection
\section{Full Sampling Algorithm}
\label{appsec:alg}
\begin{algorithm}[hbt!]
\setstretch{1.3}
\caption{Exact simulation of $\mathbf{Z}_{1:D} \sim P_\theta(\mathbf{Z}_{1:D} \mid \Zrd 
\teq Y)$}\label{alg:generalsample}

\begin{algorithmic}[1]
\STATE {\bf Input}: observation $Y \in \mathbb{Z}$,\, order $D \in \mathbb{N}$,\, rank $r \in [D]$,\, parent distribution $f_\theta$
\STATE {\bf Initialize}: $\nley_0 = \neqy_0 = \ngey_0 = 0$

\STATE {\bf Define}: $ \nley_{\textrm{max}} = r - 1$ and $\ngey_{\textrm{max}} = D - r$\\[0.5em]

\FOR{$d = 1 \dots D$}

\STATE Compute $[\pdley, \pdeqy, \pdgey]$ \hspace{8em}\texttt{// as defined in~\Cref{eq:catprob} } \\[0.5em]

\STATE Sample $C_d \sim \textrm{Categorical}\Big(\pdley,\pdeqy,\pdgey\Big) \hspace{2em} \texttt{// where } C_d \in \{\ms{<Y},\,\, \ms{=Y},\,\, \ms{>Y}\}$ \\[0.5em]

\STATE $N_d^{\ms{(c)}} \leftarrow N_{d-1}^{\ms{(c)}} + \ind\{C_d = c\} \textrm{ for } c \!\in\! \{\ms{<Y},\,\, \ms{=Y},\,\, \ms{>Y}\}$ \hspace{0.5em}\texttt{// update sufficient stats} \\[0.5em]

\STATE Sample $Z_d \sim \underset{\ms{\mathbb{Z}_{\ms{C_d}}} \hspace{1em}}{\textrm{trunc}\,f_\theta}$ \hspace{11em} \texttt{// as given in ~\Cref{theorem:sampling}}\\[0.5em]

\vspace{0.5em}
\texttt{// (Optional) assess/execute the break conditions in~\Cref{theorem:break}}
\label{alg:firstcondition}

\label{algline:cond1}
\IF{$\ndeqy \geq 1$ and $\ndley = \nley_{\textrm{max}}$ } 
\STATE $Z_{d+1},\dots, Z_{D} \iidsim \underset{\ms{[Y,\,\infty)} \hspace{1em}}{\textrm{trunc}\,f_\theta}$ \hspace{8em}\texttt{// break condition 1} \\
\STATE \textrm{\textbf{break}}\\[0.5em]
\ENDIF

\label{algline:cond2}
\IF{$\ndeqy \geq 1$ and $\ndgey = \ngey_{\textrm{max}}$ } 
\STATE $Z_{d+1},\dots, Z_{D} \iidsim \underset{\ms{(-\infty,\, Y]} \hspace{1em}}{\textrm{trunc}\,f_\theta}$ \hspace{8em}\texttt{// break condition 2} \\
\STATE \textrm{\textbf{break}}\\[0.5em]
\ENDIF

\label{algline:cond3}
\IF{$\ndeqy=\textrm{max}\big(j-\ndley,\, \ngey_{\textrm{max}}-\ndgey+1\big)$ } 
\STATE $Z_{d+1},\dots, Z_{D} \iidsim f_\theta$ \hspace{10.5em}\texttt{// break condition 3} \\
\label{algline:finalbreak}
\STATE \textrm{\textbf{break}}\\[0.5em]
\ENDIF

\ENDFOR
\STATE {\bf Output}:\,\,$\{Z_1,\dots,Z_{D}\}$
\end{algorithmic}
\end{algorithm}

\phantomsection
\section{Proofs for Section 5}
\label{supp:proofsection}

We first provide statements and proofs for auxiliary lemmas used to prove~\Cref{lemma:suffstat}. Then we provide a proof of~\Cref{lemma:suffstat} and proofs for~\Cref{theorem:sampling} and~\Cref{theorem:break}, which rely primarily on~\Cref{lemma:suffstat}.

\subsection{Auxiliary Lemmas}

\begin{lemma}\label[lemma]{lemma:indicators}
    For $1 < r < D$,\begin{align*}
        P_\theta&(\Zrd=Y \mid \mathbf{Z}_{1:D}) \\
        &=\ind\{Z_1 < Y\}P_\theta(\Zrmdm =Y \mid \mathbf{Z}_{2:D} )\\
        &\hspace{1cm}+
        \ind\{Z_1 > Y\}P_\theta(\Zrdm =Y \mid \mathbf{Z}_{2:D} )\\
        &\hspace{1cm}+
        \ind\{Z_1 = Y\}P_\theta(\Zrd=Y \mid Z_1 =Y, \mathbf{Z}_{2:D} )
    \end{align*}
    If $1 < r = D$,then
     \begin{align*}
        P_\theta&(\Zdd = Y \mid \mathbf{Z}_{1:D} ) \\
        &=\ind\{Z_1 < Y\}P_\theta(\Zdmdm = Y \mid \mathbf{Z}_{2:D} )
        +
        \ind\{Z_1 = Y\}P_\theta(\Zdd = Y \mid Z_1 =Y, \mathbf{Z}_{2:D} )
    \end{align*}
    and if $r=1$ and $r < D$ then
        \begin{align*}
        P_\theta&(\Zoned = Y \mid \mathbf{Z}_{1:D} ) \\
        &=\ind\{Z_1 > Y\}P_\theta(\Zonedm = Y \mid \mathbf{Z}_{2:D} )
        +
     \ind\{Z_1 = Y\}P_\theta(\Zoned = Y \mid Z_1 =Y, \mathbf{Z}_{2:D} )
    \end{align*}
    and if $r=D=1$
    \begin{align*}
    P_\theta(\Zoneone = Y \mid Z_1 ) = \ind\{Z_1 = Y\}  
    \end{align*}
\end{lemma}

\textit{Proof of~\Cref{lemma:indicators}.} Note that $\mathbf{Z}_{1:D}$ completely determines $
\Zrd$ and $\mathbf{Z}_{2:D}$ completely determines $Z^{\ms{(r',D-1)}}$ for $r' \in [D\tm1]$. Therefore, both sides of each statement must evaluate to $0$ or $1$. 

Begin with the $1 < r < D$ case. We will show that LHS = 1 if and only if RHS = 1.  
\begin{enumerate}
    \item Assume LHS = 1. Then $Y$ is the $r^\textrm{th}$ smallest value of the values $ \mathbf{Z}_{1:D}$. We case on  $\{Z_1 < Y, Z_1 > Y, Z_1 = Y\}$.
    \begin{enumerate}
        \item If $Z_1 < Y$, then $Y$ must be the $r\tm1^{\textrm{th}}$ smallest value of the remaining $ \mathbf{Z}_{2:D}$. Therefore, $P_\theta(\Zrmdm =Y \mid \mathbf{Z}_{2:D})=1$, so RHS = 1.
        \item If instead $Z_1 > Y$, $Y$ must be be the $r^{\textrm{th}}$ smallest value of the remaining $\mathbf{Z}_{2:D}$, so $P_\theta(\Zrdm =Y \mid \mathbf{Z}_{2:D})=1$, and RHS = 1.
        \item If $Z_1 =Y$, RHS must equal $1$ because the probability on the RHS is the same as on the left but with the added restriction $Z_1 =Y$.
    \end{enumerate}
    \item Now assume the RHS is 1. 
    \begin{enumerate}
        \item If $Z_1 < Y$, then $P_\theta(\Zrmdm =Y \mid \mathbf{Z}_{2:D})=1$. Hence $Y$ must be the $r^{\textrm{th}}$ smallest value of $\mathbf{Z}_{1:D}$ because it is larger than $Z_1$ and the $r\tm1^{\textrm{th}}$ smallest of $ \mathbf{Z}_{2:D}$. Therefore,  LHS = 1.
        \item If $Z_1 > Y$, then $P_\theta(\Zrdm =Y \mid \mathbf{Z}_{2:D})=1$,  so $Y$ must be the the $r^{\textrm{th}}$ smallest of all $ \mathbf{Z}_{1:D}$, and thus  LHS = 1.
        \item If $Z_1=Y$, then $P_\theta(\Zrd=Y \mid Z_1 =Y, \mathbf{Z}_{2:D}) = 1$, so RHS = 1.
    \end{enumerate}
\end{enumerate}
Hence, the LHS of the expression is 1 if and only if the RHS is 1. Because each side must evaluate either to $0$ or $1$, the expressions are then equivalent.

The expressions for $r=D$ and $r=1$ proceed exactly as above, noting that none of the values can be strictly greater than $Y$ when $r=D$ and strictly less than $Y$ when $r=1$. When $r=D=1$, only $Z_1=Y$ can make the probability $1$.

\begin{lemma}\label[lemma]{lemma:joint}
    Consider $d \in [1,D]$. Then 
    when $1 < r < D$,
    \begin{align*}
P_\theta(\mathbf{Z}_{1:d},\Zrd = Y)
     = f_\theta(Z_d)&\ind\{Z_d < Y\}P_\theta(\mathbf{Z}_{1:d-1},\Zrmdm =Y)\\
     \hspace{1cm}+ f_\theta(Z_d)&\ind\{Z_d > Y\}P_\theta(\mathbf{Z}_{1:d-1},\Zrdm =Y)\\
     \hspace{1cm}+ &\ind\{Z_d=Y\}P_\theta(Z_d=Y,\mathbf{Z}_{1:d-1},\Zrd=Y)\end{align*}
    and for $1< r = D$ 
    \begin{align*}
    P_\theta(\mathbf{Z}_{1:d},\Zrd = Y)
     = f_\theta(Z_d)&\ind\{Z_d < Y\}P_\theta(\mathbf{Z}_{1:d-1},\Zrmdm =Y)\\
     \hspace{1cm}+ &\ind\{Z_d=Y\}P_\theta(Z_d=Y,\mathbf{Z}_{1:d-1},\Zrd=Y)\end{align*}
    and for $r = 1$ and $r < D$
  \begin{align*}
     P_\theta(\mathbf{Z}_{1:d},\Zrd = Y)
     =  f_\theta(Z_d)&\ind\{Z_d > Y\}P_\theta(\mathbf{Z}_{1:d-1},\Zrdm =Y)\\
     \hspace{1cm}+ &\ind\{Z_d=Y\}P_\theta(Z_d=Y,\mathbf{Z}_{1:d-1},\Zrd=Y)\end{align*}
    and for $r=D=1$
    \begin{align*}P_\theta(Z_1,\Zoneone = Y) = f_\theta(Z_1)\ind\{Z_1 = Y\}\end{align*}
    
\end{lemma}
\textit{Proof of~\Cref{lemma:joint}}. Say $1 < r < D$. Then using~\Cref{lemma:indicators} on $Z_d$ we can write 
    \begin{align*}
    P_\theta(&\mathbf{Z}_{1:d},\Zrd = Y) \\
       &=\sum_{Z_{d+1}} \ldots \sum_{Z_{D}} P_\theta(\Zrd=Y \mid \mathbf{Z}_{1:D})\prod_{t=1}^D f_\theta(Z_t)\\
       &= \prod_{t=1}^D f_\theta(Z_t) \sum_{Z_{d+1} } \ldots \sum_{Z_{D}} \Big[\ind\{Z_d < Y\}P_\theta(\Zrmdm =Y \mid \mathbf{Z}_{1:d-1},\mathbf{Z}_{d+1:D})\\
       &\hspace{1cm}+ \ind\{Z_d > Y\}P_\theta(\Zrdm =Y \mid \mathbf{Z}_{1:d-1},\mathbf{Z}_{d+1:D})\\ 
       &\hspace{1cm}+ \ind\{Z_d=Y\}P_\theta(\Zrd=Y \mid Z_d=Y,\mathbf{Z}_{1:d-1},\mathbf{Z}_{d+1:D})\Big]\\
        &= f_\theta(Z_d)\ind\{Z_d < Y\}P_\theta(\mathbf{Z}_{1:d-1},\Zrmdm =Y)
         + f_\theta(Z_d)\ind\{Z_d > Y\}P_\theta(\mathbf{Z}_{1:d-1},\Zrdm =Y)\\
         &\hspace{1cm}+ \ind\{Z_d = Y\}P_\theta( Z_d=Y,\mathbf{Z}_{1:d-1},\Zrd=Y)
    \end{align*}
The expressions for the other cases proceed exactly in the same applying the appropriate cases of~\Cref{lemma:indicators}.

\begin{lemma}\label[lemma]{lemma:ladder} Consider $d \in [1,D]$. Denote $\mathbf{Z}^{\ms{(=Y)}} \subseteq \mathbf{Z}_{1:d}$ as the subset of $\mathbf{Z}_{1:d}$ which exactly equal $Y$. Define $\ndley = \sum_{t=1}^{d} \ind\{Z_t < Y\}$, $\ndeqy = \sum_{t=1}^{d} \ind\{Z_t = Y\}$, and $\ndgey = \sum_{t=1}^{d} \ind\{Z_t > Y\}$. Then 
\begin{align*}
    P(&\Zrd = Y \mid \mathbf{Z}_{1:d})\\ &= \begin{cases}
    P_\theta(Z^{\ms{(r-\ndley,D-\ndley-\ndgey)}}=Y \mid \mathbf{Z}^{\ms{(=Y)}}) &\textrm{ if }\ndley < r \textrm{ and }\ndgey < D-r+1\\0&\textrm{otherwise}
    \end{cases}
\end{align*}
\end{lemma}

\textit{Proof of~\Cref{lemma:ladder}}. Note that if $\ndley \geq r$, then $\Zrd \neq Y$ because the $r^{\textrm{th}}$ smallest of the $\mathbf{Z}_{1:D}$ will be less than $Y$, not equal. Likewise, if $\ndgey \geq D - r + 1$, then $\Zrd \neq Y$ as well because the $r^{\textrm{th}}$ smallest of $\mathbf{Z}_{1:D}$ will be greater than $Y$.

Now say $\ndley < r$ and $\ndgey < D - r + 1$. Then using~\Cref{lemma:indicators} we can write
\begin{align*}
    P(\Zrd = Y\mid \mathbf{Z}_{1:d}) &= \sum_{Z_{d+1}}\ldots\sum_{Z_D} P(\Zrd = Y \mid \mathbf{Z}_{1:D})\prod_{t=d+1}^D f_\theta(Z_t)\\
    &= \sum_{Z_{d+1}}\ldots\sum_{Z_D} P(Z^{\ms{(r-\ndley,D-\ndley-\ndgey}} = Y\mid \mathbf{Z}^{\ms{(=Y)}}, \mathbf{Z}_{d+1:D} )\prod_{t=d+1}^D f_\theta(Z_t)\\
    &= P(Z^{\ms{(r-\ndley,D-\ndley-\ndgey)} } = Y\mid \mathbf{Z}^{\ms{(=Y)}})
\end{align*}
where we have successively applied the appropriate cases of~\Cref{lemma:indicators} a total of $\ndley + \ndgey$ times. The final order statistic $Z^{\ms{(r-\ndley,D-\ndley-\ndgey)}}$ is valid because $\ndley < r$ and $\ndgey < D - r + 1$.

\begin{lemma}\label[lemma]{lemma:davidandshu}
    \citep{david1978robustness} Take $1 < r \leq  D$. Then
    \begin{align*}
        \Frmd(Y) = \Frd_{\theta}(Y) + {D \choose r-1}(F_\theta(Y))^{r-1} (1-F_\theta(Y))^{D-r+1}
    \end{align*}Also, if $1 \leq r <  D$\begin{align*}
        \Frd_{\theta}(Y) = \Frdm(Y) + {D-1 \choose r-1}(F_\theta(Y))^{r} (1-F_\theta(Y))^{D-r}
    \end{align*}
\end{lemma}

\begin{lemma} \label[lemma]{lemma:davidandshucor}
    Consider $k \in \mathbb{N}_0$ such that $1 < r < D - k - 1$. Then 
    \begin{align*}
        \Frdk(Y) = F_\theta(Y) \Frmdkm(Y) + (1-F_\theta(Y))\Frdkm(Y)
    \end{align*}
\end{lemma}

\textit{Proof of~\Cref{lemma:davidandshucor}.} This fact follows from two applications of~\Cref{lemma:davidandshu}
    \begin{align*}
        &F_\theta(Y) \Frmdkm(Y) + (1-F_\theta(Y))\Frdkm(Y)\\ &= \Frdkm(Y) + F_\theta(Y) \Big(\Frmdkm(Y) - \Frdkm(Y)\Big)\\
        &=  \Frdkm(Y) + F_\theta(Y) {D-k-1 \choose r-1}F_\theta(Y)^{r-1} (1-F_\theta(Y))^{D-k-r}\\
        &= \Frdk(Y)
    \end{align*}

\begin{lemma}\label[lemma]{lemma:bigjointbase}
    If $1 < r < D$, then
    \begin{align*}
        P_\theta(\Zrd=Y \mid Z_1 = Y)=(\Frmdm(Y) - \Frdm(Y-1))
    \end{align*}
    If $1 < r = D$, then
    \begin{align*}
        P_\theta(\Zdd = Y \mid Z_1 = Y)= \Fdmdm(Y)
    \end{align*}
    and if $r = 1$ and $r < D$
    \begin{align*}
     P_\theta(\Zoned = Y \mid Z_1 = Y)=   1-F_\theta^{\ms{(1,D-1)}}(Y-1)
    \end{align*}
\end{lemma}

\textit{Proof of~\Cref{lemma:bigjointbase}}. We use the cases of~\Cref{lemma:joint} when $d=1$ and $\mathbf{Z}_{1:d-1}=\emptyset$.
\begin{enumerate}
\item Beginning when $1 < r = D$
    \begin{align*}
        P_\theta(Z_1 < Y, \Zdd = Y) &= \sum_{Z_1=-\infty}^{Y-1} P_\theta(Z_1, \Zdd = Y)\\
        &= \sum_{Z_1=-\infty}^{Y-1} f_\theta(Z_1)P_\theta(\Zdmdm = Y)\\
        &= F_\theta(Y-1)P_\theta(\Zdmdm = Y)
    \end{align*}
    and $P_\theta(Z_1 > Y, \Zdd = Y) = 0$ so we get
    \begin{align*}
        P_\theta(Z_1 = Y, \Zdd = Y) &= P_\theta(\Zdd = Y) - P_\theta(Z_1 > Y, \Zdd = Y) - P_\theta(Z_1 < Y, \Zdd = Y)\\
        &=  P_\theta(\Zdd = Y) - F_\theta(Y-1)P_\theta(\Zdmdm = Y)\\
        &= F_\theta(Y)^D - F_\theta(Y-1)^D - F_\theta(Y-1)\Big(F_\theta(Y)^{D-1}-F_\theta(Y-1)^{D-1}\Big)\\
        &= f_\theta(Y)F_\theta(Y)^{D-1}\\
        &= f_\theta(Y)F_\theta^{\ms{(D\txm1,D\txm1)}}(Y)
    \end{align*}
Then $P_\theta(\Zdd = Y\mid Z_1 = Y) = F_\theta^{\ms{(D\txm1,D\txm1)}}(Y)$ as desired.

\item And similarly for $r = 1$ and $r < D$
     \begin{align*}
            P_\theta(Z_1 > Y, \Zoned = Y) &= \sum_{Z_1=Y+1}^{\infty} P_\theta(Z_1, \Zoned = Y)\\
            &= \sum_{Z_1=Y+1}^{\infty} f_\theta(Z_1)P_\theta(\Zonedm = Y)\\
            &= (1-F_\theta(Y))P_\theta(\Zonedm = Y)
        \end{align*}
        and $P_\theta(Z_1 < Y, \Zoned = Y) = 0$ so we get
        \begin{align*}
            P_\theta(&Z_1 = Y, \Zoned = Y) = P_\theta(\Zoned = Y) - P_\theta(Z_1 > Y, \Zoned = Y) - P_\theta(Z_1 < Y, \Zoned = Y)\\
            &=  P_\theta(\Zoned = Y) - (1-F_\theta(Y))P_\theta(\Zonedm = Y)\\
            &= (1-F_\theta(Y-1))^D - (1-F_\theta(Y))^D  - (1-F_\theta(Y))\Big((1-F_\theta(Y-1))^{D-1}-(1-F_\theta(Y))^{D-1}\Big)\\
            &=f_\theta(Y)(1-F_\theta(Y-1))^{D-1}\\
            &=f_\theta(Y)(1-F_\theta^{\ms{(1,D\txm1)}}(Y-1))
        \end{align*}
Then $P_\theta(\Zoned = Y\mid Z_1 = Y) = 1- F_\theta^{\ms{(1,D\txm1)}}(Y-1)$ as desired.
        
\item Last, for $1 < r < D$, we proceed in the same manner as above to write
\begin{align*}
    P_\theta(Z_1 < Y, \Zrd=Y) &= F_\theta(Y-1)P_\theta(\Zrmdm =Y)\\
    P_\theta(Z_1 > Y, \Zrd=Y) &= (1- F_\theta(Y))P_\theta(\Zrdm =Y)
\end{align*}
which we apply below
\begin{align*}
    P_\theta(&Z_1 = Y, \Zrd=Y) = P_\theta(\Zrd=Y) - P_\theta(Z_1 < Y, \Zrd=Y) - P_\theta(Z_1 > Y, \Zrd=Y)\\
    &= \Frd_{\theta}(Y) - \Frd_{\theta}(Y-1)\\
    &\hspace{1cm} - F_\theta(Y-1)\Big(F_\theta^{\ms{(r\txm1,D\txm1)}}(Y) - F_\theta^{\ms{(r\txm1,D\txm1)}}(Y-1)\Big)\\
     &\hspace{1cm}- (1-F_\theta(Y))\Big(\Frdm(Y) - \Frdm(Y-1)\Big)\\
    &=  \Frd_{\theta}(Y) - F_\theta(Y)F_\theta^{\ms{(r\txm1,D\txm1)}}(Y) - (1-F_\theta(Y))\Frdm(Y)\\
    &\hspace{1cm}-\Big(\Frd_{\theta}(Y-1) - F_\theta(Y-1)F_\theta^{\ms{(r\txm1,D\txm1)}}(Y-1) -(1-F_\theta(Y-1))\Frdm(Y-1)\Big)\\
    &\hspace{1cm}+f_\theta(Y)\Big(F_\theta^{\ms{(r\txm1,D\txm1)}}(Y) - \Frdm(Y-1)\Big)\\
    &= f_\theta(Y)(F_\theta^{\ms{(r\txm1,D\txm1)}}(Y) - \Frdm(Y-1))
\end{align*}
where the terms in the final step canceled via two applications of~\Cref{lemma:davidandshucor} with $k=0$ for the $Y$ and $Y-1$ terms, respectively. Then the conditional distribution is $P_\theta(\Zrd=Y \mid Z_1 = Y) = F_\theta^{\ms{(r\txm1,D\txm1)}}(Y) - \Frdm(Y-1)$ as desired, completing the proof.  
\end{enumerate}

\newpage

\subsection{Proof of Lemma 5.1} 

First note that if $\ndley \geq r$ or $\ndgey \geq D - r + 1$, then $P_\theta(\Zrd \mid \mathbf{Z}_{1:d}) = 0$ because the $r^\textrm{th}$ smallest of the full $\mathbf{Z}_{1:D}$ cannot exactly equal $Y$ if either condition is realized.

Now assume $\ndley < r$ and $\ndgey < D - r + 1$. Observe by~\Cref{lemma:ladder} that
\begin{align*}
    P_\theta(\Zrd = Y \mid \mathbf{Z}_{1:d}) &= P_\theta(Z^{\ms{(r\txm\ndley,D\txm\ndley\txm\ndgey)}} = Y \mid \mathbf{Z}^{\ms{(=Y)}})\\
    &= \tfrac{1}{P_\theta(\mathbf{Z}^{\ms{(=Y)}})}P_\theta(Z^{\ms{(r\txm\ndley,D\txm\ndley\txm\ndgey)}} = Y, \mathbf{Z}^{\ms{(=Y)}}) 
\end{align*}
where $\mathbf{Z}^{\ms{(=Y)}} \subseteq \mathbf{Z}_{1:d}$ denotes the subset of size $\ndeqy$ of the $\mathbf{Z}_{1:d}$ which exactly equal $Y$. 
We proceed by finding the following form for the joint probability $P_\theta(\mathbf{Z}^{\ms{(=Y)}}, \Zrd = Y)$:
  \begin{align*}P_\theta&(\mathbf{Z}^{\ms{(=Y)}}, \Zrd = Y)\\ &= \begin{cases}
    f_\theta(Y)^{n_2}\left(F_\theta^{\ms{(r-n_2,D-n_2)}}(Y) - F_\theta^{\ms{(r,D-n_2)}}(Y-1)\right), & \text{if } n_2< \text{min}(r, D-r+1)\\
        f_\theta(Y)^{n_2}(1- F_\theta^{\ms{(r,D-n_2)}}(Y-1)), & \text{if }  r \leq n_2< D-r+1\\
        f_\theta(Y)^{n_2}F_\theta^{\ms{(r-n_2,D-n_2)}}(Y), & \text{if } D-r+1 \leq n_2< r\\
        f_\theta(Y)^{n_2}, & \text{if }\text{max}(r, D-r+1) \leq n_2 \leq D
        \end{cases}
    \end{align*}
where we have let $\ndeqy = n_2$. Once this form is obtained, applying it for rank $r-\ndley$ and order $D - \ndley - \ndgey$ and noting that $P_\theta(\mathbf{Z}^{\ms{(=Y)}}) = f_\theta(Y)^{\ndeqy}$ completes the proof of the form of the conditional probability.
Further, this probability only depends on $\mathbf{Z}_{1:d}$ via the statistics $\ndley$, $\ndeqy$, and $\ndgey$,
showing that these statistics are indeed sufficient for $\mathbf{Z}_{1:d}$.

We prove that the joint probability takes this form via induction on $n_2$. First observe that only the first of the four cases can be achieved when $n_2=0$. The joint probability then reduces to $$P_\theta(\Zrd=Y) = F_\theta^{\ms{(r,D)}}(Y) - F_\theta^{\ms{(r,D)}}(Y-1)$$ which is indeed the PMF of a discrete order statistic. Thus the statement holds for $n_2=0$.
For $n_2=1$, the conditions respectively reduce to $1 < r < D$, $r = 1 < D$, and $D = r >1$, and $r=D=1$.
Using the appropriate cases from~\Cref{lemma:bigjointbase} gives the stated joint probability in each case.
Therefore, we have proven the base case $n_2=1$.
For the induction step, take the statement for $n_2$ as the induction hypothesis (IH), and we will work case by case to show the statement must hold for $n_2+1$.

    Before proceeding case-by-case, using~\Cref{lemma:joint}, note that we can write
    \begin{align*}
        P_\theta(Z_{d+1} < Y, \mathbf{Z}^{\ms{(=Y)}}, \Zrd=Y) &= \sum_{Z_{d+1}=-\infty}^{Y-1}P_\theta(Z_{d+1}, \mathbf{Z}^{\ms{(=Y)}}, \Zrd=Y)\\
        &= F_\theta(Y-1) P_\theta(\mathbf{Z}^{\ms{(=Y)}}, \Zrmdm =Y)\\
        P_\theta(Z_{d+1} > Y, \mathbf{Z}^{\ms{(=Y)}}, \Zrd=Y) &= \sum_{Z_{d+1}=Y+1}^{\infty}P_\theta(Z_{d+1}, \mathbf{Z}^{\ms{(=Y)}}, \Zrd=Y)\\
        &= (1-F_\theta(Y))P_\theta(\mathbf{Z}^{\ms{(=Y)}}, \Zrdm =Y)
    \end{align*}
    Using these expressions we can then write by marginalizing out $Z_{d+1}$
    \begin{align*}
    P_\theta(Z_{d+1}=Y, \mathbf{Z}^{\ms{(=Y)}}, \Zrd=Y) &=  P_\theta(\mathbf{Z}^{\ms{(=Y)}}, \Zrd=Y)\\
    &\hspace{1cm}-F_\theta(Y-1)P_\theta(\mathbf{Z}^{\ms{(=Y)}}, \Zrmdm =Y)\\
    &\hspace{1cm}-(1-F_\theta(Y))P_\theta(\mathbf{Z}^{\ms{(=Y)}}, \Zrdm =Y)
    \end{align*}
    where a form of $P_\theta(Z_{d+1}=Y, \mathbf{Z}^{\ms{(=Y)}}, \Zrd=Y)$ is what is needed to prove the statement for $n_2 + 1$. Each case will begin by using this fact, and the values of the
    probabilities $P_\theta(\mathbf{Z}^{\ms{(=Y)}}, \Zrd=Y)$, $P_\theta(\mathbf{Z}^{\ms{(=Y)}}, \Zrmdm =Y)$, and 
$P_\theta(\mathbf{Z}^{\ms{(=Y)}}, \Zrdm =Y)$ will differ according to each case and IH.
    \begin{enumerate}
        \item Consider the case when $n_2+1  < \text{min}(r,D - r + 1)$.  Then $n_2 < r - 1$ and $n_2 < D - r$, so using IH we get the following three statements
         \begin{align*}
&P_\theta(\mathbf{Z}^{\ms{(=Y)}}, \Zrd=Y) = f_\theta(Y)^{n_2} \left(F_\theta^{\ms{(r-n_2,D-n_2)}}(Y) - F_\theta^{\ms{(r,D-n_2)}}(Y-1)\right)\\
         &P_\theta(\mathbf{Z}^{\ms{(=Y)}}, \Zrmdm =Y) = f_\theta(Y)^{n_2}\left(F_\theta^{\ms{(r-n_2-1,D-n_2-1)}}(Y) - F_\theta^{\ms{(r-1,D-n_2-1)}}(Y-1)\right)\\
          &P_\theta(\mathbf{Z}^{\ms{(=Y)}}, \Zrdm =Y) = f_\theta(Y)^{n_2} \left(F_\theta^{\ms{(r-n_2,D-n_2-1)}}(Y) - F_\theta^{\ms{(r,D-n_2-1)}}(Y-1)\right)
    \end{align*}
    Hence
    \begin{align*}
        P_\theta(&Z_{d+1}=Y, \mathbf{Z}^{\ms{(=Y)}}, \Zrd=Y) \\
        &= f_\theta(Y)^{n_2} \Big[F_\theta^{\ms{(r-n_2,D-n_2)}}(Y) - F_\theta^{\ms{(r,D-n_2)}}(Y-1) \\
        &\hspace{1cm}- F_\theta(Y-1) \left(F_\theta^{\ms{(r-n_2-1,D-n_2-1)}}(Y) - F_\theta^{\ms{(r-1,D-n_2-1)}}(Y-1)\right)\\ 
        &\hspace{1cm}-(1-F_\theta(Y))\left(F_\theta^{\ms{(r-n_2,D-n_2-1)}}(Y) - F_\theta^{\ms{(r,D-n_2-1)}}(Y-1)\right)\Big]\\
        &= f_\theta(Y)^{n_2} \Big[F_\theta^{\ms{(r-n_2,D-n_2)}}(Y) - F_\theta(Y) F_\theta^{\ms{(r-n_2-1,D-n_2-1)}}(Y)\\
        &\hspace{1cm}- (1-F_\theta(Y))F_\theta^{\ms{(r-n_2,D-n_2-1)}}(Y)+  f_\theta(Y) F_\theta^{\ms{(r-n_2-1,D-n_2-1)}}(Y)\\
        &\hspace{1cm}- (F_\theta^{\ms{(r,D-n_2-1)}}(Y-1) - F_\theta(Y-1)F_\theta^{\ms{(r-1,D-n_2-1)}}(Y-1)\\
        &\hspace{1cm}- (1-F_\theta(Y-1))F_\theta^{\ms{(r,D-n_2-1)}}(Y-1) - f_\theta(Y)F_\theta^{\ms{(r,D-n_2-1)}}(Y-1))\Big]\\
        &=  f_\theta(Y)^{n_2+1} \left(F_\theta^{\ms{(r-n_2-1,D-n_2-1)}}(Y) - F_\theta^{\ms{(r,D-n_2-1)}}(Y-1)\right)
    \end{align*}
    where in the last step we applied~\Cref{lemma:davidandshucor} twice, once for the terms that depend on $Y$ and once for the terms that depend on $Y-1$, completing the proof in this case.
    \item  We will prove the second case in two sub-cases: $n_2 + 1 = r$ and $n_2 + 1 > r$
    \begin{enumerate}
        \item First, say $ n_2 + 1 < D - r + 1$ and $n_2 + 1 = r$.  Then because $n_2 < r$ but $n_2 \geq r -1$, IH gives
     \begin{align*}
        &P_\theta(\mathbf{Z}^{\ms{(=Y)}}, \Zrd=Y) = f_\theta(Y)^{n_2} \left(F_\theta^{\ms{(r-n_2,D-n_2)}}(Y) - F_\theta^{\ms{(r,D-n_2)}}(Y-1)\right)\\
         &P_\theta(\mathbf{Z}^{\ms{(=Y)}}, \Zrmdm =Y) = f_\theta(Y)^{n_2} \left(1 - F_\theta^{\ms{(r-1,D-n_2-1)}}(Y-1)\right)\\
          &P_\theta(\mathbf{Z}^{\ms{(=Y)}}, \Zrdm =Y) = f_\theta(Y)^{n_2} \left(F_\theta^{\ms{(r-n_2,D-n_2-1)}}(Y) - F_\theta^{\ms{(r,D-n_2-1)}}(Y-1)\right)
    \end{align*}
    and then proceeding as before, noting that $r-n_2=1$
    \begin{align*}
        P_\theta(&Z_{d+1}=Y,\mathbf{Z}^{\,{\scriptstyle=Y}}, \Zrd=Y) \\
        &= f_\theta(Y)^{n_2} \Big[F_\theta^{\ms{(1,D-n_2)}}(Y) -(1-F_\theta(Y))F_\theta^{\ms{(1,D-n_2-1)}}(Y) - F_\theta(Y-1)\\ 
        &\hspace{1cm}- \Big(F_\theta^{\ms{(r,D-n_2)}}(Y-1) - F_\theta(Y-1)  F_\theta^{\ms{(r-1,D-n_2-1)}}(Y-1)\\
        &\hspace{1cm}-(1-F_\theta(Y))F_\theta^{\ms{(r,D-n_2-1)}}(Y-1)\Big)\Big]\\
        &= f_\theta(Y)^{n_2} \Big[F_\theta^{\ms{(1,D-n_2)}}(Y) -(1-F_\theta(Y))F_\theta^{\ms{(1,D-n_2-1)}}(Y)\\
        &\hspace{1cm}- F_\theta(Y-1) - f_\theta(Y)F_\theta^{\ms{(r,D-n_2-1)}}(Y-1)\Big]\\
        &= f_\theta(Y)^{n_2+1} \Big(1- F_\theta^{\ms{(r,D-n_2-1)}}(Y-1)\Big)
    \end{align*}
    where we have again used~\Cref{lemma:davidandshucor} and also $F_\theta^{\ms{(1,D-n_2)}}(Y) -(1-F_\theta(Y))F_\theta^{\ms{(1,D-n_2-1)}}(Y) = F_\theta(Y)$ via form of the CDF of a minimum order statistic.
    \item  Now say that $ n_2 + 1 < D - r + 1$ and $n_2 + 1 > r$. Then IH gives instead
     \begin{align*}
        &P_\theta(\mathbf{Z}^{\ms{(=Y)}}, \Zrd=Y) = f_\theta(Y)^{n_2} \left(1 - F_\theta^{\ms{(r,D-n_2)}}(Y-1)\right)\\
         &P_\theta(\mathbf{Z}^{\ms{(=Y)}}, \Zrmdm =Y) = f_\theta(Y)^{n_2} \left(1 - F_\theta^{\ms{(r-1,D-n_2-1)}}(Y-1)\right)\\
          &P_\theta(\mathbf{Z}^{\ms{(=Y)}}, \Zrdm =Y) =  f_\theta(Y)^{n_2} \left(1 - F_\theta^{\ms{(r,D-n_2-1)}}(Y-1)\right)
    \end{align*}
    and again applying~\Cref{lemma:davidandshucor}
    \begin{align*}
        P_\theta(&Z_{d+1}=Y, \mathbf{Z}^{\ms{(=Y)}}, \Zrd=Y) \\
        &= f_\theta(Y)^{n_2} \Big[ \left(1 - F_\theta^{\ms{(r,D-n_2)}}(Y-1)\right) - F_\theta(Y-1)\left(1 - F_\theta^{\ms{(r-1,D-n_2-1)}}(Y-1)\right) \\
        &\hspace{1cm}- (1-F_\theta(Y))\left(1 - F_\theta^{\ms{(r,D-n_2-1)}}(Y-1)\right)\Big]\\
        &= f_\theta(Y)^{n_2} \Big[1- F_\theta(Y-1) - (1-F_\theta(Y)) - f_\theta(Y)F_\theta^{\ms{(r,D-n_2-1)}}(Y-1)\Big]\\
        &= f_\theta(Y)^{n_2+1} (1- F_\theta^{\ms{(r,D-n_2-1)}}(Y-1))
    \end{align*}
    \end{enumerate}
     Together, the two sub-cases give the result for $n_2 + 1 < D - r + 1$ and $n_2 + 1 \geq r$.

     \item  Now say $n_2 + 1 < r$ and $n_2 + 1 \geq D - r +1$. We will again proceed via two sub-cases: $n_2 + 1 = D - r + 1$ and $n_2 + 1 < D - r +1$.
     \begin{enumerate}
         \item First say $n_2 + 1 < r$ and  $n_2 + 1 = D - r +1$. Applying IH gives
    \begin{align*}
        &P_\theta(\mathbf{Z}^{\ms{(=Y)}}, \Zrd=Y) = f_\theta(Y)^{n_2} \left(F_\theta^{\ms{(r-n_2,D-n_2)}}(Y) - F_\theta^{\ms{(r,D-n_2)}}(Y-1)\right)\\
         &P_\theta(\mathbf{Z}^{\ms{(=Y)}}, \Zrmdm =Y) = f_\theta(Y)^{n_2} \left(F_\theta^{\ms{(r-n_2-1,D-n_2-1)}}(Y) - F_\theta^{\ms{(r-1,D-n_2-1)}}(Y-1)\right)\\
          &P_\theta(\mathbf{Z}^{\ms{(=Y)}}, \Zrdm =Y) = f_\theta(Y)^{n_2} \left(F_\theta^{\ms{(r-n_2,D-n_2-1)}}(Y)\right)
    \end{align*}
    Then again using~\Cref{lemma:davidandshucor} and noting that $r=D-n_2$
    \begin{align*}
        P_\theta(&Z_{d+1}=Y, \mathbf{Z}^{\ms{(=Y)}}, \Zrd=Y) \\
        &= f_\theta(Y)^{n_2}\Big[ F_\theta^{\ms{(r-n_2,D-n_2)}}(Y)  - F_\theta(Y-1) F_\theta^{\ms{(r-n_2-1,D-n_2-1)}}(Y)\\
        &\hspace{1cm}-(1-F_\theta(Y))\Big(F_\theta^{\ms{(r-n_2,D-n_2-1)}}(Y) \Big)\\
        &\hspace{1cm}- \Big(F_\theta^{\ms{(D\txm n_2,D\txm n_2)}}(Y-1) - F_\theta(Y-1) F_\theta^{\ms{(D\txm n_2-1,D\txm n_2-1)}}(Y-1)\Big)\Big]\\ 
        &= f_\theta(Y)^{n_2} \Big[f_\theta(Y)F_\theta^{\ms{(r-n_2-1,D-n_2-1)}}(Y)- \Big(F_\theta^{\ms{(r,r)}}(Y-1) - F_\theta(Y-1) F_\theta^{\ms{(r-1,r-1)}}(Y-1)\Big)\Big]\\ 
        &= f_\theta(Y)^{n_2+1} F_\theta^{\ms{(r-n_2-1,D-n_2-1)}}(Y)
    \end{align*} 
    where in the last step we used the form of the maximum order statistic CDF.
    \item  Now, say $n_2 + 1 < r$ and $n_2 + 1 > D - r +1$.  Then IH gives 
     \begin{align*}
        &P_\theta(\mathbf{Z}^{\ms{(=Y)}}, \Zrd=Y) = f_\theta(Y)^{n_2} \left(F_\theta^{\ms{(r-n_2,D-n_2)}}(Y)\right)\\
         &P_\theta(\mathbf{Z}^{\ms{(=Y)}}, \Zrmdm =Y) = f_\theta(Y)^{n_2} \left(F_\theta^{\ms{(r-n_2-1,D-n_2-1)}}(Y)\right)\\
          &P_\theta(\mathbf{Z}^{\ms{(=Y)}}, \Zrdm =Y) = f_\theta(Y)^{n_2} \left(F_\theta^{\ms{(r-n_2,D-n_2-1)}}(Y)\right)
    \end{align*}
    and applying~\Cref{lemma:davidandshucor} similarly gives the desired result
    \begin{align*}
        P_\theta(&Z_{d+1}=Y, \mathbf{Z}^{\ms{(=Y)}}, \Zrd=Y) \\
        &= f_\theta(Y)^{n_2} \Big[F_\theta^{\ms{(r-n_2,D-n_2)}}(Y) -(1-F_\theta(Y))F_\theta^{\ms{(r-n_2,D-n_2-1)}}(Y) \\
        &\hspace{1cm}- F_\theta(Y) F_\theta^{\ms{(r-n_2-1,D-n_2-1)}}(Y) + f_\theta(Y)F_\theta^{\ms{(r-n_2-1,D-n_2-1)}}(Y)\Big]\\
        &= f_\theta(Y)^{n_2+1} F_\theta^{\ms{(r-n_2-1,D-n_2-1)}}(Y) 
    \end{align*} 
     \end{enumerate}
    Together, the two sub-cases give the result for $n_2 + 1 < r$ and $n_2 + 1 \geq D - r +1$.
     \item For the last case, there are three sub-cases.
     \begin{enumerate}
         \item Say $n_2+1 = r = D -r + 1$. Then applying IH gets
    \begin{align*}
        &P_\theta(\mathbf{Z}^{\ms{(=Y)}}, \Zrd=Y) = f_\theta(Y)^{n_2} \left(F_\theta^{\ms{(r-n_2,D-n_2)}}(Y) - F_\theta^{\ms{(r,D-n_2)}}(Y-1)\right)\\
         &P_\theta(\mathbf{Z}^{\ms{(=Y)}}, \Zrmdm =Y) = f_\theta(Y)^{n_2} \left(1 - F_\theta^{\ms{(r-1,D-n_2-1)}}(Y-1)\right)\\
          &P_\theta(\mathbf{Z}^{\ms{(=Y)}}, \Zrdm =Y) = f_\theta(Y)^{n_2} \left(F_\theta^{\ms{(r-n_2,D-n_2-1)}}(Y)\right)
    \end{align*}
    and noting that $r - n_2 =1$ and $r = D - n_2$ gets
    \begin{align*}
        P_\theta(&Z_{d+1}=Y, \mathbf{Z}^{\ms{(=Y)}}, \Zrd=Y) \\
        &= f_\theta(Y)^{n_2} \Big[F_\theta^{\ms{(1,r)}}(Y) - F_\theta^{\ms{(r,r)}}(Y-1) \\&\hspace{1cm}-F_\theta(Y-1)\left(1 - F_\theta^{\ms{(r-1,r-1)}}(Y-1)\right)-(1-F_\theta(Y))F_\theta^{\ms{(1,r-1)}}(Y)\Big]\\
        &= f_\theta(Y)^{n_2} \Big[F_\theta^{\ms{(1,r)}}(Y)-(1-F_\theta(Y))F_\theta^{\ms{(1,r-1)}}(Y)\\&\hspace{1cm}-(F_\theta(Y-1) + F_\theta^{\ms{(r,r)}}(Y-1)-F_\theta(Y-1)F_\theta^{\ms{(r-1,r-1)}}(Y-1)\Big]\\
        &= f_\theta(Y)^{n_2+1}
    \end{align*}
    once again using the maximum and minimum CDF to simplify.
    \item  Now consider $n_2+1 > r$ and $n_2+1 = D - r + 1$. Then by IH
    \begin{align*}
        &P_\theta(\mathbf{Z}^{\ms{(=Y)}}, \Zrd=Y) = f_\theta(Y)^{n_2} \left(1 - F_\theta^{\ms{(r,D-n_2)}}(Y-1)\right)\\
         &P_\theta(\mathbf{Z}^{\ms{(=Y)}}, \Zrmdm =Y) = f_\theta(Y)^{n_2} \left(1 - F_\theta^{\ms{(r-1,D-n_2-1)}}(Y-1)\right)\\
          &P_\theta(\mathbf{Z}^{\ms{(=Y)}}, \Zrdm =Y) = f_\theta(Y)^{n_2}
    \end{align*}
    and noting $r=D-n_2$
        \begin{align*}
        P_\theta(&Z_{d+1}=Y, \mathbf{Z}^{\ms{(=Y)}}, \Zrd=Y) \\
        &= f_\theta(Y)^{n_2} \Big[1 - F_\theta^{\ms{(r,r)}}(Y-1) -F_\theta(Y-1)\left(1 - F_\theta^{\ms{(r-1,r-1)}}(Y-1)\right)-(1-F_\theta(Y))\Big]\\
        &= f_\theta(Y)^{n_2} \Big[1-F_\theta(Y-1) - 1 + F_\theta(Y)\Big]\\
        &= f_\theta(Y)^{n_2+1}
    \end{align*}
    \item Finally, consider $n_2+1 = r$ and $n_2+1 > D - r + 1$. Then by IH
    \begin{align*}
        &P_\theta(\mathbf{Z}^{\ms{(=Y)}}, \Zrd=Y) = f_\theta(Y)^{n_2} F_\theta^{\ms{(r-n_2,D-n_2)}}(Y)\\
         &P_\theta(\mathbf{Z}^{\ms{(=Y)}}, \Zrmdm =Y) = f_\theta(Y)^{n_2} \\
          &P_\theta(\mathbf{Z}^{\ms{(=Y)}}, \Zrdm =Y) = f_\theta(Y)^{n_2}F_\theta^{\ms{(r-n_2,D-n_2)}}(Y)
    \end{align*}
    and noting $r-n_2=1$
        \begin{align*}
        P_\theta(&Z_{d+1}=Y, \mathbf{Z}^{\ms{(=Y)}}, \Zrd=Y) \\
        &= f_\theta(Y)^{n_2} \Big[F_\theta^{\ms{(1,D-n_2)}}(Y) -F_\theta(Y-1)-(1-F_\theta(Y))F_\theta^{\ms{(1,D-n_2)}}(Y)\Big]\\
        &= f_\theta(Y)^{n_2} \Big[1-F_\theta(Y-1) - 1 + F_\theta(Y)\Big]\\
        &= f_\theta(Y)^{n_2+1}
    \end{align*}
    as desired.
     \end{enumerate}
    \end{enumerate}
Our sub-cases have been exhaustive, so we have proven the induction step and thus the form of the joint probability by induction, completing the proof.

\subsection{Proof of Theorem 5.2}

Due to the sufficiency shown in~\Cref{lemma:suffstat}, $P_\theta(\Zrd = Y\mid \mathbf{Z}_{1:d}, \mathbf{C}_{1:d}) = P_\theta(\Zrd = Y \mid \mathbf{C}_{1:d})$. Then observe
\begin{align*}
P_\theta(\mathbf{Z}_{1:d} \mid \mathbf{C}_{1:d}, \Zrd = Y) &\propto P_\theta(\Zrd = Y\mid \mathbf{Z}_{1:d}, \mathbf{C}_{1:d})P_\theta(\mathbf{Z}_{1:d} \mid \mathbf{C}_{1:d})P_\theta(\mathbf{C}_{1:d})\\
&\propto P_\theta(\Zrd = Y \mid \mathbf{C}_{1:d})P_\theta(\mathbf{Z}_{1:d} \mid \mathbf{C}_{1:d})\\
&\propto P_\theta(\mathbf{Z}_{1:d} \mid \mathbf{C}_{1:d})
\end{align*}

Then $P_\theta(\mathbf{Z}_{1:d} \mid \mathbf{C}_{1:d}, \Zrd = Y) =  \prod_{t=1}^d \underset{\mathbb{Z}_{C_t}}{trunc}\, f_\theta(Z_t)$ for $d \in [D]$. Now for $d=D$ we write
$$P_\theta(\mathbf{Z}_{1:D} \mid \Zrd = Y) = P_\theta(\mathbf{Z}_{1:D} \mid \mathbf{C}_{1:D}, \Zrd = Y)P_\theta(\mathbf{C}_{1:D} \mid \Zrd = Y)$$
which reveals that first sampling $\mathbf{C}_{1:D} \sim P_\theta(\mathbf{C}_{1:D} \mid \Zrd = Y)$
and then $Z_{1:D} \mid \mathbf{C}_{1:D} \sim \prod_{t=1}^D \underset{\mathbb{Z}_{C_t}}{trunc} \,f_\theta$ generates a sample from the desired $P_\theta(\mathbf{Z}_{1:D} \mid \Zrd = Y)$.

\subsection{Proof of Theorem 5.3}

First note that once one of the conditions is attained for the sufficient statistics $\mathbf{N}_{d}$, the condition must continue to hold for all statistics $\mathbf{N}_{t}$ for $t \geq d$ because $\ndeqy$ is non-decreasing in $d$. Therefore,  $P_\theta(Z_{d+1},\dots,Z_D \mid \mathbf{C}_{1:d}, \Zrd = Y) = \prod_{t=d+1}^D P_\theta(Z_{t} \mid \mathbf{C}_{1:d}, \Zrd = Y)$ and so we need only analyze the probability $P_\theta(Z_{d+1} \mid \mathbf{C}_{1:d}, \Zrd = Y)$ for each condition.

Introduce $C_{d+1}$ and marginalize:
\begin{align*}
    P_\theta(Z_{d+1} &\mid \mathbf{C}_{1:d}, \Zrd = Y)\\ = P_\theta(Z&_{d+1} \mid \mathbf{C}_{1:d}, C_{d+1} =\,{\scriptstyle<Y}, \Zrd = Y)P_\theta(C_{d+1} =\,{\scriptstyle<Y} \mid \mathbf{C}_{1:d}, \Zrd = Y)\\ &+P_\theta(Z_{d+1} \mid \mathbf{C}_{1:d}, C_{d+1} =\,{\scriptstyle=Y}, \Zrd = Y)P_\theta(C_{d+1} =\,{\scriptstyle=Y} \mid \mathbf{C}_{1:d}, \Zrd = Y)\\
    &+P_\theta(Z_{d+1} \mid \mathbf{C}_{1:d}, C_{d+1} = \,{\scriptstyle>Y}, \Zrd = Y)P_\theta(C_{d+1} = \,{\scriptstyle>Y} \mid \mathbf{C}_{1:d}, \Zrd = Y)\\
       &=\frac{f_\theta( Z_{d+1})\ind\{Z_{d+1} < Y\}}{F_\theta(Y-1)}p_{d+1}^{\ms{(<Y)}}+\ind\{Z_{d+1} = Y\}p_{d+1}^{\ms{(=Y)}}+\frac{f_\theta( Z_{d+1})\ind\{Z_{d+1} > Y\}}{1 - F_\theta(Y)}p_{d+1}^{\ms{(>Y)}}\\
    &= \frac{f_\theta(Z_{d+1})}{P_\theta(\Zrd = Y \mid \mathbf{N}_d)}\Big(\ind\{Z_{d+1} < Y\}P_\theta(\Zrd = Y \mid \mathbf{N}_d, C_{d+1} =\,{\scriptstyle<Y})\\
        &\hspace{3.7cm}+ \ind\{Z_{d+1} = Y\}P_\theta(\Zrd = Y \mid \mathbf{N}_d, C_{d+1} =\,{\scriptstyle=Y})\\
    &\hspace{3.7cm}+\ind\{Z_{d+1} > Y\}P_\theta(\Zrd = Y \mid \mathbf{N}_d, C_{d+1} =\,{\scriptstyle>Y})
    \Big)\\
        &\propto f_\theta(Z_{d+1})\Big(\ind\{Z_{d+1} < Y\}P_\theta(\Zrd = Y \mid \mathbf{N}_d, C_{d+1} =\,{\scriptstyle<Y})\\
        &\hspace{3.7cm}+ \ind\{Z_{d+1} = Y\}P_\theta(\Zrd = Y \mid \mathbf{N}_d, C_{d+1} =\,{\scriptstyle=Y})\\
        &\hspace{3.7cm}+\ind\{Z_{d+1} > Y\}P_\theta(\Zrd = Y \mid \mathbf{N}_d, C_{d+1} =\,{\scriptstyle>Y})
    \Big)
\end{align*}
where we have used~\Cref{theorem:sampling} and the form of the probabilities $\pdley$, $\pdeqy$, and $\pdgey$. For each of the three conditions, we will analyze each of the probabilities in the final expression, applying the appropriate case of the probability in~\Cref{lemma:suffstat}. It will be useful to recall that $d = \ndley + \ndeqy + \ndgey$.
\begin{enumerate}
    \item \textbf{Condition 1: (Maximal $N_d^{\ms{(<Y)}}$):} $N_d^{\ms{(=Y)}} \geq 1 $ and $N_d^{\ms{(<Y)}} = r - 1$. 

     $N_d^{\ms{(<Y)}} = r - 1$ implies $P_\theta(\Zrd = Y \mid \mathbf{N}_d, C_{d+1} =\,{\scriptstyle<Y}) = 0$. Further, both $N_d^{\ms{(=Y)}} \geq 1$ and $N_d^{\ms{(<Y)}} = r - 1$ with $d < D$ entail that $\ndgey < D - r$. Therefore, the other two probabilities are non-zero. To find their form, we consider two sub-cases: $d = D-1$ and $d < D-1$. Note as well that $N_d^{\ms{(=Y)}} \geq 1 $ and $N_d^{\ms{(<Y)}} = r - 1$ imply $r - \ndley \leq \ndeqy$.

    \begin{enumerate}
        \item First, say
        $d = D -1$. Then $\ndeqy = D - r - \ndgey$. Consider the probability $P_\theta(\Zrd = Y \mid \mathbf{N}_{d}, C_{d+1}=\,{\scriptstyle>Y})$.
        Then the new sufficient statistic is $\ndpgey = \ndgey + 1$ and so
        $\ndpeqy = D - r - \ndpgey + 1$.  Then $\ndpeqy \geq \textrm{max}(r-\ndpley,D - r - \ndpgey + 1)$. Hence $P_\theta(\Zrd = Y \mid \mathbf{N}_{d}, C_{d+1}=\,{\scriptstyle>Y}) = 1$.

        Now consider the probability $P_\theta(\Zrd = Y \mid \mathbf{N}_{d}, C_{d+1}=\,{\scriptstyle=Y})$.
        Then the new sufficient statistic is $\ndpeqy = \ndeqy + 1$. Thus,
        $\ndpeqy = D - r - \ndpgey + 1$ and so $\ndpeqy \geq \textrm{max}(r-\ndpley,D - r - \ndpgey + 1)$. Hence $P_\theta(\Zrd = Y \mid \mathbf{N}_{d}, C_{d+1}=\,{\scriptstyle>Y}) = 1$ as well.

        Because the probabilities are equal, we get
        $$P_\theta(Z_{d+1} \mid \mathbf{C}_{1:d}, \Zrd = Y) \propto f_\theta(Z_{d+1})\ind\{Z_{d+1} \geq Y\}$$
        as desired.
    \item Now say $d < D - 1$. Then $\ndley < D - \ndgey - r$.
    
    Consider the probability $P_\theta(\Zrd = Y \mid \mathbf{N}_{d}, C_{d+1}=\,{\scriptstyle>Y})$.
        Then the new sufficient statistic is $\ndpgey = \ndgey + 1$ and so
        $r - \ndpley\leq \ndpeqy < D - r - \ndpgey + 1$.  Therefore, $P_\theta(\Zrd = Y \mid \mathbf{N}_{d}, C_{d+1}=\,{\scriptstyle>Y}) = 1 - F_\theta^{\ms{(r-\ndley, D - d)}}$.

        Now consider the probability $P_\theta(\Zrd = Y \mid \mathbf{N}_{d}, C_{d+1}=\,{\scriptstyle=Y})$.
        Then the new sufficient statistic is $\ndpeqy = \ndeqy + 1$. Thus, as well,
        $r - \ndpley \leq \ndpeqy < D - r - \ndpgey + 1$ and so $P_\theta(\Zrd = Y \mid \mathbf{N}_{d}, C_{d+1}=\,{\scriptstyle>Y}) = 1 - F_\theta^{\ms{(r-\ndley, D - d)}}$.

        As before, because the probabilities are equal, $P_\theta(Z_{d+1} \mid \mathbf{C}_{1:d}, \Zrd = Y) \propto f_\theta(Z_{d+1})\ind\{Z_{d+1} \geq Y\}$ as desired.
    \end{enumerate}

    \item \textbf{Condition 2 (Maximal $\ndgey$):} $\ndeqy \geq 1 $ and $\ndgey = D - r$. 

    Both facts lead to $\ndley < r - 1$. Therefore, we know that 
    $P_\theta(\Zrd = Y \mid \mathbf{N}_d, C_{d+1} = \,{\scriptstyle>Y}) = 0$ and the other two are non-zero. Note $\ndeqy \geq D - r - \ndgey = 1$. Again consider the sub-cases $d = D -1$ and $d < D - 1$.
    \begin{enumerate}
        \item Say $d = D - 1$. Then $\ndeqy + 1 = r - \ndley$. 
        
        Consider the probability $P_\theta(\Zrd = Y \mid \mathbf{N}_{d}, C_{d+1}=\,{\scriptstyle<Y})$.
        Then the new sufficient statistic is $\ndpley = \ndley + 1$ and so $\ndpeqy = r - \ndpley$. Thus,
        $\ndpeqy \geq \textrm{max}(r-\ndpley,D - r - \ndpgey + 1)$. Hence $P_\theta(\Zrd = Y \mid \mathbf{N}_{d}, C_{d+1}=\,{\scriptstyle<Y}) = 1$.

        Now consider the probability $P_\theta(\Zrd = Y \mid \mathbf{N}_{d}, C_{d+1}=\,{\scriptstyle=Y})$.
        Then the new sufficient statistic is $\ndpeqy = \ndeqy + 1$ and so $\ndpeqy = r - \ndpley$. Hence,
        $\ndpeqy \geq \textrm{max}(r-\ndpley,D - r - \ndpgey + 1)$. Therefore $P_\theta(\Zrd = Y \mid \mathbf{N}_{d}, C_{d+1}=\,{\scriptstyle=Y}) = 1$ as well.~\looseness=-1

        Because both probabilities are equal, we yield the desired truncated probability $P_\theta(Z_{d+1} \mid \mathbf{C}_{1:d}, \Zrd = Y) \propto f_\theta(Z_{d+1})\ind\{Z_{d+1} \leq Y\}$.
        \item Say $d < D - 1$. Then $\ndeqy + 1 < r - \ndley$. 
        
        Consider the probability $P_\theta(\Zrd = Y \mid \mathbf{N}_{d}, C_{d+1}=\,{\scriptstyle<Y})$.
        Then the new sufficient statistic is $\ndpley = \ndley + 1$ and so $\ndpeqy < r - \ndpley$. Thus,
        $D - r - \ndpgey + 1 \leq \ndpeqy < r - \ndpley$. Hence $P_\theta(\Zrd = Y \mid \mathbf{N}_{d}, C_{d+1}=\,{\scriptstyle<Y}) = F_\theta^{\ms{(r-\ndpley-\ndpeqy,D-d}}$.

        Now consider the probability $P_\theta(\Zrd = Y \mid \mathbf{N}_{d}, C_{d+1}=\,{\scriptstyle=Y})$.
        Then the new sufficient statistic is $\ndpeqy = \ndeqy + 1$ and so $\ndpeqy < r - \ndpley$. Hence, again, $D - r - \ndpgey + 1 \leq \ndpeqy < r - \ndpley$ and so the probability is 
       $P_\theta(\Zrd = Y \mid \mathbf{N}_{d}, C_{d+1}=\,{\scriptstyle=Y}) = F_\theta^{\ms{(r-\ndpley-\ndpeqy,D-d}}$.
       
        Both probabilities are equal so
        $P_\theta(Z_{d+1} \mid \mathbf{C}_{1:d}, \Zrd = Y) \propto f_\theta(Z_{d+1})\ind\{Z_{d+1} \leq Y\}$.
    \end{enumerate}

    \item \textbf{Condition 3 (Enough $\ndeqy$):} $\ndeqy \geq \textrm{max}(r-\ndley, D-r-\ndgey)$
    
    First note that if $\ndley \geq r - 1$, then $\ndeqy < D - r - \ndgey - r + 1$ because $d < D$. This is a contradiction, so $\ndley < r - 1$. Further, note that if $\ndgey \geq D - r$, then $\ndeqy < r - \ndley$, again due to $d < D$. This is a contradiction as well, so $\ndgey < D - r$. Therefore, we can write the probability of interest in this case as
    \begin{align*}
        P_\theta(Z_{d+1} &\mid \mathbf{C}_{1:d}, \Zrd = Y)\\
    &\propto f_\theta(Z_{d+1})\Big(\ind\{Z_{d+1} < Y\}P_\theta(\Zrd = Y \mid \mathbf{N}_d, C_{d+1} = \,{\scriptstyle<Y})\\
    &\hspace{3.7cm}+ \ind\{Z_{d+1} = Y\}P_\theta(\Zrd = Y \mid \mathbf{N}_d, C_{d+1} = \,{\scriptstyle=Y})\\
    &\hspace{3.7cm}+\ind\{Z_{d+1} > Y\}P_\theta(\Zrd = Y \mid \mathbf{N}_d,C_{d+1} = \,{\scriptstyle>Y})
    \Big)
    \end{align*}
    Note that no matter which of the sufficient statistics is incremented, the conditions $\ndpeqy \geq r - \ndpley$ and $\ndpeqy \geq D - \ndpgey - r + 1$ hold. Therefore, each of the three probabilities are $1$. Thus, $P_\theta(Z_{d+1} \mid \mathbf{C}_{1:d}, \Zrd = Y) \propto f_\theta(Z_{d+1})$ as desired.
\end{enumerate}

\newpage

\phantomsection
\section{Additional Experimental Results}
\label{sec:addexp}

\subsection{Example: Missing Structure Hides Underdispersion}
\label{sec:example}

We provide a motivating example to illustrate both the challenges and opportunities of modeling underdispersion. We generate synthetic data from the following model 
\begin{align}
    &Y_{i,j} \indsim \textrm{MaxPois}_{\mu_{i,j}}^{\ms{(D)}} \textrm{   where   } \mu_{i,j} =\sum_{k=1}^K \theta_{i,k}\phi_{k,j} \textrm{ for }i \in [40], j\in [40]
\end{align}
where we set $K=5$ and $D=5$, and further draw the parameters from the following gamma priors $\theta_{i,k} \iidsim \Gamma(1,1)$ and $\phi_{k,j}\iidsim \Gamma(1,1)$. This generates a $40 \times 40$ matrix $\mathbf{Y}$ of counts, that are empirically marginally overdispersed $\mathbb{D}[Y_{i,j}] \approx 3.35$. We then fit to this data MaxPoisson models of the same form but with varying $D\in \{1,\dots,10\}$ and $K\in\{1,5,50\}$, where the models for $D=1$ correspond to a Poisson baseline. 

\begin{figure}
\begin{center}
\includegraphics[width=4 in]{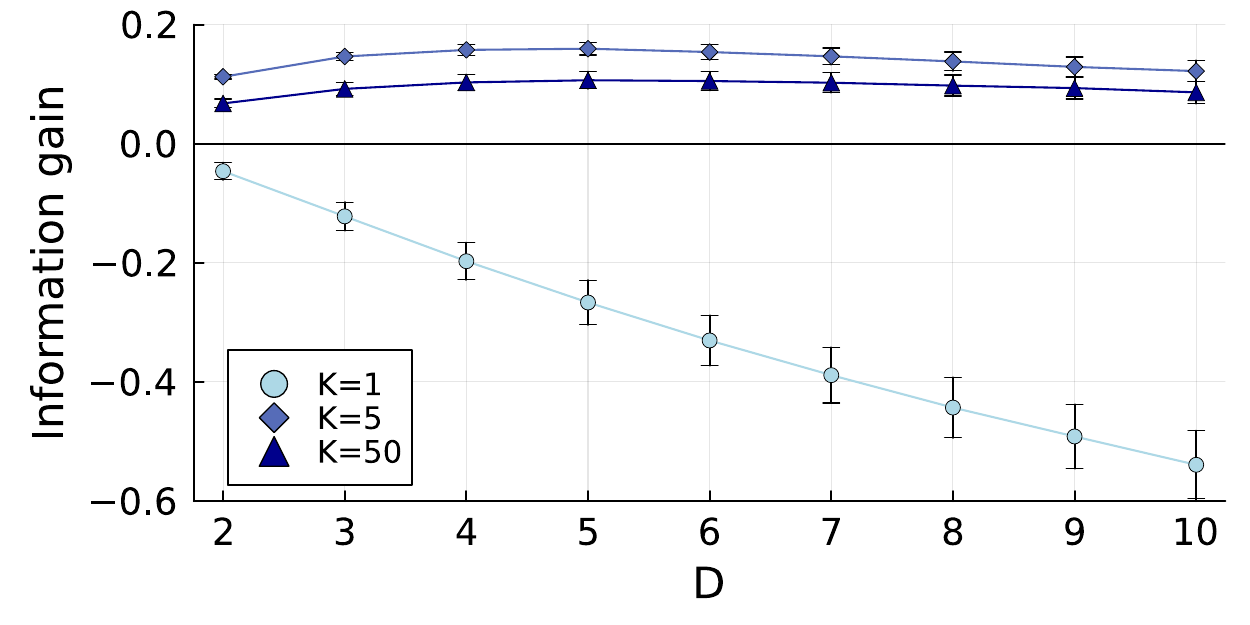}
\end{center}
\caption{A failure to condition on the correct structure masks the benefits of an underdispersed likelihood. Information gain on heldout data averaged over five ground truth datasets and five heldout masks for $D \in \{2,\dots,10\}$ and $K \in \{1,5,50\}$. We fit each model using Gibbs sampling with 5,000 iterations of warmup and then collecting $S=100$ posterior samples each thinned by ten iterations.
\label{hiden_lppd}}
\end{figure}

\Cref{hiden_lppd} displays the heldout \textit{information gain} (higher is better; see~\Cref{sec:casestudies}) for each of the underdispersed models $(D\geq2)$ over the corresponding Poisson baseline $(D=1)$ for each setting of $K$. When underdispersed models have too little structure ($K=1$), the information gain is negative, showing that underdispersion hurts performance. However, when models have sufficient structure ($K \geq 5$), the information gain is positive, indicating that the underdispersed likelihood's ability to concentrate improves predictive performance. 

\Cref{hiden_lppd} illustrates both the challenge and opportunity in modeling underdispersion. A dataset may appear overdispersed marginally, yet still be fit well by a conditionally underdispersed model. However, it is only when a model accounts for sufficient structure that the benefit of an underdispersed likelihood may be realized. Thus one may have to cycle through many underdispersed models before finding one that fits.

\subsection{Case Study II: Additional Results}
\label{sec:addexpcovid}

\cref{table:COVIDinfo} shows information gain over a Poisson model for a range of underdispersed models with latent dimension $K \in \{1,5,10\}$ and order $D_{i,t}$ either inferred with a prior structure described in~\Cref{sec:covid} or fixed with all $D_{i,t}\in\{1,3,5,7,9\}$. The model with inferred $D_{i,t}$ and $K=10$ achieves the best performance. For models with fixed $D_{i,t}$, the information gain improves as the fixed $D_{i,t}$ increases until it gets too large.
\begin{table}[h]
\begin{center}
\begin{tabular}{lccc}

\toprule
$D$ & $K=1$ & $K=5$ & $K=10$ \\
\midrule
$D_{i,t}$ inferred & 
$0.351 \pm 0.002$ & 
$0.361 \pm 0.002$ & 
$\textbf{0.373} \pm 0.004$ \\
$D_{i,t} = 3$ & 
$0.208 \pm 0.014$ & 
$0.237 \pm 0.011$ & 
$0.236 \pm 0.011$ \\
$D_{i,t} = 5$ & 
$0.252 \pm 0.022$ & 
$0.323 \pm 0.017$ & 
$0.326 \pm 0.016$ \\
$D_{i,t} = 7$ & 
$0.226 \pm 0.030$ & 
$0.343 \pm 0.025$ & 
$0.350 \pm 0.023$ \\
$D_{i,t} = 9$ & 
$0.166 \pm 0.038$ & 
$0.334 \pm 0.029$ & 
$0.333 \pm 0.031$ \\
\bottomrule
\end{tabular}
\end{center}
\caption{Information gain ($\uparrow$) over the Poisson model $(D_{i,t}=1)$ for underdispersed models with fixed $D_{i,t}$ and the model with heterogeneous $D_{i,t} \in \{1,3,\dots,D_\textrm{max}\}$. All models improve over the Poisson, with the heterogeneous model performing best.}
\label{table:COVIDinfo}
\end{table}

\subsection{Case Study III: Additional Results}
\label{sec:addexpbirds}

We first built a non-regression variant of the model described in~\Cref{sec:birds} to test whether an underdispersed likelihood leads to improved heldout predictive performance as compared to a Poisson likelihood. As in~\Cref{sec:birds}, we assumed the number of bird species $j$ observed at sampling site $i$ is drawn
\begin{align}
     Y_{i,j} &\sim \textrm{MaxPois}_{\mu_{i,j}}^{\ms{(D_{i,j})}} \textrm{ where } \mu_{i,j} \defeq \sum_{k=1}^K \theta_{i,k}\phi_{kj}
\end{align}
We then assumed the order $D_{i,j}$ is drawn from a shifted binomial:
\label{eq:birdbinomial}
\begin{align*}
(D_{i,j}-1) &\sim \textrm{Binomial}\Big(D_{\textrm{max}}-1,\,\, \textrm{logit}^{-1}\big(\sum_{q=1}^Q \tau_{i,q}\beta_{q,j}\big)\Big)
\end{align*}
and assumed simple priors on the remaining parameters: $\theta_{i,k} \iidsim \Gamma(1,1)$, $\phi_{k,j} \iidsim \Gamma(1,1)$, $\tau_{i,q} \iidsim \mathcal{N}(0,\,1)$, $\beta_{q,j} \iidsim \mathcal{N}(0,\,1)$. All priors yield closed-form complete conditionals after augmentation; see~\Cref{sec:MCMC}.

We set $K=10$ and $D_{\textrm{max}} = 5$. We then use five train-test splits that hold out a random 5\% of the data, and calculate information gain over the Poisson baseline ($D_{i,j}=1$).

\begin{table}
\footnotesize
\begin{center}
\begin{tabular}{lccccc}
\toprule
subset of heldout $i,j$ &all & $\mathbb{E}[D_{i,j}] \in [1,2)$ & $\mathbb{E}[D_{i,j}] \in [2,3)$ & $\mathbb{E}[D_{i,j}] \in [3,4)$ & $\mathbb{E}[D_{i,j}] \in [4,5)$ \\
\midrule
info gain (IG) $(\uparrow)$ & $0.004 \pm .013$ &
$0.001 \pm .004$ &
$\textbf{0.013} \pm .009$ &
$\textbf{0.067} \pm 0.046$ &
$\textbf{0.444} \pm 0.272$ \\
\bottomrule
\end{tabular}
\end{center}
\caption{Information gain compared to a Poisson baseline for a model $K = 10$, for subsets of heldout data points according to the posterior mean of $D_{i,j}$. While the underdispersed model does not improve predictive performance overall, prediction is improved for values in the sampling site-species matrix which are assigned large $D_{i,j}$.\label{table:birds_lppd}}
\end{table}

\Cref{table:birds_lppd} gives the results where, unlike in previous studies, the underdispersed model does not show significant gain over the Poisson model when aggregating over all held out test points. However, we find differences when disaggregating performance across heldout data points according to the posterior mean of $D_{i,j}$, where we find that larger values of inferred $D_{i,j}$, corresponding to more concentrated predictive distributions, are associated with significant gains in predictive performance over the Poisson model. These differences in predictive improvement motivated building the regression model presented in~\Cref{sec:birds} in order to explore heterogeneity in dispersion across bird species.
\newpage

\phantomsection
\section{MCMC Inference}
\label{sec:MCMC}

We present all necessary complete conditional updates to implement Gibbs sampling algorithms for the models considered in the paper. First, we introduce some preliminary facts which
are routinely referenced.

\subsection{Preliminaries}
\begin{fact}\label[fact]{fact:poisadd} (\textbf{Poisson Additivity})
    If $y_i \sim \textrm{Pois}(\theta_i)$ are independent Poisson-distributed random variables, then $\sum_{i=1}^N y_i \sim Pois\left(\sum_{i=1}^N \theta_i\right)$.
\end{fact} 
\begin{fact}\label[fact]{fact:poisthin} (\textbf{Multinomial-Poisson Thinning})
    If $y_i \sim \textrm{Pois}(\theta_i)$ are independent Poisson-distributed random variables, then $(y_1,\dots,y_N) \sim \textrm{Mult}\left(\sum_{i=1}^N y_i, \left(\frac{\theta_1}{\sum_{i=1}^N \theta_i},\dots,\frac{\theta_N}{\sum_{i=1}^N \theta_n}\right)\right)$.
\end{fact} 
\begin{fact}\label[fact]{fact:nbadd}(\textbf{Negative Binomial Additivity})
    If $y_i \sim \textrm{NB}(\alpha_i,p)$ are independent negative binomial-distributed random variables, then $\sum_{i=1}^N y_i\sim NB\left(\sum_{i=1}^N \alpha_i, p\right)$.
\end{fact} 
\begin{fact}\label[fact]{fact:nbmagic} (\textbf{Negative Binomial Augmentation})
    The following two generative processes are equivalent ways of generating from the bivariate distribution $P(y, \ell \mid \alpha, p)$
    \begin{align*}
        &\ell \mid y \sim \textrm{\textrm{CRT}}(y,\alpha), \hspace{,5cm} y \sim \textrm{NB}(\alpha,p)\\
        &y \mid \ell \sim \textrm{SL}(\ell,p),\hspace{.5cm}\ell \sim \textrm{Pois}\left(\alpha\log\left(\frac{1}{1-p}\right)\right)
    \end{align*}
    where a Chinese restaurant table (\textrm{CRT}) random variable has probability mass function $P_\theta(\ell \mid y,\alpha) = \frac{\Gamma(\alpha)}{\Gamma(y+\alpha)} | s(y,\ell)|\alpha^\ell$ and a sum-logarithmic (SL) random variable has $P_\theta(y \mid \ell,p) = \frac{p^y \ell! s(y,\ell)|}{y! (\ln(\frac{1}{1-p}))^\ell}$. $s(.,.)$ denotes the Stirling number of the first kind.
\end{fact} 

\newpage

\subsection{Illustrative Negative Binomial Model}

First, we provide the model and updates for the illustrative example in~\Cref{fig:postpredNB_choice}. For each dataset $Y_1,\dots,Y_{N}$, we assume each $Y_i$ is generated as:
\begin{align*}
    &Y_i \iidsim \textrm{NB}_{\alpha,p}^{\ms{(r,D)}}\hspace{1cm}\textrm{ for }i \in [N]\\
    &\alpha \sim \Gamma(a,b)\hspace{.5cm}p \sim\textrm{Beta}(e,f)
\end{align*}
To generate~\Cref{fig:postpredNB_choice}, we set $a=2$, $b=0.01$, $e=f=1$, $D=5$, $r=3$ and $N=1000$. Inference proceeds as follows.

\textit{Order statistic augmentation}: For each $i \in [N]$, using~\Cref{alg:generalsample}, with parent distribution $f_\theta = \textrm{NB}_{\alpha,p}$, rank $r$ and order $D$,
$$Z_i = \sum_{d=1}^D Z_{i,d}, \textrm{ where } \mathbf{Z}_{i,1:D}\sim P_{\alpha,p}(\mathbf{Z}_{i,1:D} \mid \Zrd = Y_i)$$

\textit{Sampling $\alpha$}: For each $i \in [N]$, sample $Z_i' \sim \textrm{CRT}(Z_i,Dr)$. Using~\Cref{fact:nbmagic}, $Z_i'$ is marginally Poisson distributed with rate $D\log(1/(1-p))\alpha$. Then by gamma-Poisson conjugacy,$$\alpha \sim \Gamma\left(a + \sum_{i=1}^N Z_i',b+ND\log\left(\frac{1}{1-p}\right)\right)$$

\textit{Sampling $p$}: By~\Cref{fact:nbadd}, $Z_i  \sim NB(D\alpha,p)$, so via beta-negative binomial conjugacy,
$$p \sim \textrm{Beta}\left (e + \sum_{i=1}^N Z_{i}, f + ND\alpha\right) $$
\newpage

\subsection{Toy Example Model}
\label{appsec:toy}

We present updates for the example matrix factorization model described in~\Cref{sec:example}.

\textit{Order statistic augmentation}: For each $i \in [40]$ and $j \in [40]$, using~\Cref{alg:generalsample},
$$Z_{i,j} = \sum_{d=1}^D Z_{i,j,d}\textrm{ where }\mathbf{Z}_{i,j,1:D} \sim P_{\mu_{i,j}}(\mathbf{Z}_{i,j,1:D} \mid \Zdd = Y_{i,j})$$

\textit{Sampling the latent subcounts}: Via~\Cref{fact:poisadd}, $Z_{i,j} \sim \textrm{Pois}\left(D\sum_{k=1}^K \theta_{i,k}\phi_{k,j}\right)$ and so using~\Cref{fact:poisthin},
$$Z_{i,j,1}',\dots,Z_{i,j,K}' \sim\textrm{Mult}\left(Z_{i,j},\Bigg[\frac{\theta_{i,1}\phi_{1,j}}{\sum_{k=1}^K \theta_{i,k}\phi_{k,j}},\dots,\frac{\theta_{i,K}\phi_{K,j}}{\sum_{k=1}^K \theta_{i,k}\phi_{k,j}}\Bigg]\right)$$

\textit{Sampling $\theta_{i,k}$}: For each $i \in [40]$ and $k \in [K]$, via gamma-Poisson conjugacy,
$$\theta_{i,k} \sim \Gamma\left(1 + \sum_{j=1}^{40} Z_{i,j,k}', 1+D\sum_{j=1}^{40} \phi_{k,j}\right)$$

\textit{Sampling $\phi_{k,j}$}: For each $j \in [40]$ and $k \in [K]$, via gamma-Poisson conjugacy,
$$\phi_{k,j} \sim \Gamma\left(1 + \sum_{i=1}^{40} Z_{i,j,k}', 1+D\sum_{i=1}^{40} \theta_{i,k}\right)$$

\newpage

\subsection{Case Study I: Frontier Flight Times}

The updates to the models described in~\Cref{sec:flights} differ slightly based on whether $D$ is fixed or varies by route $k$. We present updates for both versions below.

\subsubsection{Fixed $D$}
The full model is
\begin{align*}
   Y_i &\indsim \textrm{MedPois}^{\ms{(D)}}_{\mu_i} \textrm{  where  }\mu_i \defeq a_{\textrm{orig}[i]} + b_{\textrm{dest}[i]} + c_{\textrm{route}[i]}\, \textrm{dist}_{\textrm{route}[i]} &&\textrm{ for } i \in [N]\\
    c_k &\iidsim \Gamma(1,1) &&\textrm{ for } k \in [K]\\
   a_{j} &\iidsim \Gamma(1,1)\textrm{ and } b_j \iidsim \Gamma(1,1) &&\textrm{ for } j \in [A]
\end{align*}
with $N$ flights, $K$ routes, and $A$ airports. Inference on latent parameters can proceed as follows.

\textit{Order statistic augmentation}: For each $i \in [N]$, using~\Cref{alg:generalsample}, $$Z_{i} = \sum_{d=1}^D Z_{i,d} \textrm{ where }\mathbf{Z}_{i,1:D} \sim P_{\mu_i}(\mathbf{Z}_{i,1:D}\mid Z^{\ms{(\lceil D/2\rceil,D)}} = Y_i)$$

\textit{Sample latent subcounts}: For each route $k \in [K]$, form $Z_k = \sum_{i=1}^N \ind\{\textrm{route}[i] = k\}Z_{i}$ and $\mu_k = \sum_{i=1}^N \ind\{\textrm{route}[i] = k\}\mu_{i}$. Then using~\Cref{fact:poisthin},
$$Z_k^{(1)},Z_k^{(2)},Z_k^{(3)} \sim \textrm{Mult}\left(Z_k, \Bigg[\frac{a_{\textrm{orig}[k]}}{\mu_k},\frac{b_{\textrm{dest}[k]}}{\mu_k},\frac{\textrm{dist}_{\textrm{orig}[k],\textrm{dest}[k]}c_{k}}{\mu_k}\Bigg]\right)$$
Further, denote $N_k=\sum_{i=1}^N\ind\{\textrm{route}[i] = k\}$.

\textit{Sampling $a_j$ and $b_j$}: For each airport $j \in [A]$, via gamma-Poisson conjugacy, 
$$a_j \sim \Gamma\left(1+ \sum_{k=1}^K Z^{(1)}_k \ind\{\textrm{orig}[k] = j\}, 1 + D\sum_{k=1}^K N_k \ind\{\textrm{orig}[k] = j\}\right)$$
and 
$$b_j \sim \Gamma\left(1+ \sum_{k=1}^K Z^{(2)}_k \ind\{\textrm{dest}[k] = j\}, 1 + D\sum_{k=1}^K N_k \ind\{\textrm{dest}[k] = j\}\right)$$

\textit{Sampling $\mu_k$}: For each route $k$, via gamma-Poisson conjugacy,
$$c_k \sim \Gamma\left(1+ Z^{(3)}_k, 1 + D \textrm{dist}_{\textrm{orig}[k],\textrm{dest}[k]}N_k \}\right)$$

\subsubsection{Route-specific $D_k$}
The full model is now
\begin{align*}
   Y_i &\indsim \textrm{MedPois}^{\ms{(D_{\textrm{route}[i]})}}_{\mu_i} \textrm{  where  }\mu_i \defeq a_{\textrm{orig}[i]} + b_{\textrm{dest}[i]} + c_{\textrm{route}[i]}\, \textrm{dist}_{\textrm{route}[i]} &&\textrm{ for }i \in [N]\\
   D_k &\iidsim \textrm{OddBinomial}(D_{\textrm{max}},\, \rho) \textrm{ and } c_k \iidsim \Gamma(1,1) &&\textrm{ for } k \in [K]\\
   a_{j} &\iidsim \Gamma(1,1)\textrm{ and } b_j \iidsim \Gamma(1,1) &&\textrm{ for } j \in [A]\\
   \rho &\sim \textrm{Beta}(1,1)
\end{align*}

with $N$ flights, $K$ routes, and $A$ airports. Inference on latent parameters can proceed as follows.

\textit{Order statistic augmentation}: For each $i \in [N]$, using~\Cref{alg:generalsample}, $$Z_{i} = \sum_{d=1}^{D_{\textrm{route}[i]}} Z_{i,d} \textrm{ where }\mathbf{Z}_{i,1:D_{\textrm{route}[i]}} \sim P_{\mu_i}(\mathbf{Z}_{i,1:D_{\textrm{route}[i]}}\mid Z^{\ms{(\lceil D_{\textrm{route}[i]}/2\rceil,D_{\textrm{route}[i]})}} = Y_i)$$

\textit{Sample latent subcounts}: For each route $k \in [K]$, form $Z_k = \sum_{i=1}^N \ind\{\textrm{route}[i] = k\}Z_{i}$ and $\mu_k = \sum_{i=1}^N \ind\{\textrm{route}[i] = k\}\mu_{i}$. Then using~\Cref{fact:poisthin},
$$Z_k^{(1)},Z_k^{(2)},Z_k^{(3)} \sim \textrm{Mult}\left(Z_k, \Bigg[\frac{a_{\textrm{orig}[k]}}{\mu_k},\frac{b_{\textrm{dest}[k]}}{\mu_k},\frac{\textrm{dist}_{\textrm{orig}[k],\textrm{dest}[k]}c_{k}}{\mu_k}\Bigg]\right)$$
Further, denote $N_k$ as $\sum_{i=1}^N\ind\{\textrm{route}[i] = k\}$.

\textit{Sampling $a_j$ and $b_j$}: For each airport $j$, via gamma-Poisson conjugacy, 
$$a_j \sim \Gamma\left(1+ \sum_{k=1}^K Z^{(1)}_k \ind\{\textrm{orig}[k] = j\}, 1 + \sum_{k=1}^K D_k N_k \ind\{\textrm{orig}[k] = j\}\right)$$
and 
$$b_j \sim \Gamma\left(1+ \sum_{k=1}^K Z^{(2)}_k \ind\{\textrm{dest}[k] = j\}, 1 + \sum_{k=1}^K D_k N_k \ind\{\textrm{dest}[k] = j\}\right)$$

\textit{Sampling $\mu_k$}: For each route $k$, via gamma-Poisson conjugacy,
$$c_k \sim \Gamma\left(1+ Z^{(3)}_k, 1 + D_k \textrm{dist}_{\textrm{orig}[k],\textrm{dest}[k]}N_k \}\right)$$

\textit{Sampling $D_k$}: Define $d \in \{1,3,\dots,D_{max}\}$. For each route $k \in [K]$, sample according to the categorical distribution, 
$$P_\theta(D_k = d) \propto {\tfrac{D_{max} - 1}{2}  \choose \tfrac{d-1}{2}} p^{\tfrac{d-1}{2}}(1-p)^{\tfrac{D_{max}-d}{2}}\prod_{i=1}^{N_k} \textrm{MedPois}^{\ms{(d})}_{\mu_i}(Y_i)$$

\textit{Sampling $\rho$}: Via beta-binomial conjugacy,
$$\rho \sim \textrm{Beta}\left(1+\tfrac{1}{2}\left(\sum_{k=1}^K D_k - K\right), 1 + \tfrac{1}{2}\left(KD_{max} - \sum_{k=1}^K D_k\right)\right)$$

\subsection{Case Study II: Cumulative COVID-19 Case Counts}

Again, the updates for the models from~\Cref{sec:covid} vary depending on whether $D$ is fixed. We present both versions.

\subsubsection{Fixed $D$}
The full model is
\begin{align*}
    Y_{i,t} &\sim \textrm{MedPois}^{\ms{(D)}}_{\mu_{i,t}} \hspace{0.25em}\textrm{ where }\hspace{0.25em}\mu_{i,t} \,\defeq\, Y_{i,t-1} \,\,+ \,\,  \log (\textrm{pop}_i)\Big(\varepsilon + \alpha\sum_{k=1}^K \theta_{i,k}\phi_{k,t}\Big)\\
    \phi_{k,t} &\sim \Gamma(a^{\ms{(\phi)}} +b^{\ms{(\phi)}}\phi_{k,t-1},\,\,b^{\ms{(\phi)}}) \textrm{ for }t \in \{2,\dots,T\}\textrm{ and }
\phi_{k,1} \sim \Gamma(1,1)\\
\theta_{i,k} &\iidsim \Gamma(a^{\ms{(\theta)}},b^{\ms{(\theta)}}),\,\, \varepsilon \sim \Gamma(e,f),\,\, \alpha \sim \Gamma(g,h)
\end{align*}
for county $i \in [C]$ and time $t \in [T]$. Inference can proceed as follows.

\textit{Order statistic augmentation}: For each $i \in [C]$ and $t \in [T]$, using~\Cref{alg:generalsample}, 
$$Z_{i,t} = \sum_{d=1}^D Z_{i,t,d} \textrm{ where } \mathbf{Z}_{i,t,1:D} \sim P_{\mu_{i,t}}(\mathbf{Z}_{i,t,1:D} \mid Z^{\ms{(\lceil D/2\rceil,D)}})$$

\textit{Sample latent subcounts}: For each $i \in [C]$ and $t \in [T]$, using~\Cref{fact:poisthin},
$$Z_{i,t,1}',\dots Z_{i,t,K}',Z_{i,t}',Z_{i,t}'' \sim\textrm{Mult}\left(Z_{i,t}, \left[\alpha \log(\textrm{pop}_i)\frac{\theta_{i,1}\phi_{t,1}}{\mu_{i,t}},\dots,\alpha \log(\textrm{pop}_i)\frac{\theta_{i,K}\phi_{t,K}}{\mu_{i,t}},\frac{Y_{i,t\txm1}}{\mu_{i,t}},\frac{\log(\textrm{pop}_i)\varepsilon}{\mu_{i,t}}\right]\right)$$

\textit{Sample $\alpha$}: Using gamma-Poisson conjugacy,
$$\alpha \sim \Gamma\left(g + \sum_{i=1}^C\sum_{t=1}^T\sum_{k=1}^K Z_{i,t,k}', h + D \sum_{i=1}^C\sum_{t=1}^T\sum_{k=1}^K \log(\textrm{pop}_i)\theta_{i,k}\phi_{k,t}\right)$$

\textit{Sample $\varepsilon$}: Also via gamma-Poisson conjugacy,
$$\varepsilon \sim \Gamma\left(e + \sum_{i=1}^C \sum_{t=1}^T Z_{i,t}'', f + DT\sum_{i=1}^C \log(\textrm{pop}_i)\right)$$

\textit{Sample $\theta_{i,k}$}: For each $i \in [C]$ and $k \in [K]$, using gamma-Poisson conjugacy,
$$\theta_{i,k} \sim \Gamma\left(a^{\ms{(\theta)}} + \sum_{t=1}^T Z_{i,t, k}', b^{\ms{(\theta)}} + \alpha D  \log(\textrm{pop}_i)\sum_{t=1}^T \phi_{k,t}\right)$$

\textit{Sample $\phi_{k,t}$}: We use an adaptation of the  data augmentation strategy developed in~\citet{schein2016pgds}. For each $k \in [K]$, the procedure proceeds as follows.

Initialize $\ell_{k,T\txp1} = q_{k,T\txp1} = 0$. Then for each $t \in \{T,\dots,2\}$,
\begin{align*}
    &q_{k,t} = \log\left(1+ \tfrac{D\alpha}{b^{\ms{(\phi)}}}\sum_{i=1}^C \log(\textrm{pop}_i)\theta_{i,k} + q_{k,t\txp1}\right)\\
    &\ell_{k,t}' \sim \textrm{CRT}(\sum_{i=1}^C Z_{i, t,k}' + \ell_{k,t\txp1}, b^{\ms{(\phi)}} \phi_{k,t\txm1} + a^{\ms{(\phi)}})\\
    &\ell_{k,t} \sim \textrm{Binomial}\left(\ell_{k,t}', \tfrac{b^{\ms{(\phi)}}\phi_{k,t\txm1}}{b^{\ms{(\phi)}}\phi_{k,t\txm1} + a^{\ms{(\phi)}}}\right)
\end{align*}

Then, using~\Cref{fact:nbmagic}, each $\ell_{k,t}$ is marginally Poisson-distributed, so by gamma-Poisson conjugacy, 
$$\phi_{k,1} \sim \Gamma\left(1 + \sum_{i=1}^C Z_{i,1,k}' + \ell_{k,2}, 1 + D\alpha \sum_{i=1}^C \log(\textrm{pop}_i)\theta_{i,k} + b^{\ms{(\phi)}}q_{k,2}\right)
$$ and for $t \in \{2,\dots,T\}$
$$\phi_{k,t } \sim \Gamma\left(a^{\ms{(\phi)}} + b^{\ms{(\phi)}}\phi_{k,t\txm1} + \sum_{i=1}^C Z_{i,t,k}' + \ell_{k,t\txp1}, b^{\ms{(\phi)}} + D\alpha \sum_{i=1}^C \log(\textrm{pop}_i)\theta_{i,k} + b^{\ms{(\phi)}}q_{k,t+1}\right)$$

\subsubsection{County-Time-specific $D_{i,t}$}

Now the full model is
\begin{align*}
    Y_{i,t} &\sim \textrm{MedPois}^{\ms{(D_{i,t})}}_{\mu_{i,t}} \hspace{0.25em}\textrm{ where }\hspace{0.25em}\mu_{i,t} \,\defeq\, Y_{i,t-1} \,\,+ \,\,  \log (\textrm{pop}_i)\Big(\varepsilon + \alpha\sum_{k=1}^K \theta_{i,k}\phi_{k,t}\Big)\\
    D_{i,t} &\sim \textrm{OddBinomial}\left(D_{\textrm{max}},\, \rho_{i,t}\right) \textrm{ where } \rho_{i,t} \defeq \textrm{logit}^{-1}\big(\sum_{q=1}^Q \beta_{i,q}\tau_{q,t}\big)\\
   \phi_{k,t} &\sim \Gamma(a^{\ms{(\phi)}} +b^{\ms{(\phi)}}\phi_{k,t-1},\,\,b^{\ms{(\phi)}}) \,\,\,\textrm{ and }\,\,\,
\tau_{q,t} \sim \mathcal{N}(a^{\ms{(\tau)}}+b^{\ms{(\tau)}}\tau_{q,t-1},\,\nicefrac{1}{\lambda_t^{\ms{(\tau)}}})\\
\phi_{k,1} &\sim \Gamma(1,1)\textrm{ and } \tau_{q,1} \sim \mathcal{N}(0,\nicefrac{1}{\lambda_1^{\ms{(\tau)}}}),\,\, \lambda^{\ms{(\tau)}}_t \iidsim \Gamma(1,1)\\
\theta_{i,k} &\iidsim \Gamma(a^{\ms{(\theta)}} ,\,\,b^{\ms{(\theta)}}), \varepsilon \sim \Gamma(e,\,f),\,\, \alpha \sim \Gamma(g,\,h)\\
\beta_{i,q} &\indsim \mathcal{N}(0,\nicefrac{1}{\lambda^{\ms{(\beta)}}_i}),\,\, \lambda^{\ms{(\beta)}}_i \iidsim \Gamma(1,1)
\end{align*}
for county $i \in [C]$ and time $t \in [T]$. Inference can proceed as follows.

\textit{Order statistic augmentation}: For each $i \in [C]$ and $t \in [T]$, using~\Cref{alg:generalsample},
$$Z_{i,t} = \sum_{d=1}^{D_{i,t}} Z_{i,t,d} \textrm{ where } \mathbf{Z}_{i,t,1:D_{i,t}} \sim P_{\mu_{i,t}}(\mathbf{Z}_{i,t,1:D_{i,t}} \mid Z^{\ms{(\lceil D_{i,t}/2\rceil,D_{i,t})}})$$

\textit{Sample latent subcounts}: For each $i \in [C]$ and $t \in [T]$, using~\Cref{fact:poisthin},
$$Z_{i,t,1}',\dots Z_{i,t,K}',Z_{i,t}',Z_{i,t}'' \sim\textrm{Mult}\left(Z_{i,t}, \left[\alpha \log(\textrm{pop}_i)\frac{\theta_{i,1}\phi_{1,t}}{\mu_{i,t}},\dots,\alpha \log(\textrm{pop}_i)\frac{\theta_{i,K}\phi_{K,t}}{\mu_{i,t}},\frac{Y_{i,t\txm1}}{\mu_{i,t}},\frac{\log(\textrm{pop}_i)\varepsilon}{\mu_{i,t}}\right]\right)$$

\textit{Sample $\alpha$}: Using gamma-Poisson conjugacy,
$$\alpha \sim \Gamma\left(g + \sum_{i=1}^C\sum_{t=1}^T\sum_{k=1}^K Z_{i,t,k}', h +  \sum_{i=1}^C\sum_{t=1}^T\sum_{k=1}^K D_{i,t} \log(\textrm{pop}_i)\theta_{i,k}\phi_{k,t}\right)$$

\textit{Sample $\varepsilon$}: Also via gamma-Poisson conjugacy,
$$\varepsilon \sim \Gamma\left(e + \sum_{i=1}^C\sum_{t=1}^T Z_{i,t}'', f + \sum_{i=1}^C \sum_{t=1}^TD_{i,t}\log(\textrm{pop}_i)\right)$$

\textit{Sample $\theta_{i,k}$}: For each $i \in [C]$ and $k \in [K]$, using gamma-Poisson conjugacy,
$$\theta_{i,k} \sim \Gamma\left(a^{\ms{(\theta)}} + \sum_{t=1}^T Z_{i,t,k}', b^{\ms{(\theta)}} +   \log(\textrm{pop}_i)\alpha\sum_{t=1}^T D_{i,t}\phi_{k,t}\right)$$

\textit{Sample $\phi_{k,t}$}: As before, we use the data augmentation scheme from~\citet{schein2016pgds}. For each $k \in [K]$, complete the following procedure.

Initialize $\ell_{k,T\txp1} = q_{k,T\txp1} = 0$. Then for each $t \in \{T,\dots,2\}$,
\begin{align*}
    &q_{k,t} = \log\left(1+ \tfrac{\alpha}{b^{\ms{(\phi)}}}\sum_{i=1}^C D_{i,t} \log(\textrm{pop}_i)\theta_{i,k} + q_{k,t\txp1}\right)\\
    &\ell_{k,t}' \sim \textrm{CRT}\left(\sum_{i=1}^C Z_{i,t,k}' + \ell_{k,t\txp1}, b^{\ms{(\phi)}} \phi_{k,t\txm1} + a^{\ms{(\phi)}}\right)\\
    &\ell_{k,t} \sim \textrm{Binomial}\left(\ell_{k,t}', \tfrac{b^{\ms{(\phi)}}\phi_{k,t\txm1}}{b^{\ms{(\phi)}}\phi_{k,t\txm1} + a^{\ms{(\phi)}}}\right)
\end{align*}
Then, using gamma-Poisson conjugacy,
$$\phi_{k,1} \sim \Gamma\left(1 + \sum_{i=1}^C Z_{i, 1, k} + \ell_{k,2}, 1 + \alpha \sum_{i=1}^C D_{i,t} \log(\textrm{pop}_i)\theta_{i,k} + b^{\ms{(\phi)}}q_{k,2}\right)
$$ and for $t \in \{2,\dots,T\}$
$$\phi_{k,t } \sim \Gamma\left(a^{\ms{(\phi)}} + b^{\ms{(\phi)}}\phi_{k,t\txm1} + \sum_{i=1}^C Z_{i, t, k}' + \ell_{k,t\txp1}, b^{\ms{(\phi)}} + \alpha \sum_{i=1}^C D_{i,t} \log(\textrm{pop}_i)\theta_{i,k} + b^{\ms{(\phi)}}q_{k,t+1}\right)$$

\textit{Sample $D_{i,t}$}: Define $d \in \{1,3,\dots,D_{max}\}$. For each $D_{i,t}$, sample according to the categorical distribution, 
$$P(D_{i,t} = d) \propto {\tfrac{D_{max} - 1}{2}  \choose \tfrac{d-1}{2}} p^{\tfrac{d-1}{2}}(1-p)^{\tfrac{D_{max}-d}{2}} \textrm{MedPois}^{\ms{(d})}_{\mu_{i,t}}(Y_{i,t})$$

\textit{Introduce $\omega_{i,t}$}: The updates to each $\beta_{i,q}$ and $\tau_{q,t}$ can proceed via P\'olya-gamma augmentation~\citep{polson2013bayesian}. 
For each $D_{i,t}$, sample $\omega_{i,t} \sim \textrm{PG}\left(\tfrac{D_{max} - 1}{2}, \sum_{q=1}^Q \tau_{i,q}\beta_{q,t}\right)$ as a draw from the P\'olya-gamma distribution. 

For the updates below, denote $\boldsymbol{\beta} \in \mathbb{R}^{C \times Q}$ and $\boldsymbol{\tau} \in \mathbb{R}^{Q \times T}$ as matrices of parameters $\tau_{i,q}$ and $\tau_{q,t}$ respectively. Further, define $\boldsymbol{\beta}_i$ as row of $\boldsymbol{\beta}$ and $\boldsymbol{\tau}_t$ as a column of $\boldsymbol{\tau}$.

\textit{Sample $\boldsymbol{\beta}_i$}: 
For each $i \in [C]$, define $\boldsymbol{\Omega}_i = \textrm{Diag}(\omega_{i,1},\dots,\omega_{i,T})$. Then compute $\mathbf{V}_i = \boldsymbol{\tau} \boldsymbol{\Omega}_i\boldsymbol{\tau}^T + \lambda_{i}^{\ms{(\beta)}}\mathbf{I}_Q$ where $\mathbf{I}_Q$ is the identity matrix. Define $\mathbf{D}_i = \begin{bmatrix}
    D_{i,1}&\dots&D_{i,T}
\end{bmatrix}$. Then compute $\boldsymbol{\kappa}_i = \nicefrac{1}{2}\mathbf{D}_i - \nicefrac{D_{max}}{4} - \nicefrac{1}{4}$. Finally, sample $\boldsymbol{\beta}_i \sim \textrm{MVN}\left(V_i \boldsymbol{\tau} \boldsymbol{\kappa}_i, V_i\right)$.

\textit{Sample $\boldsymbol{\tau}_t$}: For each $t \in [T]$, define $\boldsymbol{\Omega}_t = \textrm{Diag}(\omega_{1,t},\dots,\omega_{C,t})$. Then compute $\mathbf{V}_t = \boldsymbol{\beta}^T \boldsymbol{\Omega}_t\boldsymbol{\beta} + \left(\lambda_{t}^{\ms{(\tau)}} + \big(b^{\ms{(\tau)}}\big)^2\lambda_{t+1}^{\ms{(\tau)}}\right)\mathbf{I}_Q$ for $t \neq T$. For $t = T$, compute instead $\mathbf{V}_T = \boldsymbol{\beta}^T \boldsymbol{\Omega}_T\boldsymbol{\beta} + \lambda_{T}^{\ms{(\tau)}}\mathbf{I}_Q$.  Define $\mathbf{D}_t = \begin{bmatrix}
    D_{1,t}&\dots&D_{C,t}
\end{bmatrix}$. Then compute $\boldsymbol{\kappa}_t = \nicefrac{1}{2}\mathbf{D}_t - \nicefrac{D_{max}}{4} - \nicefrac{1}{4}$. Then update
\begin{align*}
  \boldsymbol{\tau}_1 &\sim \textrm{MVN}\left(V_1 \boldsymbol{\beta}^T \boldsymbol{\kappa}_1 + \lambda_{1}^{\ms{(\tau)}}b^{\ms{(\tau)}}\mathbf{I}_Q  + \lambda_{2}^{\ms{(\tau)}}b^{\ms{(\tau)}}\mathbf{I}_Q(\boldsymbol{\tau}_{2} - a^{\ms{(\tau)}}\ind), V_1\right)  &&\textrm{ for }t=1\\
  \boldsymbol{\tau}_t &\sim \textrm{MVN}\left(V_t \boldsymbol{\beta}^T \boldsymbol{\kappa}_t + \lambda_{t}^{\ms{(\tau)}}\mathbf{I}_Q(a^{\ms{(\tau)}}\ind + b^{\ms{(\tau)}}\boldsymbol{\tau}_{t\txm 1}) + \lambda_{t+1}^{\ms{(\tau)}}b^{\ms{(\tau)}}\mathbf{I}_Q(\boldsymbol{\tau}_{t\txp 1} - a^{\ms{(\tau)}}\ind), V_t\right) &&\textrm{ for }1< t<T\\
  \boldsymbol{\tau}_T &\sim \textrm{MVN}\left(V_T \boldsymbol{\beta}^T \boldsymbol{\kappa}_T + \lambda_{T}^{\ms{(\tau)}}\mathbf{I}_Q(a^{\ms{(\tau)}}\ind + b^{\ms{(\tau)}}\boldsymbol{\tau}_{T\txm 1}), V_T\right) &&\textrm{ for }t=T
\end{align*}
where here $\ind$ denotes the all-ones vector.

\textit{Sample $\lambda^{(\beta)}_i$}: For each $i \in [C]$, via normal-gamma conjugacy,
$$\lambda^{(\beta)}_i \sim \Gamma\left(1 + \nicefrac{Q}{2}, 1 + \nicefrac{1}{2}\sum_{q=1}^Q \beta_{i,q}^2\right)$$

\textit{Sample $\lambda^{(\tau)}_t$}: For each $t \in [T]$, via normal-gamma conjugacy,
\begin{align*}
    \lambda^{(\tau)}_1 &\sim \Gamma\left(1 + \nicefrac{Q}{2}, 1 + \nicefrac{1}{2}\sum_{q=1}^Q (\tau_{q,t}-a^{\ms{(\tau)}})^2\right)&&\textrm{ for }t=1\\
    \lambda^{(\tau)}_t &\indsim \Gamma\left(1 + \nicefrac{Q}{2}, 1 + \nicefrac{1}{2}\sum_{q=1}^Q (\tau_{q,t}-a^{\ms{(\tau)}}-b^{\ms{(\tau)}}\tau_{q,t\txm1})^2\right)&&\textrm{ for }t \in \{2,\dots,T\}\\
\end{align*}
\subsection{Case Study III: Finnish Bird Abundance}

\subsubsection{No Covariates}

We first present complete updates for the version of the model in~\Cref{sec:addexpbirds}, which lacks covariates. We omit the updates for the baseline comparsion where $D_{i,j}=1$ is fixed, which are similar. The full model is
\begin{align*}
    Y_{i,j} &\sim \textrm{MaxPois}_{\mu_{i,j}}^{\ms{(D_{i,j})}} \textrm{ where } \mu_{i,j} \defeq \sum_{k=1}^K \theta_{i,k}\phi_{kj}\\
    (D_{i,j}-1) &\sim \textrm{Binomial}\Big(D_{\textrm{max}}-1,\,\, \textrm{logit}^{-1}\big(\sum_{q=1}^Q \tau_{i,q}\beta_{q,j}\big)\Big)\\
    \theta_{i,k} &\sim \Gamma(a^{\ms{(\theta)}},b^{\ms{(\theta)}}),\hspace{.5cm}\phi_{k,j} \sim \Gamma(a^{\ms{(\phi)}},b^{\ms{(\phi)}})\\
    \tau_{i,q} &\iidsim \mathcal{N}(0,\,1),\hspace{.5cm} \beta_{q,j} \iidsim \mathcal{N}(0,1)
\end{align*}
for sampling site $i \in [N]$ and species $j \in [M]$.

\textit{Order statistic augmentation}: For each $i\in [N]$ and $j\in [M]$, using~\Cref{alg:generalsample}, $$Z_{i,j} = \sum_{d=1}^{D_{i,j}} Z_{i,j,d} \textrm{ where }\mathbf{Z}_{i,j,1:D_{i,j}} \sim P_{\mu_{i,j}}(\mathbf{Z}_{i,1:D_{i,j}}\mid Z^{\ms{(D_{i,j},D_{i,j}})} = Y_{i,j})$$

\textit{Sample latent subcounts}: For each $i\in [N]$ and $j\in [M]$, using~\Cref{fact:poisthin},
$$Z_{i,j,1}',\dots Z_{i,j,K}'\sim\textrm{Mult}\left(Z_{i,j}, \left[ \frac{\theta_{i,1}\phi_{1,j}}{\mu_{i,j}},\dots,\frac{\theta_{i,K}\phi_{K,j}}{\mu_{i,j}}\right]\right)$$

\textit{Sample $\theta_{i,k}$}: For each $i \in [N]$ and $k \in [K]$, using gamma-Poisson conjugacy,
$$\theta_{i,k} \sim \Gamma\left(a^{\ms{(\theta)}} + \sum_{j=1}^M Z_{i,j,k}', b^{\ms{(\theta)}} + \sum_{j=1}^M D_{i,j}\phi_{k,j}\right)$$

\textit{Sample $\phi_{k,j}$}: For each $j \in [M]$ and $k \in [K]$, using gamma-Poisson conjugacy,
$$\phi_{k,j} \sim \Gamma\left(a^{\ms{(\phi)}} + \sum_{i=1}^N Z_{i, j, k}', b^{\ms{(\phi)}} +  \sum_{i=1}^N D_{i,j}\theta_{i,k}\right)$$

\textit{Sample $D_{i,j}$}: Define $d \in \{1,2,\dots,D_{max}\}$. For each $i\in [N]$ and $j\in [M]$, sample according to the categorical distribution,
$$P_\theta(D_{i,j} = d) \propto {D_{max} - 1  \choose d-1} \rho_{i,j}^{d-1}(1-\rho_{i,j})^{D_{max}-d} \textrm{MaxPois}_{\mu_{i,j}}^{\ms{(d)}}(Y_{i,j})$$
where $\rho_{i,j} = \textrm{logit}^{-1}\left(\sum_{q=1}^Q \tau_{i,q}\beta_{q,j}\right)$.

\textit{Introduce $\omega_{i,j}$}: The updates to each $\tau_{i,q}$ and $\beta_{j,q}$ proceed via P\'olya-gamma augmentation~\citep{polson2013bayesian}. 
For each $i\in [N]$ and $j\in [M]$, sample $\omega_{i,j} \sim \textrm{PG}\left(D_{max} - 1, \sum_{q=1}^Q \tau_{i,q}\beta_{q,j}\right)$ as a draw from the P\'olya-gamma distribution. 

For the updates below, denote $\boldsymbol{\tau} \in \mathbb{R}^{N \times Q}$ and $\boldsymbol{\beta} \in \mathbb{R}^{Q \times M}$ as matrices of parameters $\tau_{i,q}$ and $\beta_{q,j}$ respectively. Further, define $\boldsymbol{\tau}_i$ as row of $\boldsymbol{\tau}$ and $\boldsymbol{\beta}_j$ as a column of $\boldsymbol{\beta}$.

\textit{Sample $\boldsymbol{\tau}_i$}: For each $i$, define $\boldsymbol{\Omega}_i = \textrm{Diag}(\omega_{i1},\dots,\omega_{i,M})$. Then compute $\mathbf{V}_i = \boldsymbol{\beta} \boldsymbol{\Omega}_i\boldsymbol{\beta}^T +\mathbf{I}_Q$ where $\mathbf{I}_Q$ is the identity matrix. Define $\mathbf{D}_i = \begin{bmatrix}
    D_{i,1}&\dots&D_{i,M}
\end{bmatrix}$. Then compute $\boldsymbol{\kappa}_i = \mathbf{D}_i - \nicefrac{D_{max}}{2} - \nicefrac{1}{2}$. Then update $\boldsymbol{\tau}_i \sim \textrm{MVN}\left(\mathbf{V}_i \boldsymbol{\beta} \boldsymbol{\kappa}_i, \mathbf{V}_i\right)$.

\textit{Sample $\boldsymbol{\beta}_j$}: For each $j$, define $\boldsymbol{\Omega}_j = \textrm{Diag}(\omega_{1,j},\dots,\omega_{N,j})$. Then compute $\mathbf{V}_j = \boldsymbol{\tau}^T \boldsymbol{\Omega}_j\boldsymbol{\tau} + \mathbf{I}_Q$. Define $\mathbf{D}_j = \begin{bmatrix}
    D_{1,j}&\dots&D_{N,j}
\end{bmatrix}$. Then compute $\boldsymbol{\kappa}_j = \mathbf{D}_j - \nicefrac{D_{max}}{2} - \nicefrac{1}{2}$. Then update $\boldsymbol{\beta}_j \sim \textrm{MVN}\left(\mathbf{V}_j \boldsymbol{\tau}^T \boldsymbol{\kappa}_j, \mathbf{V}_j\right)$.

\subsubsection{Covariates}

The updates to the model with covariates in~\Cref{sec:birds} are almost equivalent to those for the model above after replacing $\tau_{i,q}$ with covariate $X_{i,q}$ and holding it fixed. We additionally assume more structure on the prior for each $\beta_{q,j}$. The full model is
\begin{align*}
    Y_{i,j} &\sim \textrm{MaxPois}_{\mu_{i,j}}^{\ms{(D_{i,j})}} \textrm{ where } \mu_{i,j} \defeq \sum_{k=1}^K \theta_{i,k}\phi_{kj}\\
    (D_{i,j}-1) &\sim \textrm{Binomial}\Big(D_{\textrm{max}}-1,\,\, \textrm{logit}^{-1}\big(\sum_{q=1}^Q X_{i,q}\beta_{q,j}\big)\Big)\\
    \theta_{i,k} &\sim \Gamma(a^{\ms{(\theta)}},b^{\ms{(\theta)}}),\hspace{.5cm}\phi_{k,j} \sim \Gamma(a^{\ms{(\phi)}},b^{\ms{(\phi)}})\\
    \beta_{q,j} &\indsim \mathcal{N}(m_q,\,\nicefrac{1}{\lambda_q}),\hspace{.5cm} \lambda_q \iidsim \Gamma(1,1),\hspace{.5cm}m_q \iidsim \mathcal{N}(0,1)
\end{align*}
for sampling site $i \in [N]$ and species $j \in [M]$.

\textit{Order statistic augmentation}: For each $i\in [N]$ and $j\in [M]$, using~\Cref{alg:generalsample}, $$Z_{i,j} = \sum_{d=1}^{D_{i,j}} Z_{i,j,d} \textrm{ where }\mathbf{Z}_{i,j,1:D_{i,j}} \sim P_{\mu_{i,j}}(\mathbf{Z}_{i,1:D_{i,j}}\mid Z^{\ms{(D_{i,j},D_{i,j}})} = Y_{i,j})$$

\textit{Sample latent subcounts}: For each $i\in [N]$ and $j\in [M]$, using~\Cref{fact:poisthin},
$$Z_{i,j,1}',\dots Z_{i,j,K}'\sim\textrm{Mult}\left(Z_{i,j}, \left[ \frac{\theta_{i,1}\phi_{1,j}}{\mu_{i,j}},\dots,\frac{\theta_{i,K}\phi_{K,j}}{\mu_{i,j}}\right]\right)$$

\textit{Sample $\theta_{i,k}$}: For each $i \in [N]$ and $k \in [K]$, using gamma-Poisson conjugacy,
$$\theta_{i,k} \sim \Gamma\left(a^{\ms{(\theta)}} + \sum_{j=1}^M Z_{i,j,k}', b^{\ms{(\theta)}} + \sum_{j=1}^M D_{i,j}\phi_{k,j}\right)$$

\textit{Sample $\phi_{k,j}$}: For each $j \in [M]$ and $k \in [K]$, using gamma-Poisson conjugacy,
$$\phi_{k,j} \sim \Gamma\left(a^{\ms{(\phi)}} + \sum_{i=1}^N Z_{i, j, k}', b^{\ms{(\phi)}} +  \sum_{i=1}^N D_{i,j}\theta_{i,k}\right)$$

\textit{Sample $D_{i,j}$}: Define $d \in \{1,2,\dots,D_{max}\}$. For each $i\in [N]$ and $j\in [M]$, sample according to the categorical distribution 
$$P_\theta(D_{i,j} = d) \propto {D_{max} - 1  \choose d-1} \rho_{i,j}^{d-1}(1-\rho_{i,j})^{D_{max}-d} \textrm{MaxPois}_{\mu_{i,j}}^{\ms{(d)}}(Y_{i,j})$$
where $\rho_{i,j} = \textrm{logit}^{-1}\left(\sum_{q=1}^Q X_{i,q}\beta_{q,j}\right)$.

\textit{Introduce $\omega_{i,j}$}: The updates to $\beta_{j,q}$ proceed via P\'olya-gamma augmentation~\citep{polson2013bayesian}. 
For each $i\in [N]$ and $j\in [M]$, sample $\omega \sim \textrm{PG}\left(D_{max} - 1, \sum_{q=1}^Q X_{i,q}\beta_{q,j}\right)$ as a draw from the P\'olya-gamma distribution. 

For the updates below, denote $\boldsymbol{\beta} \in \mathbb{R}^{Q \times M}$ as the matrix of parameters  $\beta_{q,j}$ and $\mathbf{X}^{N\times Q}$ as the covariate matrix. Further, define $\boldsymbol{\beta}_j$ as a column of $\boldsymbol{\beta}$ and denote $\mathbf{m} = \begin{bmatrix}
    m_1&\dots&m_Q
\end{bmatrix}$.

\textit{Sample $\boldsymbol{\beta}_j$}: For each $j \in [M]$, define $\boldsymbol{\Omega}_j = \textrm{Diag}(\omega_{1,j},\dots,\omega_{N,j})$ and $\boldsymbol{\Lambda} = \textrm{Diag}(\lambda_{1},\dots,\lambda_{Q})$. Then compute $\mathbf{V}_j = \boldsymbol{\tau}^T \boldsymbol{\Omega}_j\boldsymbol{\tau} + \boldsymbol{\Lambda}$. Define $\mathbf{D}_j = \begin{bmatrix}
    D_{1,j}&\dots&D_{N,j}
\end{bmatrix}$. Then compute $\boldsymbol{\kappa}_j = \mathbf{D}_j - \nicefrac{D_{max}}{2} - \nicefrac{1}{2}$. Then update $\boldsymbol{\beta}_j \sim \textrm{MVN}\left(\mathbf{V}_j \boldsymbol{X}^T \boldsymbol{\kappa}_j + \boldsymbol{\Lambda} \mathbf{m}, \mathbf{V}_j\right)$.

\textit{Sample $\lambda_q$}: For each $q \in [Q]$, via normal-gamma conjugacy,
$$\lambda_q \sim \Gamma\left(1 + \nicefrac{M}{2}, 1 + \nicefrac{1}{2}\sum_{j=1}^M (\beta_{q,j} - m_q)^2\right)$$

\textit{Sample $m_q$}: For each $q \in [Q]$, via normal-normal conjugacy,
$$m_q \indsim \mathcal{N}\left(\frac{\lambda_q}{1 + \lambda_q M}\sum_{j=1}^M \beta_{j,q}, (1+M\lambda_q)^{-1}\right)$$

\subsection{Case Study IV: RNA-sequencing}

The full model is
\begin{align*}
    Y_{i,j} &\sim \textrm{MedNB}_{\alpha_{i,j},\, p_j}^{\ms{(D_{i,j})}} \hspace{0.25em} \textrm{ where } \alpha_{i,j} \defeq \sum_{k=1}^K \theta_{i,k}\phi_{k,j}\\
    D_{i,j} &\sim \textrm{OddBinomial}\left(D_{\textrm{max}},\,\rho_{i,j}\right)  \hspace{0.25em}\textrm{ where } \rho_{i,j} \defeq \textrm{logit}^{-1}\big(c_j + \sum_{q=1}^Q \tau_{i,q}\beta_{q,j}\big)\\
    p_j &\iidsim \textrm{Beta}(1,1),\,\,\, \theta_{i,k} \iidsim \Gamma(a^{\ms{(\theta)}},b^{\ms{(\theta)}}),\,\,\,\phi_{k,j} \iidsim \Gamma(a^{\ms{(\phi)}},b^{\ms{(\phi)}})\\
    \tau_{i,q} &\iidsim \mathcal{N}(0,\nicefrac{1}{\lambda^{\ms{(\tau)}}_i}),\,\,\,    \beta_{q,j} \iidsim \mathcal{N}(0,\nicefrac{1}{\lambda^{\ms(\beta)}_j})\\ \lambda^{\ms{(\beta)}}_j &\sim \Gamma(1,1),\,\,\, \lambda^{\ms{(\tau)}}_i \sim \Gamma(1,1),\,\,\, c_j \iidsim \mathcal{N}(c,\sigma^2)
\end{align*}
for gene $j \in [M]$ and subject $i \in [N]$. Inference in this model can proceed as follows.

\textit{Order statistic augmentation}: For each $i\in [N]$ and $j\in [M]$, using~\Cref{alg:generalsample}, $$Z_{i,j} = \sum_{d=1}^{D_{i,j}} Z_{i,j,d} \textrm{ where }\mathbf{Z}_{i,j,1:D_{i,j}} \sim P_{\alpha_{i,j},p_j}(\mathbf{Z}_{i,1:D_{i,j}}\mid Z^{\ms{(\lceil D_{i,j}/2\rceil,D_{i,j})}} = Y_{i,j})$$ 

\textit{Sample first latent subcounts}: For each $Z_{i,j}$, sample $Z_{i,j}' \sim \textrm{CRT}(Z_{i,j},D_{i,j} \alpha_{i,j})$. Using~\Cref{fact:nbmagic}, $Z_{i,j}'$ is marginally Poisson distributed with rate $D_{i,j}\log(1/(1-p_j))\alpha_{i,j}$. 

\textit{Sample second latent subcounts}: For each $i \in [N]$ and $j \in [M]$, using~\Cref{fact:poisthin},
$$Z_{i,j,1}'',\dots Z_{i,j,K}'' \sim\textrm{Mult}\left(Z_{i,j}', \left[ \frac{\theta_{i,1}\phi_{1,j}}{\alpha_{i,j}},\dots,\frac{\theta_{i,K}\phi_{K,j}}{\alpha_{i,j}}\right]\right)$$

\textit{Sample $\theta_{i,k}$}: For each $i \in [N]$ and $k \in [K]$, using gamma-Poisson conjugacy,
$$\theta_{i,k} \sim \Gamma\left(a^{\ms{(\theta)}} + \sum_{j=1}^M Z_{i,j,k}'', b^{\ms{(\theta)}} + \sum_{j=1}^M D_{i,j}\phi_{k,j}\right)$$

\textit{Sample $\phi_{k,j}$}: For each $j \in [M]$ and $k \in [K]$, using gamma-Poisson conjugacy,
$$\phi_{k,j} \sim \Gamma\left(a^{\ms{(\phi)}} + \sum_{i=1}^N Z_{i,j,k}'', b^{\ms{(\phi)}} +  \sum_{i=1}^N D_{i,j}\theta_{i,j}\right)$$

\textit{Sample $D_{i,j}$}: Define $d \in \{1,3,\dots,D_{max}\}$. For each $i\in [N]$ and $j\in [M]$, sample according to the categorical distribution 
$$P_\theta(D_{i,j} = d) \propto {\tfrac{D_{max} - 1}{2}  \choose d-1} \rho_{i,j}^{\tfrac{d-1}{2}}(1-\rho_{i,j})^{\tfrac{D_{max}-d}{2}} \textrm{MedNB}^{\ms{(d})}_{\alpha_{i,j},p_j}(Y_{i,j})$$

\textit{Introduce $\omega_{i,j}$}: The parameters $\tau_{i,q}$ and $\beta_{q,j}$ will be updated via P\'olya-gamma augmentation~\citep{polson2013bayesian}. 
For each $i \in [N]$ and $j \in [M]$, sample $\omega_{i,j} \sim \textrm{PG}\left(\tfrac{D_{max} - 1}{2}, c_j + \sum_{q=1}^Q \tau_{i,q}\beta_{q,j}\right)$ as a draw from the P\'olya-gamma distribution. 

For the updates below, denote $\boldsymbol{\tau} \in \mathbb{R}^{N \times Q}$ and $\boldsymbol{\beta} \in \mathbb{R}^{Q \times M}$ as matrices of parameters $\tau_{i,q}$ and $\beta_{q,j}$ respectively. Further, define $\boldsymbol{\tau}_i$ as row of $\boldsymbol{\tau}$ and $\boldsymbol{\beta}_j$ as a column of $\boldsymbol{\beta}$.

\textit{Sample $\boldsymbol{\tau}_i$}: For each $i$, define $\boldsymbol{\Omega}_i = \textrm{Diag}(\omega_{i1},\dots,\omega_{i,M})$. Then compute $\mathbf{V}_i = \boldsymbol{\tau} \boldsymbol{\Omega}_i\boldsymbol{\tau}^T + \lambda^{\ms{(\tau)}}_i\mathbf{I}_Q$ where $\mathbf{I}_Q$ is the identity matrix. Define $\mathbf{D}_i = \begin{bmatrix}
    D_{i,1}&\dots&D_{i,M}
\end{bmatrix}$. Then compute $\boldsymbol{\kappa}_i = \mathbf{D}_i - \nicefrac{D_{max}}{4} - \nicefrac{1}{4} - \boldsymbol{h}_i$ where $\boldsymbol{h}_i =  \begin{bmatrix}
    c_1\omega_{i,1}&\dots&c_M \omega_{i,M}
\end{bmatrix}$. Then update $\boldsymbol{\tau}_i \sim \textrm{MVN}\left(\mathbf{V}_i \boldsymbol{\beta} \boldsymbol{\kappa}_i, \mathbf{V}_i\right)$.

\textit{Sample $\boldsymbol{\beta}_j$}: For each $j$, define $\boldsymbol{\Omega}_j = \textrm{Diag}(\omega_{1,j},\dots,\omega_{N,j})$. Then compute $\mathbf{V}_j = \boldsymbol{\tau}^T \boldsymbol{\Omega}_j\boldsymbol{\tau} + \lambda^{\ms{(\beta)}}_j\mathbf{I}_Q$. Define $\mathbf{D}_j = \begin{bmatrix}
    D_{1,j}&\dots&D_{N,j}
\end{bmatrix}$. Then compute $\boldsymbol{\kappa}_j = \mathbf{D}_j - \nicefrac{D_{max}}{4} - \nicefrac{1}{4} - c_j \boldsymbol{\omega}_j$ 
where $\boldsymbol{\omega}_j = \begin{bmatrix}
   \omega_{1,j},\dots,\omega_{N,j}
\end{bmatrix}$. Then update $\boldsymbol{\beta}_j \sim \textrm{MVN}\left(\mathbf{V}_j \boldsymbol{\tau}^T \boldsymbol{\kappa}_j, \mathbf{V}_j\right)$.

\textit{Sample $\lambda^{\ms{(\tau)}}_i$}: For each $i \in [N]$, via normal-gamma conjugacy,
$$\lambda^{\ms{(\tau)}}_i \indsim \Gamma\left(1 + \nicefrac{Q}{2}, 1 + \nicefrac{1}{2}\sum_{q=1}^Q \tau_{i,q}^2\right)$$

\textit{Sample $\lambda^{\ms{(\beta)}}_j$}: For each $j \in [M]$, via normal-gamma conjugacy,
$$\lambda^{\ms{(\beta)}}_j \sim \Gamma\left(1 + \nicefrac{Q}{2}, 1 + \nicefrac{1}{2}\sum_{q=1}^Q \beta_{q,j}^2\right)$$

\textit{Sample $c_j$}: Define $\mathbf{k}_j = \nicefrac{1}{2}\mathbf{D}_j - \nicefrac{D_{max}}{4} - \nicefrac{1}{4}$, $v_j = \nicefrac{1}{\left(\nicefrac{1}{\sigma^2} + \sum_{i=1}^N \omega_{i,j}\right)}$ and $m_j= \frac{v_jc}{\sigma^2} + \sum_{i=1}^N \left(k_{i,j} - \omega_{i,j} - \sum_{q=1}^Q \tau_{i,q}\beta_{q,j}\right)$, and then sample
$c_j \sim \mathcal{N}(m_j,v_j)$.

\end{document}